\documentclass[longauth]{aa} % for the long lists of affiliations 

\usepackage[citecolor=blue]{hyperref}

\usepackage{natbib}
\usepackage{graphicx}
\usepackage{txfonts}
\usepackage{amsmath}
\usepackage{mwe}
\usepackage{booktabs}
\usepackage{longtable}

\hypersetup{colorlinks,allcolors=blue}

\begin{document}

\authorrunning{Gozaliasl et al.}
\titlerunning{Brightest Group Galaxies}

\title{COSMOS Brightest Group Galaxies - III: evolution of stellar ages}

\author{
            G.~Gozaliasl\inst{1,2}\thanks{email: ghassem.gozaliasl@aalto.fi}\texorpdfstring{\href{https://orcid.org/0000-0002-0236-919X}{\protect\includegraphics{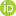}}}{}, 
            A.~Finoguenov\inst{2}\texorpdfstring{\href{https://orcid.org/0000-0002-4606-5403}{\protect\includegraphics{ORCID-iD_icon-16x16.png}}}{},
            A.~Babul\inst{3,4}\texorpdfstring{\href{https://orcid.org/0000-0003-1746-9529}{\protect\includegraphics{ORCID-iD_icon-16x16.png}}}{},
            O.~Ilbert\inst{13}\texorpdfstring{\href{https://orcid.org/0000-0002-7303-4397}{\protect\includegraphics{ORCID-iD_icon-16x16.png}}}{},
            M.~Sargent\inst{5}\texorpdfstring{\href{https://orcid.org/0000-0003-1033-9684}{\protect\includegraphics{ORCID-iD_icon-16x16.png}}}{},
            E.~Vardoulaki\inst{6}\texorpdfstring{\href{https://orcid.org/0000-0002-4437-1773}{\protect\includegraphics{ORCID-iD_icon-16x16.png}}}{},
            A.~L.~Faisst\inst{7}\texorpdfstring{\href{https://orcid.org/0000-0002-9382-9832}{\protect\includegraphics{ORCID-iD_icon-16x16.png}}}{},
            Z. Liu~ \inst{14,15,16}\texorpdfstring{\href{https://orcid.org/0000-0002-9252-114X}{\protect\includegraphics{ORCID-iD_icon-16x16.png}}}{},
            M. ~Shuntov\inst{8,9}\texorpdfstring{\href{https://orcid.org/0000-0002-7087-0701}{\protect\includegraphics{ORCID-iD_icon-16x16.png}}}{},
            O.~Cooper\inst{10,11}\texorpdfstring{\href{https://orcid.org/0000-0003-3881-1397}{\protect\includegraphics{ORCID-iD_icon-16x16.png}}}{},
            K. ~Dolag\inst{8,9}\texorpdfstring{\href{https://orcid.org/0000-0003-1750-286X}{\protect\includegraphics{ORCID-iD_icon-16x16.png}}}{},
            S.~Toft\inst{17,18}\texorpdfstring{\href{https://orcid.org/0000-0003-3631-7176}{\protect\includegraphics{ORCID-iD_icon-16x16.png}}}{}, 
            G. E.~Magdis\inst{8,9,12}\texorpdfstring{\href{https://orcid.org/0000-0002-4872-2294}{\protect\includegraphics{ORCID-iD_icon-16x16.png}}}{},
            G.~Toni\inst{19,20,21}\texorpdfstring{\href{https://orcid.org/0009-0005-3133-1157}{\protect\includegraphics{ORCID-iD_icon-16x16.png}}}{},
            B.~Mobasher\inst{22}{},
            R.~Barr\'e\inst{3}{},
            W.~Cui\inst{23,24}\texorpdfstring{\href{https://orcid.org/0000-0002-2113-4863}{\protect\includegraphics{ORCID-iD_icon-16x16.png}}}{},
            D.~Rennehan\inst{25}\texorpdfstring{\href{https://orcid.org/0000-0002-1619-8555}{\protect\includegraphics{ORCID-iD_icon-16x16.png}}}{}
             }%,
\institute{
%1
 Department of Computer Science, Aalto University, PO Box 15400, Espoo, FI-00 076, Finland
 \and
%2
Department of Physics, University of Helsinki, P. O. Box 64, FI-00014 Helsinki, Finland 
\and
% 3
Department of Physics and Astronomy, University of Victoria, BC V8X 4M6, Canada
\and
%4
Infosys Visiting Chair Professor, Indian Institute of Science, Bangalore 560012, India
\and
%5
International Space Science Institute (ISSI), Hallerstr. 6, CH-3012 Bern, Switzerland
%6
\and
Thüringer Landessternwarte, Sternwarte 5, 07778 Tautenburg, Germany
\and
%7
Caltech/IPAC, 1200 E. California Blvd. Pasadena, CA 91125, USA
\and
%8
Cosmic Dawn Center (DAWN), Denmark
%9
\and
Niels Bohr Institute, University of Copenhagen, Jagtvej 128, 2200 Copenhagen, Denmark
%10 
\and
 NSF Graduate Research Fellow
%11
\and
The University of Texas at Austin, 2515 Speedway Boulevard Stop C1400, Austin, TX 78712, USA
%12
\and
 DTU-Space, Technical University of Denmark, Elektrovej 327, DK2800 Kgs. Lyngby, Denmark
 \and
 %13
 Laboratoire d'Astrophysique de Marseille, 38 rue Frederic Joliot Curie, 13388 Marseille, France
 %14
 \and
 Kavli Institute for the Physics and Mathematics of the Universe (Kavli IPMU, WPI), UTIAS, Tokyo Institutes for Advanced Study, University of Tokyo, Chiba, 277-8583, Japan
 %15
 \and
Department of Astronomy, School of Science, The University of Tokyo, 7-3-1 Hongo, Bunkyo, Tokyo 113-0033, Japan
%16
\and 
Center for Data-Driven Discovery, Kavli IPMU (WPI), UTIAS, The University of Tokyo, Kashiwa, Chiba 277-8583, Japan
\and
%117
Universitäts-Sternwarte, Fakultät für Physik, Ludwig-Maximilians-Universität München, Scheinerstr.1, 81679 München, Germany
\and
%18
Max-Planck-Institut für Astrophysik, Karl-Schwarzschild-Straße 1, 85741 Garching, Germany
\and 
%19
Dipartimento di Fisica e Astronomia, Alma Mater Studiorum Universit\`a di Bologna, via Gobetti 93/2, 40129 Bologna, Italy
\and
%20
INAF - Osservatorio di Astrofisica e Scienza dello Spazio di Bologna, via Gobetti 93/3, 40129 Bologna, Italy
\and
%21
Zentrum für Astronomie, Universität Heidelberg, Philosophenweg 12, 69120 Heidelberg, Germany
%22
\and
Department of Physics and Astronomy, University of California Riverside, Pierce Hall, Riverside, CA 92521, USA 
%23
\and
Departamento de Física Teórica, M-8, Universidad Autónoma de Madrid, Cantoblanco E-28049, Madrid, Spain
%24
\and
Centro de Investigación Avanzada en Física Fundamental (CIAFF), Universidad Autónoma de Madrid, Cantoblanco, E-28049 Madrid, Spain
%25
\and
Center for Computational Astrophysics,
Flatiron Institute, 162 Fifth Ave, New York, NY 10010, USA
}

 \date{Received February 10, 2024; accepted ---}

  \abstract{The unique characteristics of the brightest group galaxies (BGGs) serve as a link in the evolutionary continuum between galaxies such as the Milky Way and the more massive BCGs found in dense clusters. This research investigates the evolution of stellar properties of BGGs over cosmic time ($z = 0.08-1.30$), extending from our prior studies \citep[][Paper I and Paper II]{gozaliasl2016brightest,gozaliasl2018brightest}. We analyze data of 246 BGGs selected from our X-ray galaxy group catalog within the COSMOS field, examining stellar age, mass, star formation rate (SFR), specific SFR (sSFR), and halo mass. We compare observations with the Millennium and Magneticum simulations. Additionally, we investigate whether stellar properties vary with the projected offset from the X-ray peak or the hosting halo center. We evaluated the accuracy of SED-derived stellar ages using a mock galaxy catalog, finding a mean absolute error of around one Gyr. Interestingly, observed BGG age distributions exhibit a bias towards younger intermediate ages compared to both semi-analytical models and the magneticum simulation. Our analysis of stellar age versus mass unveils intriguing trends with a positive slope, hinting at complex evolutionary pathways across redshifts. We observe a negative correlation between stellar age and SFR across all redshift ranges. We employ a cosmic-time-dependent main sequence framework to identify star-forming BGGs and find that approximately 20\% of BGGs in the local universe continue to exhibit characteristics typical of star-forming galaxies, with this proportion increasing to 50\% at $z=1.0$. Our findings support an inside-out formation scenario for BGGs, where older stellar populations reside near the X-ray peak, and younger populations at larger offsets indicate ongoing star formation. The observed distribution of stellar ages, particularly for lower-mass BGGs in the range of $10^{10-11} M_\odot$, deviates from the constant ages predicted by the models across all stellar mass ranges and redshifts. This discrepancy aligns with current models' known limitations in accurately capturing galaxies' complex star formation histories.}
\maketitle

\section{Introduction} 
\label{sect:intro} 
Over the last decade, there has been a notable shift in the focus of galaxy evolution research towards galaxy groups as key environments to investigate the impact of galaxies on their surroundings \citep[e.g.,][]{yang2007galaxy,robotham2011galaxy,overzier2016realm,Lovisari2021, Einasto2023arXiv231101868E}. Galaxy groups, which are the smallest structures in the cosmic hierarchy, generally contain a few bright galaxies, the central ones known as Brightest Group Galaxies (BGGs). Unlike extensive studies of brightest cluster galaxies (BCGs) in rich clusters \citep{webb2015star,webb2017detection,mcdonald2016star,bonaventura2017red,radovich2020amico,castignani2022star}, BGGs have received less attention in previous research.
 The evolving properties of BGGs may not mirror those of BCGs, which emphasizes the need for dedicated investigations. This study particularly focuses on the central, brightest, and most massive members of galaxy groups, exploring how they evolve and the corresponding signatures in group properties. With up to half of galaxies residing in groups locally, understanding the impact of the group environment on galaxy evolution is crucial. Positioned at the bottom of their host halo's gravitational potential well, BGGs are anticipated to undergo numerous mergers and tidal encounters with other group-member galaxies during their evolution. Observational evidence supports merger-driven size growth and recent accretion events in galaxy halos. The simulation results indicate that these interactions play a role in the stellar mass build-up of BGGs and might cause changes in their kinematic properties \citep{Gozaliasl2014a, Tacchella2019MNRAS.487.5416T, Jackson2020MNRAS.497.4262J, Jung2022MNRAS.515...22J, Oppenheimer2021Univ....7..209O,Loubser2022, Olivares2022,Lagos2022}.
 
Within the larger framework of simulations that examine the characteristics of central galaxies, numerous studies \citep{Schaye2015MNRAS.446..521S, Clauwens2018MNRAS.478.3994C, Dave2019MNRAS.486.2827D, Tacchella2019MNRAS.487.5416T, Davison2020MNRAS.497...81D, Pulsoni2020A&A...641A..60P} have focused on exploring trends across a broad range of galaxies, covering 2 to 3 orders of magnitude in stellar mass. Although these studies include BGGs among their samples, their primary emphasis is on a wide variety of galactic systems. In particular, only a limited number of studies have specifically explored the evolution of BGGs over cosmic time,  including stellar mass growth, star formation rates (SFR), and morphological changes \citep{Ragone-Figueroa2013MNRAS.436.1750R, Ragone-Figueroa2018, Ragone-Figueroa2020MNRAS.495.2436R, leburn2014MNRAS.441.1270L, Martizzi2014MNRAS.443.1500M, Remus2017Galax...5...49R, Nipoti2017MNRAS.467..661N, Pillepich2018MNRAS.475..648P, Rennehan2020MNRAS.493.4607R, Jackson2020MNRAS.497.4262J,  Henden2020MNRAS.498.2114H, Bassini2020A&A...642A..37B, Marini2021MNRAS.507.5780M}, with additional insights available in Section 4 of the recent review article by \cite{Oppenheimer2021Univ....7..209O}. More recently, according to \cite{Jung2022MNRAS.515...22J}, we observed diverse properties of BGGs in high-resolution Romulus simulations at z = 0, including early- and late-type galaxies, varied kinematics, and morphology transformations due to interactions, despite a slight underrepresentation of quenched BGGs; interactions may disrupt disc structure and quench star formation, yet gas-rich mergers also trigger star formation and rejuvenation. The AGN feedback contributes to a decrease in SFR, but rarely results in complete quenching (\citealt{Tremmel2019,Jung2022MNRAS.515...22J}; see references therein). 

For the BCGs, the common scenario is their formation at high redshift ($z\sim5$), followed by evolution via dry mergers, resulting in an increase in mass and size over cosmic time \citep[e.g.][]{collins2009early,stott2011little}.  However, observations suggest that there is a distribution of star formation rates in BCGs up to $z \sim 2$ \citep{webb2015star}.  \cite{Rennehan2020MNRAS.493.4607R} theorize that this is due to a distribution of core collapse times on each mass scale, leading to a distribution in SFRs.  Groups should therefore have a distribution of SFRs based on their assembly history. In the rarest overdense regions, the SFRs should be highest in the early Universe, especially at the high mass end of the group scale \citep{Rennehan2024manhattanarxiv}.

 Deep and wide multi-wavelength observations demonstrate a strong evolution in the stellar properties of BGGs below $ z\sim1.5 $, resulting in a BGG magnitude evolving by at least $ \sim+1 $ mag \citep{Gozaliasl2014a, Gozaliasl2014gap}, stellar mass growing by a factor of 1-2  \citep{burke2013growth,lidman2012evidence,lin2013stellar,laporte2013growth,de2007hierarchical,gozaliasl2018brightest}, a significant drop in star formation activities, and increasing the fraction of passive BGGs with decreasing redshift with a corresponding decrease in the fraction of star-forming BGGs \citep{gozaliasl2016brightest}. Recently, using the SDSS group catalog, \citet{Einasto2023arXiv231101868E} found that the total optical luminosity of the group ($L_r$) is a key factor in categorizing groups and their central galaxies. They also found that the connectivity of groups to the cosmic web is an important factor in shaping the evolution of the BGG properties \citep[and references therein]{Einasto2023arXiv231101868E}.

It has been noted that stellar metallicity and the abundance of $\mathrm{\alpha}$ elements exhibit an increase with the galaxy mass, the central velocity dispersion \citep{bernardi2003early}, or the dynamical mass \citep{chang2006colours,la2014spider}. Although earlier studies did not identify substantial correlations between stellar age and mass \citep{trager2000stellar,kuntschner2002early}, subsequent research has revealed a positive correlation between stellar age and mass, particularly at lower masses \citep{gallazzi2006ages,gallazzi2014charting,munoz2015framework}. 

Moreover, the stellar age, metallicity, [$\mathrm{\alpha}/$ Fe], and color of galaxies correlate with halo mass, local galaxy number density, and the distance from the center of galaxy groups or clusters \citep{thomas2005epochs,bernardi2003early,clemens2009history,cooper2010galaxy,pasquali2010ages,la2014spider}. Galaxies in low-density environments tend to be younger than those in high-density environments by approximately 1-2 Gyr \citep{trager2000stellar,kuntschner2002early,terlevich2002catalogue,thomas2005epochs,clemens2009history}. 

\cite{pasquali2010ages} discovered that central BGGs tend to be younger and have lower metal content compared to satellites with similar stellar mass, with these differences becoming more evident at lower stellar masses. \cite{la2014spider} also suggested that velocity dispersion, which serves as a proxy for mass,  influences the stellar population characteristics of central early-type galaxies (ETGs). In particular, central ETGs situated in massive halos (halo masses exceeding $10^{12.5} M_\odot$) show younger stellar ages, reduced $\mathrm{\alpha}/\mathrm{Fe}$ ratios, and increased internal reddening in comparison to those located in low-mass halos (halo masses below $10^{12.5} M_\odot$). \cite{Edwards_2019} investigated the ages of 23 local BCGs and their displacement from the cluster center using integral spectroscopy. The findings revealed a swift and early formation of BCG cores. The outer regions and intercluster light (ICL) were formed more recently through minor mergers, which resulted in younger ages and lower metallicities compared to BCG cores. Velocity-dispersion profiles typically show either a rising or flat trend, indicating that recent star formation is a minor component associated with the cool-core of the hosting cluster. When examining studies on BGG stellar populations, there is a significant lack of comprehensive research into the total stellar population of BGGs.

Previous studies have predominantly examined the stellar characteristics of central galaxies in dense clusters, varied samples of groups and clusters, or incomplete and frequently biased samples of low-redshift groups. 

To establish our sample, we used images from more than 40 filters by combining deep ($AB\sim 25-26$) and multiwavelength ($0.25\mu\mathrm{}{m}-24\mu\mathrm{m}$) data along with a comprehensive spectroscopic dataset in COSMOS \citep{scoville2007cosmic,laigle2016cosmos2015,casey2023}. Our methodology integrates all available XMM-Newton and Chandra observations, allowing us to define the group center with a remarkable precision of approximately 5 arcseconds using Chandra imaging. For group identification, we used various methods including red-sequence analysis and galaxy overdensity using both spectroscopic and photometric redshifts, in addition to X-ray selection. This diverse approach ensures a robust selection of galaxy groups. Importantly, our COSMOS dataset is the most comprehensive data among all astronomical surveys, significantly improving the reliability and depth of our analyzes \citep[see][]{gozaliasl2019,gozaliasl2020,Bahar2024}.

Our sample spans a halo mass range from $M_{200}\sim 10^{12.85} $ to $ 10^{14.2} M_{\odot} $ over approximately 9 billion years ($z\sim 0.1-1.5$), covering a substantial portion of the age of the Universe. This paper is the third in a series investigating the evolution of the BGG stellar properties through a comprehensive analysis that combines deep observations and numerical simulations. In \citet[][Paper I]{gozaliasl2016brightest}, we studied the distribution and evolution of the relations $M_*$, sSFR, $M_*-M_{200}$, and $SFR-M_*$. In particular, we found evidence for significant growth of stellar mass by a factor of $\sim2$ and the existence of a distinct population of BGGs with $M_*\sim10^{10.5} M_\odot$ that may be actively star-forming and young. However, existing models struggle to accurately predict the presence of this population. In \citet[][Paper II]{gozaliasl2018brightest}, we investigated the distribution and evolution of the contribution of BGGs to the total baryon of groups within $R_{200}$ ($f^{BGG}_{b,200}$) and explored how the distribution for low mass groups $M_{200} < 10^{13.5}$ is different from that of massive groups and how $f^{BGG}_{b,200}$ grows by a factor of $\sim2.5$. Another aspect of this study was the influence of the evolution of halo mass and ongoing star formation on the accumulation of BGG mass, assuming a constant SFR since $z=1$. Additionally, we analyzed the stellar-to-halo mass relationship and compared our results with various simulations.

Subsequently, \citep{gozaliasl2019} measured both the observed and modeled intrinsic scatter in the stellar mass of BGGs at fixed halo masses along with its evolutionary patterns. Furthermore, we analyzed the displacement of BGGs from the group's X-ray peak employing deep Chandra imaging in COSMOS. Our results in \citep{gozaliasl2020} disclosed that BGGs in low-mass groups with $M_{200} < 10^{13.5}$ demonstrate significantly high proper velocities in comparison to the group velocity dispersion, suggesting dynamical instability within these systems \citep{gozaliasl2019, gozaliasl2020}.

The primary aim of this research is to examine the distribution and evolution of the stellar ages of BGGs. We analyze the connections between stellar age, stellar mass, SFR, and halo mass for a sample of 246 BGGs, selected either by spectroscopic (80\%) or photometric redshift (20\%) from our updated, well-regulated dataset of X-ray selected groups \citep{gozaliasl2019}. Moreover, we will contrast our observational data with the prediction of the semi-analytic model (SAM) by \citet[][hereafter H15]{henriques2015galaxy} and employ the magneticum hydrodynamical simulation \citep{dolag2016}.

The structure of this paper is as follows. Section 2 details the data, sample definition, and methodology. Section 3 covers results including the distribution of BGG stellar age, their evolution over cosmic time, stellar age in relation to stellar mass, SFR distribution, and how stellar age correlates with halo mass. Section 4 presents our discussion. Section 5 summarizes the key findings and conclusions.
Throughout our analysis, we use the cosmological parameters for the dark energy density and the total dark matter density specified as: $(\Omega_{\Lambda}, \Omega_{M}) = (0.7, 0.3)$, with the Hubble constant defined as 70 km s$^{-1}$ Mpc$^{-1}$. We adjusted the data from the simulations if the cosmological parameters differed. \cite{chabrier2003} IMF is used in all data sets, including observations and simulations. This uniform application of the IMF ensures consistent estimation of stellar masses, facilitating accurate comparisons between observations and simulations. This consistency is essential for assessing galaxy ages, given the fundamental role of stellar mass in determining their evolutionary stage.

\section{Sample selection and methodology}
\subsection{Data and samples of BGGs}
\subsubsection{COSMOS galaxy groups and samples' definition}
COSMOS stands out as a unique survey, benefitting from the synergy of deep ($AB\sim 25-26$) and multi-wavelength ($0.25 \mu m-24 \mu m$) data \citep{scoville2007cosmic}. In this study, we used the catalogs provided by \cite{laigle2016cosmos2015}, which provide valuable information on photometric redshifts and the physical properties of galaxies.

\begin{figure}
\includegraphics[width=0.458\textwidth]{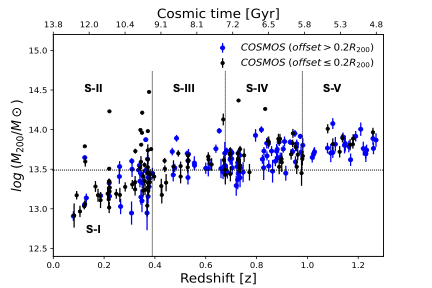}
\caption{The halo mass of X-ray galaxy groups ($M_{200}$) as a function of redshift. Galaxy groups have been identified in the COSMOS field. The color shows the split of the data based on the ratio of the offset of the BCG/BGG position from the X-ray center of halos to the $R_{200}$ radius.}\label{mhz}
\end{figure}

Our investigation relies on a catalog of 246 X-ray groups of galaxies detected within a 2 $deg^{2}$ area of the COSMOS field \citep{gozaliasl2019}. The groups span a redshift range of $0.08\leq z < 1.5$ and exhibit a mass range of $M_{200}=8\times10^{12}-3\times10^{14}\;M_{\odot}$. Although high-mass systems within this mass range may straddle the boundary between groups and clusters, for this study, we categorize them as groups. Figure \ref{mhz} illustrates the halo mass of the groups ($M_{200}$) within $R_{200}$, defined as the radius delineating a sphere with a mean interior density 200 times the critical density as a function of redshift. To ensure accurate and reliable mass measurements, we employ a robust weak lensing calibration method \citep{leauthaud2009weak}, as detailed in \cite{gozaliasl2019}, to infer the values $M_{200}$ for our galaxy groups. A substantial portion of the groups ($71\%$) fall within a halo mass range of $13.50 < \log\left(\frac{M_{200}}{M_{\odot}}\right) \le 14.02$. This particular mass regime corresponds precisely to the transition zone between massive clusters and low-mass groups, which constitutes the primary focus of this study. 
In line with the approach taken in \cite{gozaliasl2016brightest, gozaliasl2018brightest}, we have categorized galaxy groups into five subsamples (referred to as S-I to S-V) based on the halo mass and redshift plane, as shown in Fig. \ref{mhz}. The characteristics of these sub-samples are presented in Table \ref{tab1:subsample}. For further details on the properties of the groups, as well as their detection and physical properties, we refer to \cite{gozaliasl2019}.

Subsequently, we shall use the designations S-I through S-V to denote these subsamples. Within these subsamples, we focus on the most massive galaxies within the characteristic radius of groups known as $R_{200}$. Most massive galaxies are typically central most luminous group galaxies, so we often refer to them as BGGs.

Drawing inspiration from a body of work that encompasses recent investigations into dynamically evolved galaxy groups, such as fossil galaxy groups \citep{raouf2014ultimate,raouf2016evolution,khosroshahi2017galaxy,gozaliasl2019}, we undertake a comprehensive analysis of the spatial offset of BGGs from the centroid of the group halo, defined by the peak of the X-ray emission. This offset tends to be minimal in dynamically evolved groups, but assumes larger values for younger or less dynamically evolved groups.

As elaborated in \citep{gozaliasl2019}, we quantify this offset by assessing the projected separation between the position of the BGGs and the peak of the group's X-ray emission ($d_{sep}$). The BGG offset is characterized as a parameter corresponding to the ratio of this angular separation to the group's $R_{200}$, expressed as $d_{sep}/R_{200}$.

Using this parameter, we establish a distinction between two categories of BGGs: those positioned close to the peak of the X-ray emission (e.g., offset $\leq 0.2 R_{200}$) and those located outside the core of the group (e.g., offset $\geq 0.2 R_{200}$). It is important to acknowledge that the selection of the threshold $0.2R_{200}$ is somewhat arbitrary; however, it has proven effective in segregating BGGs within the core of groups, typically characterized by dynamical relaxation, from those outside the core. This threshold was chosen considering the resolution of the X-ray survey and the average distance within the sample, allowing for a meaningful differentiation between these two categories. 
In essence, we conduct a comprehensive analysis of the data extracted from five distinct sub-samples of BGGs (denoted as S-I, S-II, S-III, S-IV, and S-V), which are categorized as follows:
\begin{enumerate}

\item Minimal offset: The central dominant BGGs are characterized by a minimal offset from the group center or X-ray peak position (offset $\leq 0.2R_{200}$).

\item Displacement indication: BGGs with a significant offset indicate their displacement from the group's core ($0.2R_{200} <$ offset $\leq R_{200}$).

\item Comprehensive assessment: A comprehensive assessment encompassing all BGGs, regardless of their offset.    

\end{enumerate}

Fig. \ref{mhz} illustrates the halo mass-redshift relationship for groups hosting centrally dominant and offset BGGs, represented by black and blue data points, respectively. According to the definition outlined above, 48\% of BGGs are classified as central, while 52\% are considered offset.  
\subsubsection{Semi-Analytic Models and Magneticum Simulation}

To facilitate a robust comparison of the observations with model predictions, we contrast our findings with the predictions generated by SAM developed by \citet[][hereafter H15]{henriques2015galaxy} and the Magneticum hydrodynamical simulation \citep{dolag2016}.

A comprehensive summary of the key characteristics of this SAM can be found in previous work \citep{gozaliasl2018brightest}. For an in-depth understanding of this model, we refer interested readers to the original publications of \cite{henriques2013simulations} and \cite{henriques2015galaxy}.

For our analysis, we used data from the H15 SAM. Specifically, we randomly select 5000 BCGs within halos falling within specified ranges of halo mass and redshift, corresponding to categories S-I through S-V. We then compare the predictions of this model with our observational findings in Section 3.
 
While the Magneticum hydrodynamical simulation \citep{dolag2016} consistently predicts the stellar mass function with observational data across various redshift intervals, the simulated stellar masses of BCGs tend to be notably higher than their observed counterparts (however, see \citealt{rennehan2023three} for a recent counterexample). This discrepancy can be attributed, in part, to the need for more efficient AGN (Active Galactic Nuclei) feedback in the simulations. A discrepancy also arises because the estimates of stellar masses for BCGs in the simulations encompass the BCG and ICL components. Distinguishing between these two components is a complex undertaking. Based on a dynamic separation of these two stellar components, simulations suggest that the stellar mass associated with the BCG accounts for approximately 45\% of the total stellar mass, encompassing both BCG and ICL, as indicated by \cite{dolag2010dynamical} and \cite{Remus2017Galax...5...49R}. However, it should be noted that the observed contribution of the ICL to the light emanating from the BCG is greater and is subject to variations depending on the magnitude threshold applied \citep{cui2013characterizing}. Therefore, for our analysis, we assume that the observed stellar mass fraction of the simulated BCG corresponds to approximately 70\% of the total stellar mass (comprising both BCG and ICL), as inferred from the simulations. A short summary of the simulation properties can be found in \cite{gozaliasl2019}. 
\begin{table*}[ht]
       \caption{Characteristics of galaxy group subsamples selected using the COSMOS 2019 X-ray group catalog \citep{gozaliasl2019}.} 
           \centering
    \begin{tabular}{lcccccc}
    \hline
        Subsample & $ \log(M_{200}/M_{\odot})_{range} $ & $N_{groups}$ & $z_{median}$ & Med. cosmic time [Gyr] & $z_{range}$ \\ 
        \hline
        \midrule
        S-I   & [12.85, 13.50] & 59 & 0.31 & 10.16 & [0.08, 0.40] \\ 
        S-II  & [13.50, 14.10] & 22 & 0.34 & 9.89  & [0.08, 0.40] \\ 
        S-III & [13.50, 14.10] & 38 & 0.61 & 7.84  & [0.40, 0.70] \\ 
        S-IV  & [13.50, 14.10] & 59 & 0.88 & 6.37  & [0.70, 1.0]  \\ 
        S-V   & [13.50, 14.10] & 29 & 1.16 & 5.26  & [1.0, 1.30]  \\ 
        \hline
    \end{tabular}
    \label{tab1:subsample}
\end{table*}

\begin{table*}
\caption{Characteristics of the spectroscopic redshift samples.  Only the most secure spectroscopic redshifts (those with a flag between 3 and 4) are considered. The redshift range, median redshift, and apparent magnitude in the band are provided for each selected sample.}

\begin{tabular}{llllll}
\hline
Spectroscopic Survey Reference	& Instrument/telescope & $ N_b$ & $z_{med}$ & $z_{range}$ & $i^{+}_{med}$\\
\hline
zCOSMOS-bright \citep{lilly2007zcosmos}&	VIMOS/VLT	&8608&	0.48&	[0.02, 1.19]&	21.6\\	
\cite{comparat20150}&	FORS2/VLT&	788	&0.89&	[0.07, 3.65]&	22.6	\\
P. Capak et al. (in preparation),\cite{kartaltepe2010multiwavelength}&	DEIMOS/Keck II	& 2022 &	 0.93& 	[0.02, 5.87]&	23.2\\	
\cite{roseboom2012fmos} &	FMOS/Subaru	&26&	1.21&	[0.82, 1.50]	&22.5	\\	
\cite{onodera2012deep}&	MOIRCS/Subaru	&10	&1.41&	[1.24, 2.09]	&23.9	\\	
FMOS-COSMOS \citep{silverman2015fmos}&	FMOS/Subaru	&178	&1.56 &	[1.34, 1.73]&23.5\\
WFC3-grism \citep{krogager2014spectroscopic} &	WFC3/HST&	11	&2.03	&[1.88, 2.54]	&25.1\\	
zCOSMOS deep (S. Lilly et al. 2016, in preparation)&	VIMOS/VLT&	767&	2.11	&[1.50, 2.50]&	23.8\\
MOSDEF \citep{kriek2015mosfire}	&MOSFIRE/Keck I	&80&	2.15	&[0.80, 3.71]&	24.2\\	
M. Stockmann et al. (in preparation), \cite{zabl2015emission}	&XSHOOTER/VLT&	14&	2.19	&[1.98, 2.48]	&22.2\\
VUDS \citep{le2015vimos}&	VIMOS/VLT&	998	&2.70&	[0.10, 4.93]&	24.6\\
DEIMOS 10K \citep{hasinger2018} & DEIMOS/Keck II &  6617 & 1 \& 4 & [0.00, 6.00]& 23\\
\hline
\end{tabular}\label{tab1:specz}
\end{table*}
\subsection{Method and physical parameter measurements }
\subsubsection{ Photometric redshift measurement and SED fitting with \text{\small Le Phare} code}
Approximately $81\%$ of BGGs have a spectroscopic redshift, derived from extensive observations of COSMOS field as listed in Table \ref{tab1:specz}.
 When we have galaxies with no spectroscopic redshifts, we use photometric redshifts presented in the COSMOS2015 catalog \citep{laigle2016cosmos2015}. For some of our BGGs that have not been determined as photos in COSMOS2015, we apply the earlier COSMOS photoz catalogs presented in \cite{2009ApJ...690.1236I, mccracken2012ultravista, ilbert2013mass}. All of these catalogs use the SED fitting method and apply the \text{\small Le Phare} code to measure the photometric redshifts and stellar masses with the $\chi^2$ template fitting method. The details of the method can be found in \cite{2009ApJ...690.1236I, ilbert2013mass}.

We selected the COSMOS2015 catalog as presented by \citep{laigle2016cosmos2015} due to its consistent data set throughout our research, similar to \citet[Paper I \& II]{gozaliasl2016brightest,gozaliasl2018brightest}. More than $80\%$ of our BGGs possess spectroscopic redshifts. Compared with COSMOS2020 \citep{Weaver2022}, no notable differences were identified for our BGGs, including photoz measurements where $99\%$ matched or other properties of the galaxy. The differences pertain to low-mass galaxies, which are not included in this study. Moreover, as indicated in the research by \cite{toni2024}, some bright and very extended BGGs at low and intermediate redshifts may lack photometry in the COSMOS2020 catalog. Thus, to study BGGs in the range $z\sim 0.1-1.5$ and maintain consistency throughout our investigations, we opted for the COSMOS2015 catalog.

The COSMOS2015 catalog, derived from the UltraVISTA-DR2 survey, offers photometric redshifts and stellar masses for more than 500,000 sources, using $YJHK_s$ data for object detection and supplementing with 31-band data for improved photometric redshift estimation \citep{laigle2016cosmos2015}. Remarkably, it achieves a photometric redshift accuracy of $\sigma_{\Delta z/(1+z_s)} = 0.007$ with a catastrophic failure fraction of $\eta = 0.5\%$ compared to secure spectroscopic redshifts such as zCOSMOS-bright \citep{laigle2016cosmos2015}. Covering $0.46\; \mathrm{deg}^2$ Ultra-deep and $0.92\; \mathrm{deg}^2$ deep UltraVISTA surveys, it reaches $90\%$ completeness for stellar masses up to $10^{10}M_{\odot}$ at $z = 4.0$ \citep{laigle2016cosmos2015}. To ensure completeness, bluer objects and those at higher redshifts are detected using a $\chi^2$ sum of $YJHK_s$ and Subaru SUPRIME-CAM broad band $z^{++}$  images \citep{laigle2016cosmos2015}. For comprehensive identification of groups across the entire $\sim2\;\mathrm{deg}^2$ COSMOS field, we utilized both the COSMOS2015 catalog and the i-band selected v.2 catalog of photometric redshifts by \cite{2009ApJ...690.1236I} and \cite{mccracken2012ultravista}, along with photometric redshifts of X-ray sources from the catalog by \cite{marchesi2016chandra}.

\subsubsection{Physical Properties: absolute magnitudes, stellar masses, SFR and stellar age}

The calculation of the correction term $k$ \citep{1968ApJ...154...21O} is based on the selection of the most appropriate template, a critical factor that contributes to the systematic error in determining the absolute magnitudes and color of the rest of the frame. To address this, \cite{laigle2016cosmos2015} follow the methodology detailed in Appendix A of \cite{ilbert2006accurate}. To mitigate the uncertainties introduced by \textit{k}-correction, the rest-frame luminosity is calculated at a specific wavelength $\lambda$ based on the apparent magnitude $m_{obs}$ observed recorded in the filter closest to $\lambda (1 + z)$. This approach reduces the reliance on the best-fit SED to determine absolute magnitude, while making it more sensitive to observational anomalies that affect $m_{obs}$. Consequently,  \cite{laigle2016cosmos2015} limits the code's consideration to broad bands for $m_{obs}$ and bands with a systematic offset of less than 0.1 mags as derived from the photometric redshift.

Stellar mass calculations are performed using {\sc LePhare}, adopting the same methodology as described in \cite{Ilbert2015}. These estimates rely on a library of synthetic spectra generated via the Stellar Population Synthesis (SPS) model by \cite{bruzual2003stellar}, assuming an Initial Mass Function (IMF) following \cite{chabrier2003}. Two distinct star formation histories (SFH) are explored: 1) exponentially declining SFH,
\begin{equation}\label{eq:sfr_declining}
\mathrm{SFH} = \exp\left(-\frac{t}{\tau}\right),
\end{equation}

and 2) delayed SFH, 
\begin{equation}\label{eq:sfr_delayed}
\mathrm{SFH} = \tau^{-2} t \exp\left(-\frac{t}{\tau}\right)
\end{equation}

Here $\mathrm{\tau}$ is the time scale, which ranges from 0.1 Gyr to 30 Gyr. Additionally,   two metallicities, one solar and one half-solar, are considered. The emission lines are incorporated following the approach in \cite{2009ApJ...690.1236I}, and \cite{laigle2016cosmos2015} introduce two attenuation curves for consideration: the starburst curve of \cite{clasetti2000} and a curve with a slope of $\lambda^{-0.9}$ \citep[Appendix A of][]{2013A&A...558A..67A}. The permissible range of values of $E(B-V)$ extends to 0.7. The assignment of the mass is based on the median of the marginalized probability distribution function (PDF).
For precision and to address uncertainties associated with SFR estimates derived from template fitting, see \cite{Ilbert2015,2015ApJ...801...80L}.

To determine the mass-weighted stellar age, we follow the method described in Section 6 of the study by \cite{wuyts2011star}. In our analysis, age is a critical parameter, and it's essential to note that two commonly used definitions of galaxy age exist in the literature, the first one being defined as the time since the onset of star formation,
\begin{equation}
\mathrm{age} = t_{\mathrm{obs}} - t_{\mathrm{form}},
\label{eq:age}
\end{equation}

and the second one is a measure of the age of the bulk of the stars,
\begin{eqnarray}
\mathrm{Age}_w = \frac{\int_{t_{\mathrm{form}}}^{t_{\mathrm{obs}}} SFR(t - t_{\mathrm{form}}) (t_{\mathrm{obs}} - t) \, \mathrm{d}t}{\int_{t_{\mathrm{form}}}^{t_{\mathrm{obs}}} SFR(t - t_{\mathrm{form}}) \, \mathrm{d}t}.
\label{eq:agew}
\end{eqnarray}

For exponentially declining star formation histories, $\mathrm{Age}_w$ ranges between 0.5 $\times$ age ($\mathrm{\tau} = \infty$) and 1 $\times$ age ($\mathrm{\tau} = 0$). Depending on the template model fitted for a galaxy, we employ either an exponentially declining SFH given by Eq. \ref{eq:sfr_declining} or a delayed SFH represented by Eq. \ref{eq:sfr_delayed} in Eq. \ref{eq:agew}. In the study by \cite{laigle2016cosmos2015}, 12 template models from \cite{bruzual2003stellar} were utilized. Among these, 8 models adopted the exponentially declining SFH, while the remaining 4 models employed the delayed SFH.

 Following \cite{wuyts2011star}, throughout this paper, we refer to the weighted age as $Age_w$.

 \subsection{Modeling scaling relations}\label{sec:linmix}
We employ the \texttt{ linmix3} routine proposed by \cite{kelly2007some} for fitting, which performs a Bayesian analysis. The \texttt{linmix} routine employs three regression parameter equations to calculate line fits. Let the independent variable be denoted as $\xi$ and the dependent variable as $\eta$; $\xi$ and $\eta$ are also termed the 'covariate' and the 'response,' respectively. We assume that $\xi$ forms a random vector of $n$ data points drawn from a specific probability distribution. The dependent variable $\eta$ is based on $\xi$ according to the conventional additive model.

\begin{equation}\label{eq:linmix_eta}
    \eta_i = \alpha + \beta \xi_i + \epsilon,
\end{equation}

Here, $(\alpha, \beta)$ represent regression coefficients, where $\alpha$ is the intercept, $\beta$ is the slope, and $\epsilon$ accounts for the intrinsic random scatter in $\eta_i$ regarding the regression relationship. We assume $\epsilon$ to follow a normal distribution with zero mean and variance $\sigma^2$. Although we do not observe the actual values of $(\xi, \eta)$, we instead observe measurable values $(x, y)$, subject to errors. These measured values are related to the actual values through the following.

\begin{equation}\label{eq:linmix x}
    x_i = \xi_i + x_{\mathrm{err, i}},
\end{equation}

where $x_i$ represents the data points with associated errors $x_{\mathrm{err, i}}$.

\begin{equation}\label{eq:linmix y}
    y_i = \eta_i + y_{\mathrm{err, i}},
\end{equation}

Here, $y_{\mathrm{err, i}}$ denotes the error in $y_i$ (both data), and

\begin{equation}\label{eq:linmix sigma}
    \sigma^2 = \mathrm{Var}(\epsilon),
\end{equation}

reflects the variance. For a detailed understanding of the Bayesian method's application to account for measurement errors in linear regression of astronomical data, refer to \cite{kelly2007some}. In our analysis, we utilize the \texttt{linmix} equation \ref{eq:linmix_eta}. The best-fit parameters, detailed in subsequent tables, are derived by fitting Eq. \ref{eq:linmix_eta} to $\ln(y)$ versus $\ln(x)$, while the corresponding figures are presented using a base-10 logarithmic scale ($\log_{10}$).If necessary, we also apply normalization. As an example, in Section \ref{sec:age-mass-relation}, where we investigate the fitting of $\ln(\mathrm{Age}_{w})$ versus $\ln(M_{*})$, the stellar mass is normalized to $10^{11} M_{\odot}$, a typical stellar mass of BGGs in low-mass halos (S-I).

\begin{figure}[ht!]
  \centering
  \includegraphics[width=0.45\textwidth]{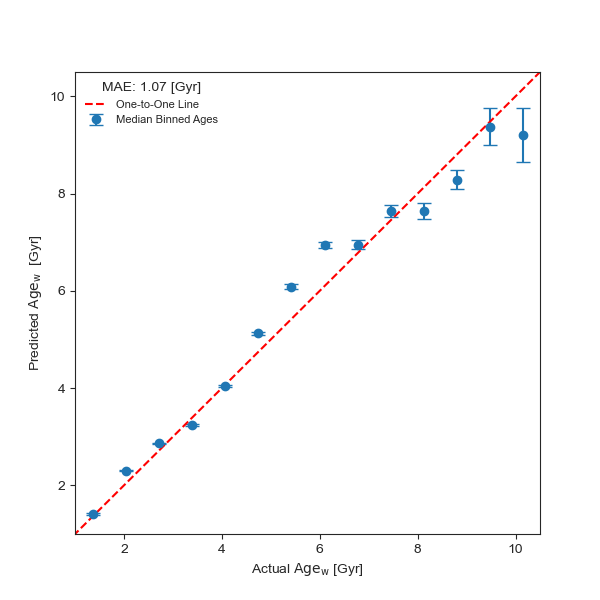}
  \caption[The one-to-one comparison of true age and predicted age.]{The one-to-one comparison of the true age and the predicted age using SED fitting techniques shows moderate agreement between the two ages for all populations of galaxies spanning cosmic time.}
  \label{Fig:one-one-age}
\end{figure}
\section{ Results} \label{age}
\subsection{Stellar age distribution}   \label{age_dist}

\begin{figure*}[ht!] 
\centering
  \includegraphics[width=0.45\textwidth]{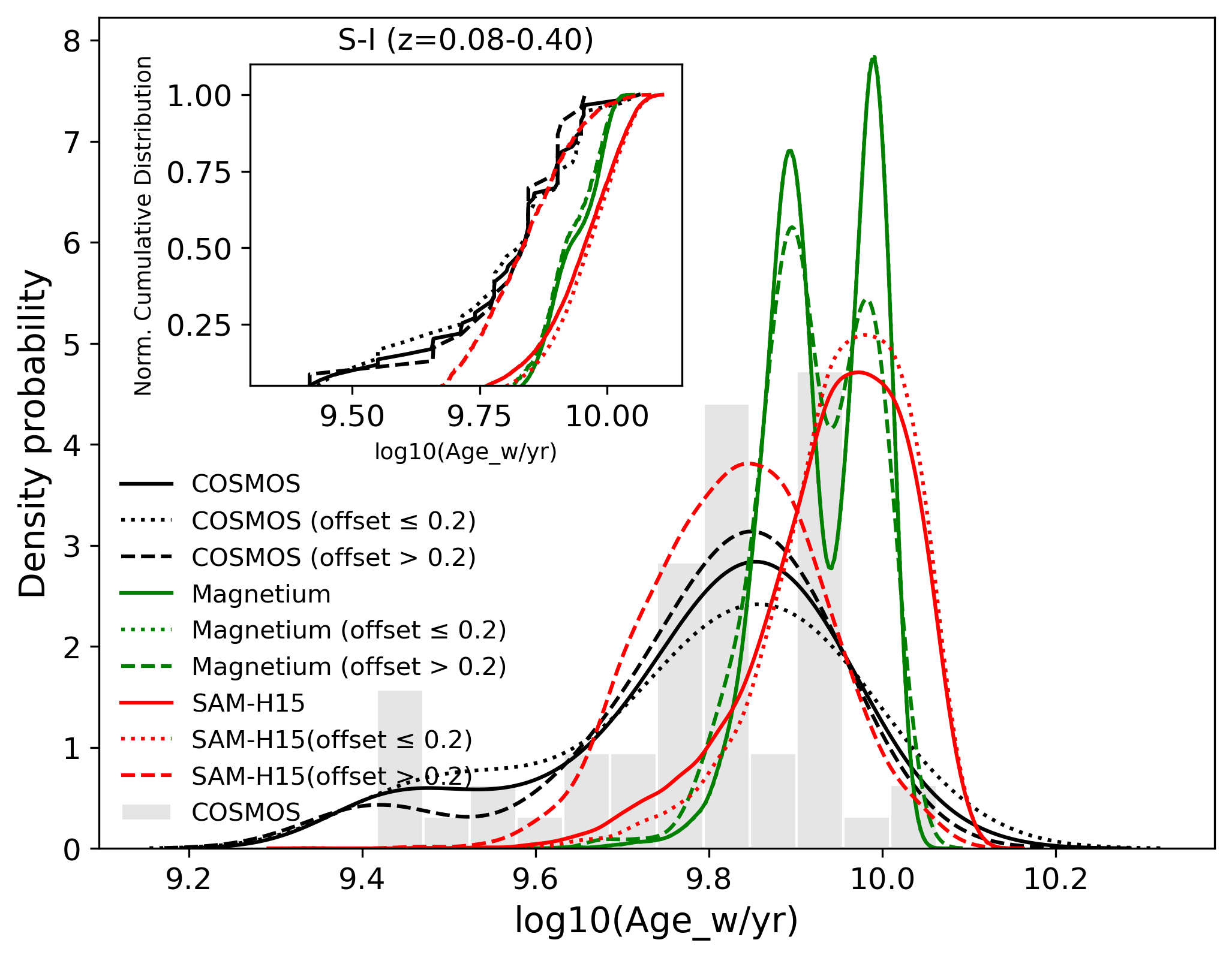}
 \includegraphics[width=0.45\textwidth]{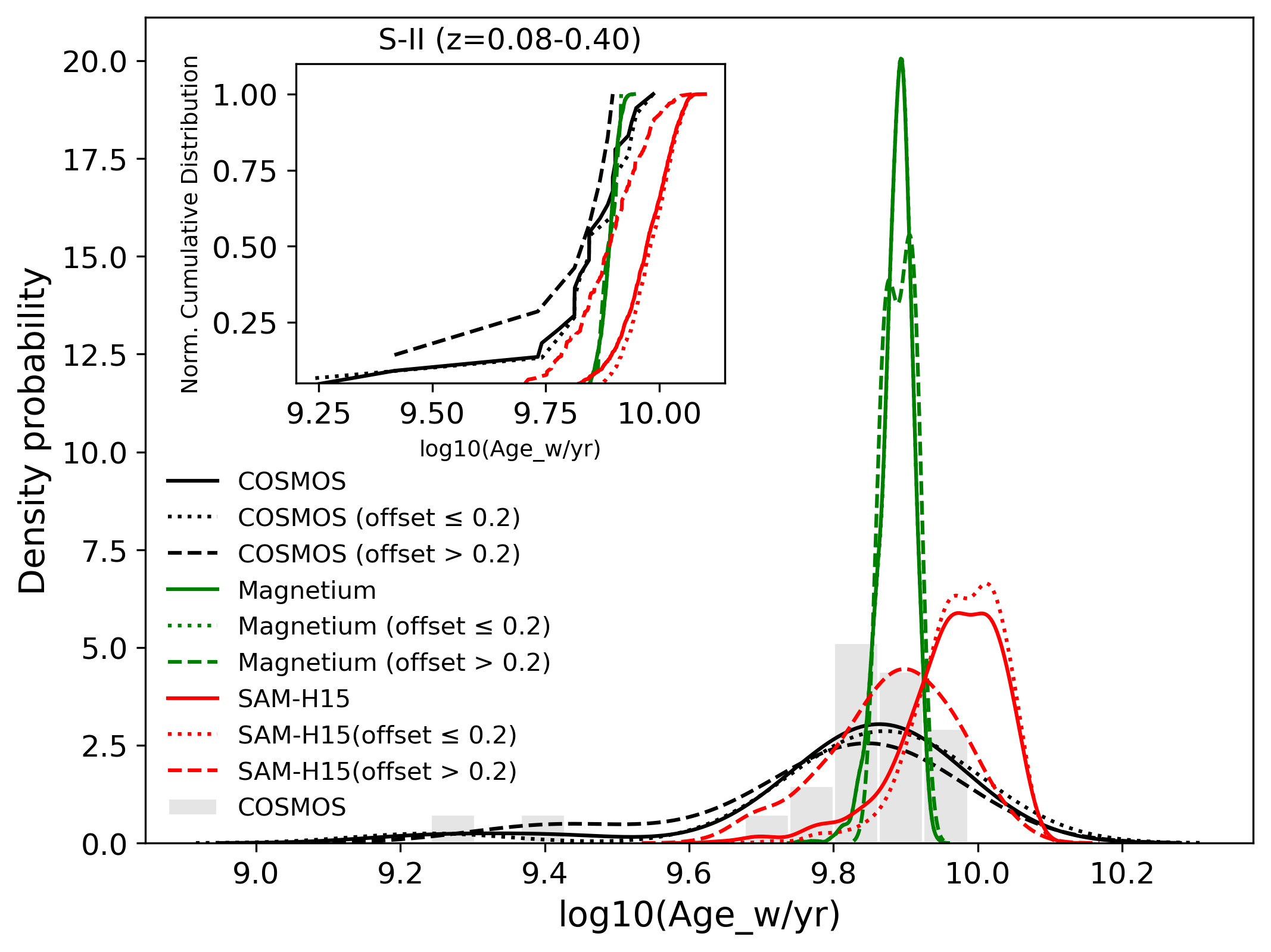}
  \includegraphics[width=0.45\textwidth]{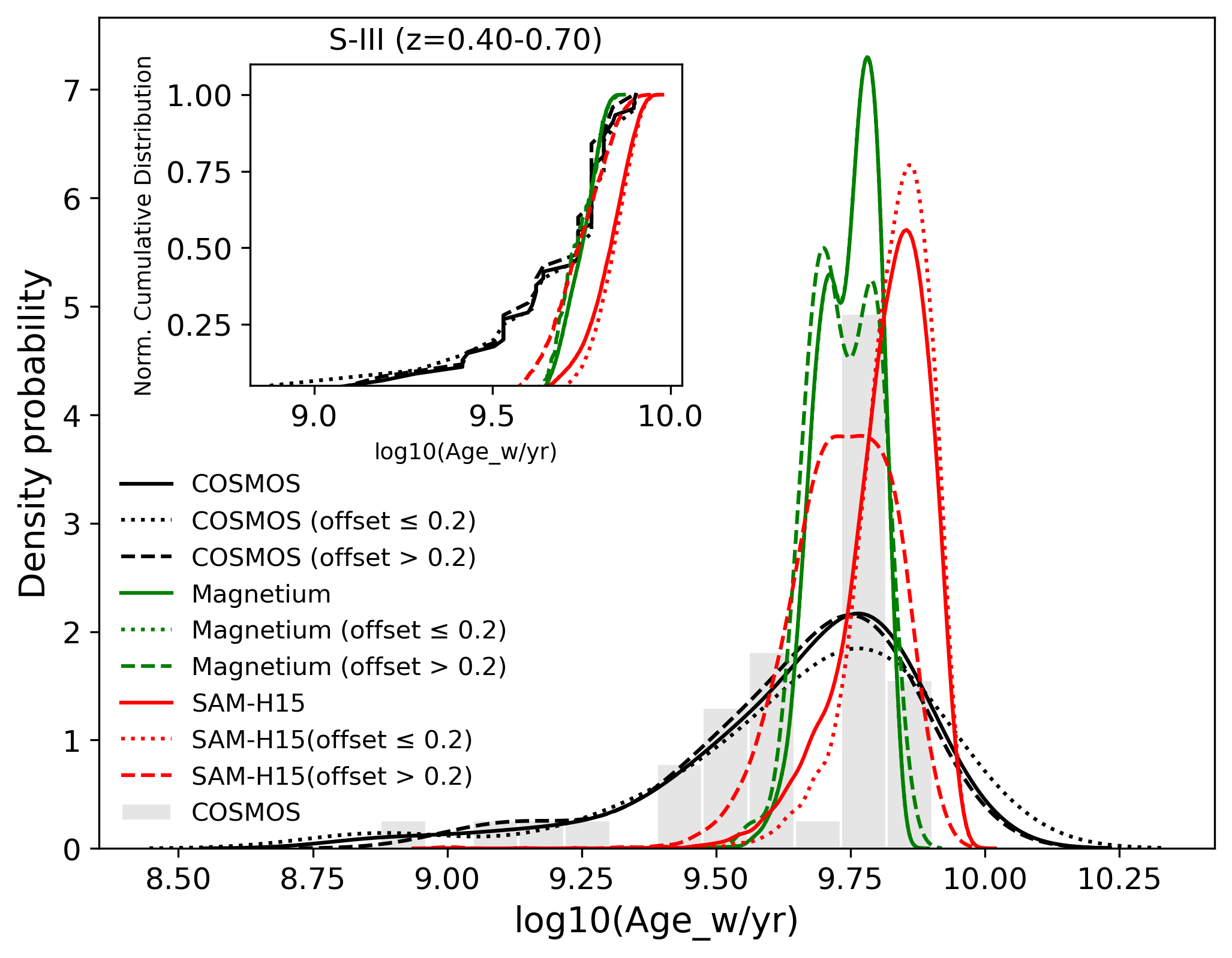}
  \includegraphics[width=0.45\textwidth]{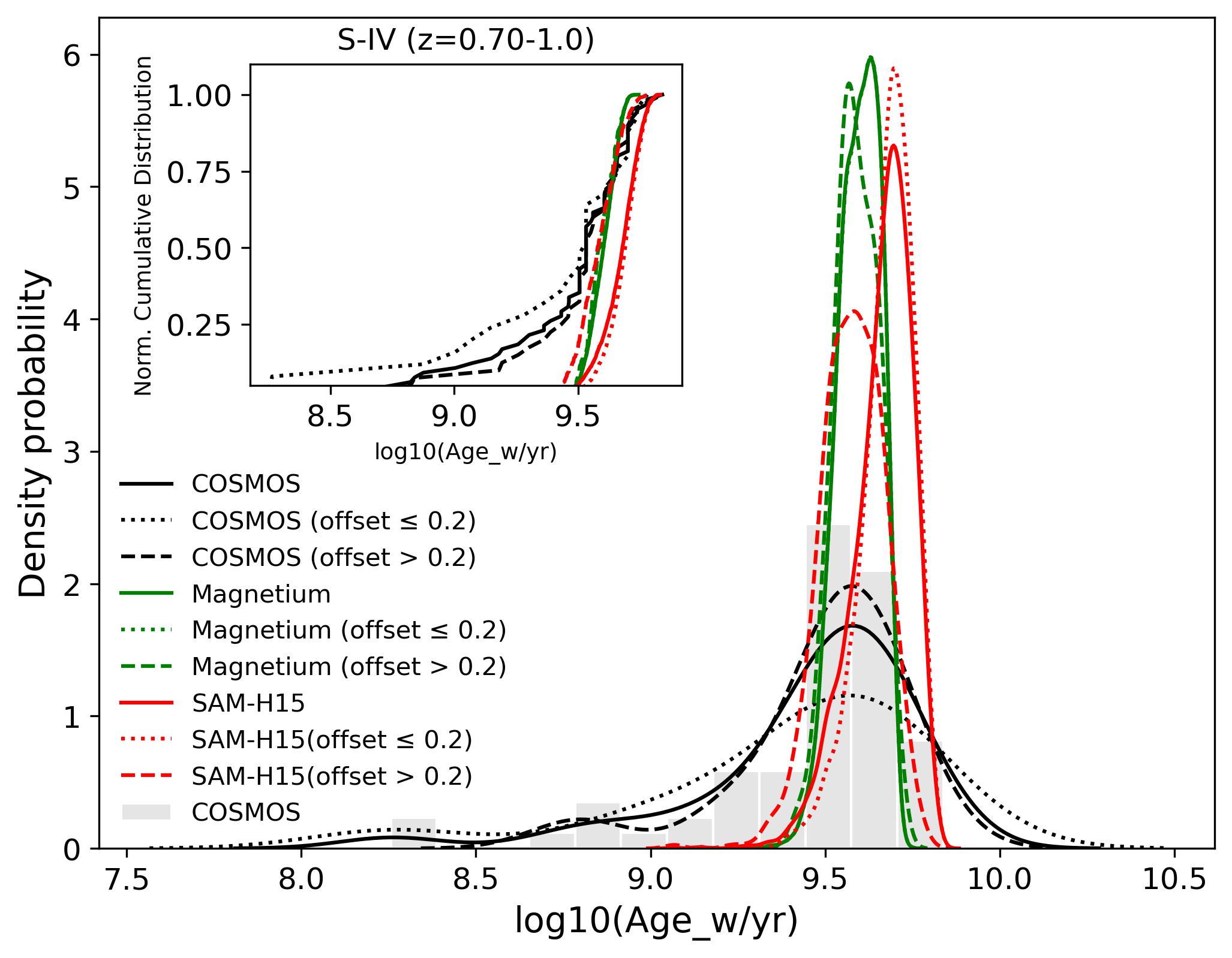}
  \includegraphics[width=0.45\textwidth]{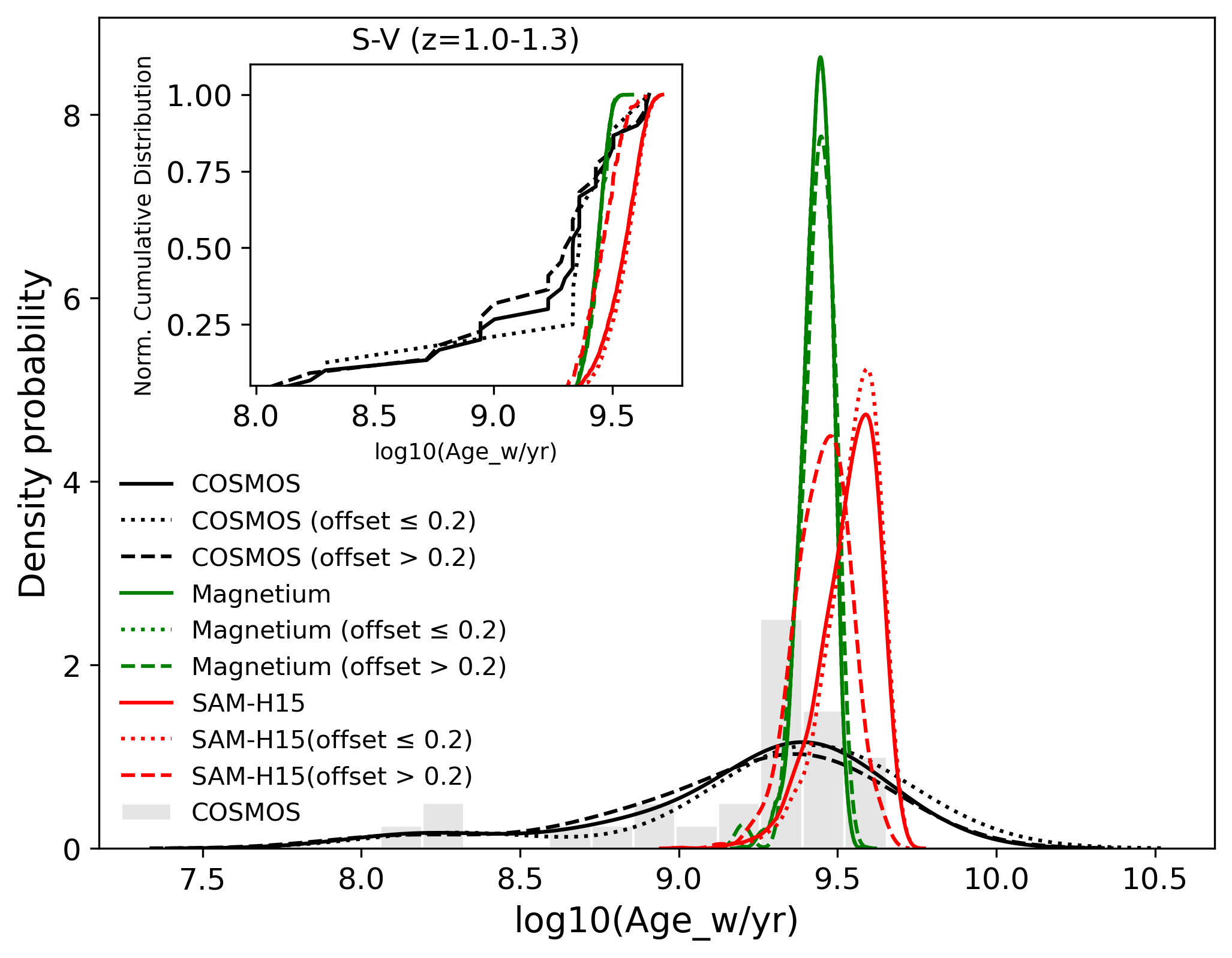}
    \includegraphics[width=0.45\textwidth]{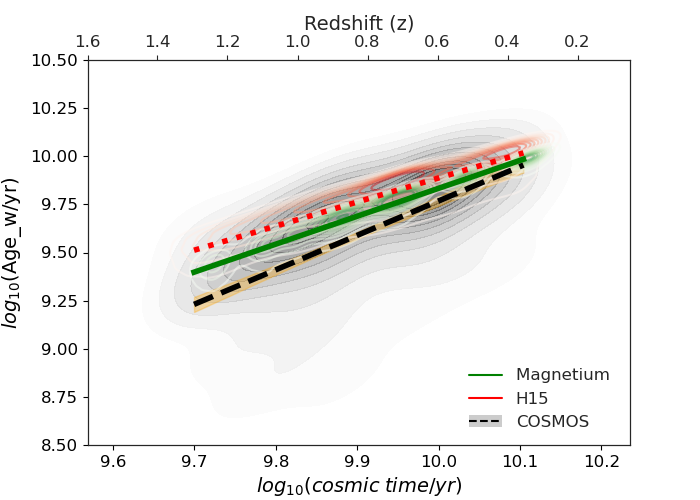}
 \caption[]{The distribution of $log\; (\mathrm{Age_w/yr})$ of BGGs in the observations (COSMOS), Magneticum simulation, and H15 SAM. The smoothed Kernel Density Estimation (KDE) distributions for BGG sub-samples are categorized by their proximity to the X-ray center: those with offset $\leq 0.20R_{200}$ are represented by the dashed black curve. In contrast, those with offset $>0.20R_{200}$ are shown with a dotted black curve. The gray histogram represents the age distribution of all BGGs in COSMOS. To offer a comprehensive comparison, the smoothed stellar age distribution of BGGs predicted by H15 and Magneticum is portrayed with red and green lines, respectively. The bottom right panel showcases density maps of $log(\mathrm{Age_w/yr})$ over cosmic time for all BGGs in both observations and models. The cumulative distribution of the age of all BGGs, central dominants, and offset BGGs in the observations (COSMOS) and H15 SAM, and Magneticum simulation are displayed with similar colors and line styles within each panel of the figure as a subplot. }
  \label{Fig:age-dist-z} 
\end{figure*}
In Paper I \citep{gozaliasl2016brightest}, we discovered a star-forming and potentially young population of BGGs. In this study, our objective is to validate the existence of these young BGGs by examining their stellar age. We also aim to assess whether existing models can accurately predict the presence of such a population. We will evaluate the model stellar age of galaxies using a weighting relative to their stellar mass, a method adopted by \citet[][section 6.1]{wuyts2011star}.

To ensure the accuracy of our stellar age measurements for galaxies, we used a mock catalog of galaxies derived from H15 SAM and evaluated their properties using \text{\small Le Phare}. Figure \ref{Fig:one-one-age} illustrates the one-to-one comparison between the true age and the predicted age obtained by the SED fitting technique by \text{\small Le Phare}. Each point represents the mean ages in a given true age bin, with error bars corresponding to the standard deviation of the predicted age in each bin.

As seen in Fig. \ref{Fig:one-one-age}, our measured stellar age indicates a marginal agreement with the actual age of galaxies.
We calculated the Mean Absolute Error (MAE) and found that the estimated stellar ages, using SED fitting, deviate from the actual values by approximately 1.26 Gyr on average. We also determined fractional errors in the stellar age estimation, which are found to increase from 20\% in the oldest ages to 40\% in the youngest ages.
We applied a stellar mass cut of $10^{10} M_{\odot}$ for our BGGs in our observational data and the models. This mass cut helps ensure the reliability of our analyses by focusing on galaxies with more consistent and well-constrained measurements. We note that there are only two BGGs with a stellar mass below the cut in the S-I and S-V subsamples, and even without the mass cut, our results remain unaffected. 

Figure \ref{Fig:age-dist-z} presents a comparison of $\log(\mathrm{Age_w/yr})$ distributions of BGG using observations from the COSMOS survey, the Magneticum simulation, and the H15 SAM for subsamples S-I to S-V. The solid, dashed, and dotted black curves correspond to the smoothed Kernel Density Estimation (KDE) distribution of BGGs located within $R_{200}$, and a distance with $\leq 0.20R_{200}$, and $>0.20R_{200}$ from the X-ray peak/center of the groups, respectively. The gray histogram represents the overall age distribution of all BGGs in the COSMOS survey. To provide a comprehensive comparison, we also depict the smoothed stellar age distributions of BGGs predicted by the H15 SAM (in red) and the Magneticum simulation (in green) with lines. We include normalized cumulative distribution plots of the age of all BGGs, central dominants, and offset BGGs  in observations (COSMOS) and the H15 SAM and Magneticum simulation. These cumulative distributions are displayed with similar colors and line styles within each subplot, facilitating a comparison of the age distributions across different datasets and models. Furthermore, in the bottom right panel, density maps of $\log(\mathrm{Age_w/yr})$ over cosmic time are displayed for all BGGs in both observations and models. 

The key findings obtained from examining Fig. \ref{Fig:age-dist-z} are::
\begin{itemize}
    \item On average, our findings reveal a marginal agreement between observations and simulations. With the advent of forthcoming large-scale surveys, the potential for substantial statistical improvements in the accuracy of the stellar age measurement using the SED fitting and Spectral analysis of galaxies becomes increasingly promising.
    \item The age distribution observed across all subsamples tends to show a pronounced skewness towards younger ages. This is evident in the normalized cumulative distribution of stellar ages presented in various panels. Generally, observed BGGs appear younger, while Magneticum's BGGs exhibit intermediate ages, and H15 features the oldest ones.
    \item In the observations and the Magneticum simulation, we observe no significant distinctions in the age distributions for BGGs with various offsets. However, we do notice a contrast between offset BGGs in H15, which tend to be younger than their central counterparts.

    \item In the case of the low-mass and low-redshift subsample (S-I) and our highest-redshift, most massive groups (S-V), the model predictions exhibit disparities with the observed data. However, for the other subsamples (S-II to S-IV), the models consistently align with the observed stellar ages of BGGs. 
    
    BGGs within low-mass groups may function as intermediaries, linking low-density environments to the highly massive structures characteristic of rich galaxy clusters. It is plausible that environmental effects have a lesser impact on BGGs in these low-mass groups compared to those in very massive halos, which are subject to significant gravitational potential wells and environmental factors such as ram pressure. This hypothesis finds support in our earlier research, as exemplified in \cite{Gozaliasl2014gap, gozaliasl2016brightest,gozaliasl2020}, where we documented intriguing dynamics between BGGs in low-mass groups. Remarkably, despite their central locations within the systems, they dynamically exhibit an unrestrained nature with respect to the dispersion of the group velocity \citep[see Fig. 3.0 and 4.0 in][for visual representations]{gozaliasl2020}. These observations encompass ongoing star formation and even late-type galaxies reminiscent of our Milky Way \citep[see Fig. A.7 in][for visual representations]{Gozaliasl2014gap}.

    Furthermore, the Magneticum simulation validates the presence of two separate populations of BGGs in low-mass and low-redshift groups (S-I), similar to the dual peaks observed in the stellar age distribution.
\item In the bottom right panel of Fig. \ref{Fig:age-dist-z}, we present the Kernel Density Estimation (KDE) density map of BGGs' age as a function of cosmic time and redshift, revealing an overlap between modeled and observed BGG ages spanning approximately the past nine billion years of the Universe's age. We utilize the \texttt{linmix} regression technique to fit the natural logarithm of $\mathrm{\ln(Age_w/yr)}$ against the natural logarithm of cosmic time. Table \ref{tab:age_cosmic_age} provides the results of the best linear fit for the age of BGGs versus cosmic time for different subsamples, including observations from COSMOS and simulations (Magneticum hydrodynamical simulation and H15 SAM). The trends in both simulations and observations show a positive correlation between the age of BGGs and cosmic time, suggesting a general aging of BGGs over time. COSMOS observations exhibit a very strong linear relationship, whereas simulations (Magneticum and H15 SAM) also show positive correlations, but with slightly different slopes and intercepts. The intrinsic scatter is generally low in all cases.
\end{itemize} 
       
\begin{table*}[h!]
  \centering
   \caption{The relationship between $\mathrm{\ln(Age_w/yr)}$ and $\mathrm{\ln(\text{cosmic age})}$ for BGGs from COSMOS, H15 SAM, and Magneticum simulation is modeled using Eq. \ref{eq:linmix_eta}. The columns, from left to right, provide information on the sub-sample name, intercept, slope, and intrinsic scatter in stellar age.
}
  \label{tab:sample-data}
  \begin{tabular}{ccccccc}
\hline
 Sample & $\alpha$ & $\beta$ & $\sigma$ \\
\hline
   COSMOS & $-18.6878 \pm 3.7954$ & $1.7884 \pm 0.1665$ & $0.6335 \pm 0.0312$ \\
 H15 & $-6.0038 \pm 0.0834$ & $1.2495 \pm 0.0036$ & $0.1466 \pm 0.0008$ \\
 Magneticum & $-10.7332 \pm 0.0405$ & $1.4495 \pm 0.0018$ & $0.0637 \pm 0.0005$ \\
\hline
  \end{tabular}\label{tab:age_cosmic_age}
\end{table*}

\begin{table*}
\centering
\caption{Best linear fit parameters for the relationship between $\ln(\mathrm{Age_w/yr})$ and $\ln(\mathrm{stellar\; mass})$ for BGGs in different subsamples (S-I to S-V). $\alpha$, $\beta$, and $\sigma$ represent the intercept, slope, and intrinsic scatter in stellar age, respectively, related to the regression relationship described in Section \ref{sec:linmix} and Eq. \ref{eq:linmix_eta}. The stellar mass of galaxies has been normalized to $10^{11} M_{\odot}$.}
\begin{tabular}{ccccc}
\hline
Sample & Subsamples & $
\alpha$ & $\beta$ & $\sigma$ \\
\hline
COSMOS & S-I & 22.502 $\pm$ 0.045 & 0.208 $\pm$ 0.056 & 0.325 $\pm$ 0.037 \\
COSMOS & S-II & 22.425 $\pm$ 0.109 & 0.229 $\pm$ 0.098 & 0.362 $\pm$ 0.066 \\
COSMOS & S-III & 21.922 $\pm$ 0.101 & 0.526 $\pm$ 0.110 & 0.421 $\pm$ 0.058 \\
COSMOS & S-IV & 21.471 $\pm$ 0.110 & 0.577 $\pm$ 0.144 & 0.694 $\pm$ 0.075 \\
COSMOS & S-V & 21.181 $\pm$ 0.112 & 0.830 $\pm$ 0.126 & 0.593 $\pm$ 0.093 \\
\hline
H15 & S-I & 22.869 $\pm$ 0.002 & -0.061 $\pm$ 0.004 & 0.171 $\pm$ 0.002 \\
H15 & S-II & 22.957 $\pm$ 0.006 & -0.001 $\pm$ 0.008 & 0.138 $\pm$ 0.004 \\
H15 & S-III & 22.614 $\pm$ 0.003 & -0.002 $\pm$ 0.005 & 0.152 $\pm$ 0.002 \\
H15 & S-IV & 22.262 $\pm$ 0.005 & -0.009 $\pm$ 0.008 & 0.169 $\pm$ 0.003 \\
H15 & S-V & 21.981 $\pm$ 0.006 & -0.038 $\pm$ 0.010 & 0.183 $\pm$ 0.004 \\
\hline
Magneticum & S-I & 22.712 $\pm$ 0.001 & 0.055 $\pm$ 0.002 & 0.002 $\pm$ 0.001 \\
Magneticum & S-II & 22.753 $\pm$ 0.007 & 0.007 $\pm$ 0.004 & 0.003 $\pm$ 0.001 \\
Magneticum & S-III & 22.409 $\pm$ 0.004 & 0.015 $\pm$ 0.002 & 0.074 $\pm$ 0.001 \\
Magneticum & S-IV & 22.068 $\pm$ 0.007 & 0.016 $\pm$ 0.004 & 0.093 $\pm$ 0.002 \\
Magneticum & S-V & 21.641 $\pm$ 0.013 & 0.043 $\pm$ 0.007 & 0.034 $\pm$ 0.007 \\
\hline
\end{tabular}
\label{tab3:age-mass}
\end{table*}
We observe that the BGG ages we obtained closely match those of high-resolution hydrodynamical simulation of  galaxy groups like Romulus and Simba. The full details of this study will be presented in the forthcoming work by Barre et al. in preparation.
\subsection{The relation between stellar age and stellar mass} \label{sec:age-mass-relation}
Although most previous studies have primarily focused on galaxies as holistic entities or the low-redshift group galaxies, our current investigation delves into exploring the stellar age-stellar mass relationship in four different redshift bins of our sample of BGGs (S-I to S-V) over the past nine billion years ($z=0.08-1.3$), covering a significant portion of the Universe's age.

Figure \ref{Fig:sm_age} illustrates the complex interplay between galaxy stellar mass and age, providing a comprehensive perspective on the 2D KDE distribution across various subsamples (S-I to S-V and the combined subsamples S-II to S-V) for BGGs in the COSMOS field, Magneticum, and H15 models at different redshifts and halo mass bins.

A discernible increase in the age of BGGs towards lower redshifts (low-$z$) is observed.  Remarkably, the evolutionary pattern displays variations across distinct stellar masses and redshift ranges. In the case of S-I, the stellar age of observed BGGs increases with stellar mass, while it remains constant for BGGs within S-II. In contrast, for S-III and S-V, a positive correlation is observed between the stellar age and stellar mass of BGGs. However, a distinct trend for BGGs in S-IV is observed after combining all BGGs in groups across four redshift bins (S-II to S-IV, as seen in the bottom right panel of Fig. \ref{Fig:sm_age}), we observe a clear positive correlation between the age and mass of BGGs.

Except for low-mass BGGs, the Magneticum simulation aligns with observations for S-I and S-II within the observational uncertainty. However, the trend is less pronounced and the BGGs exhibit a higher mass compared to the observations, particularly for S-III to S-V at z=0.4-1.3. In contrast, the BGGs in the H15 SAM exhibit no discernible dependence between stellar mass and stellar age across all subsamples despite some indications of decreasing ages with increased stellar masses.

 The uniformity of stellar ages at low $z$ contrasts significantly with the wider range observed at high $z$. Most of the stellar mass centers around $10^{11} M_\odot$ with a small variation of $\pm 0.5$ dex. A noteworthy trend is the emergence of an older galaxy population towards lower $z$ bins, especially pronounced at higher stellar masses ($\log(M_{*}) > 11.3$). This observation suggests the importance of dry mergers in forming a population of older, massive galaxies. Interestingly, this phenomenon is absent in S-IV, starts to appear in S-III, and becomes more pronounced in S-II. The existence of galaxies with large $M_{*}$ but already mature age implies that such configurations cannot be achieved through star formation alone. 

An essential point for discussion centers around identifying the transitional point at which we observe a shift from star formation-dominated growth to merger-driven growth. Understanding this transition, particularly concerning epochs and stellar masses, is crucial. Although mergers involving star formation remain crucial for both channels, the data suggest distinctive patterns, especially at high-mass ends.

Regarding simulations, Magneticum seems to produce galaxies with final stellar masses at $z>1$, enabling them to evolve over time. Conversely, the H15 SAM does not form massive galaxies, resulting in missing tracks. This necessitates further investigation into the operating mechanisms in Magneticum at various epochs and stellar masses.

The interplay between observed trends and simulated outputs raises intriguing questions about the underlying mechanisms governing galaxy evolution, particularly the contrasting behaviors of different models and their implications for our understanding of the evolution of BGGs.

In conclusion, considering the complex interplay of stellar age and mass depicted in Figure \ref{Fig:sm_age}, an alternative scenario put forward by \cite{Rennehan2020MNRAS.493.4607R} presents a fascinating explanation for the formation of highly massive central galaxies. Traditionally, it was thought that these galaxies accumulated most of their stellar mass fairly late in cosmic history, mainly at low redshifts. However, the discovery of protoclusters at high redshifts has questioned this belief, indicating that they might have formed much earlier than previously thought. Recent studies using hydrodynamical and dark-matter simulations have illuminated this process, showing that a portion of massive BGGs/BCGs gains its stellar mass through rapid mergers and bursts of star formation at high redshifts in very short timescales, roughly 1 Gyr. This swift assembly, especially in highly overdense protoclusters, leads to the existence of massive blue elliptical galaxies at redshifts greater than 1.5, suggesting a downsizing effect where BCGs in the most massive clusters form earlier than those in smaller clusters. Subsequent dry mergers mainly contribute to the accumulation of intracluster light and stellar halos. The ongoing observations by the James Webb Space Telescope (JWST) in our COSMOS-Web survey \citep{casey2023} offer a promising avenue to further investigate and confirm this intriguing hypothesis.

\subsubsection{The scaling relation of the stellar age and stellar mass}
\begin{figure*}[h]
 \centering
 \includegraphics[width=0.3\textwidth]{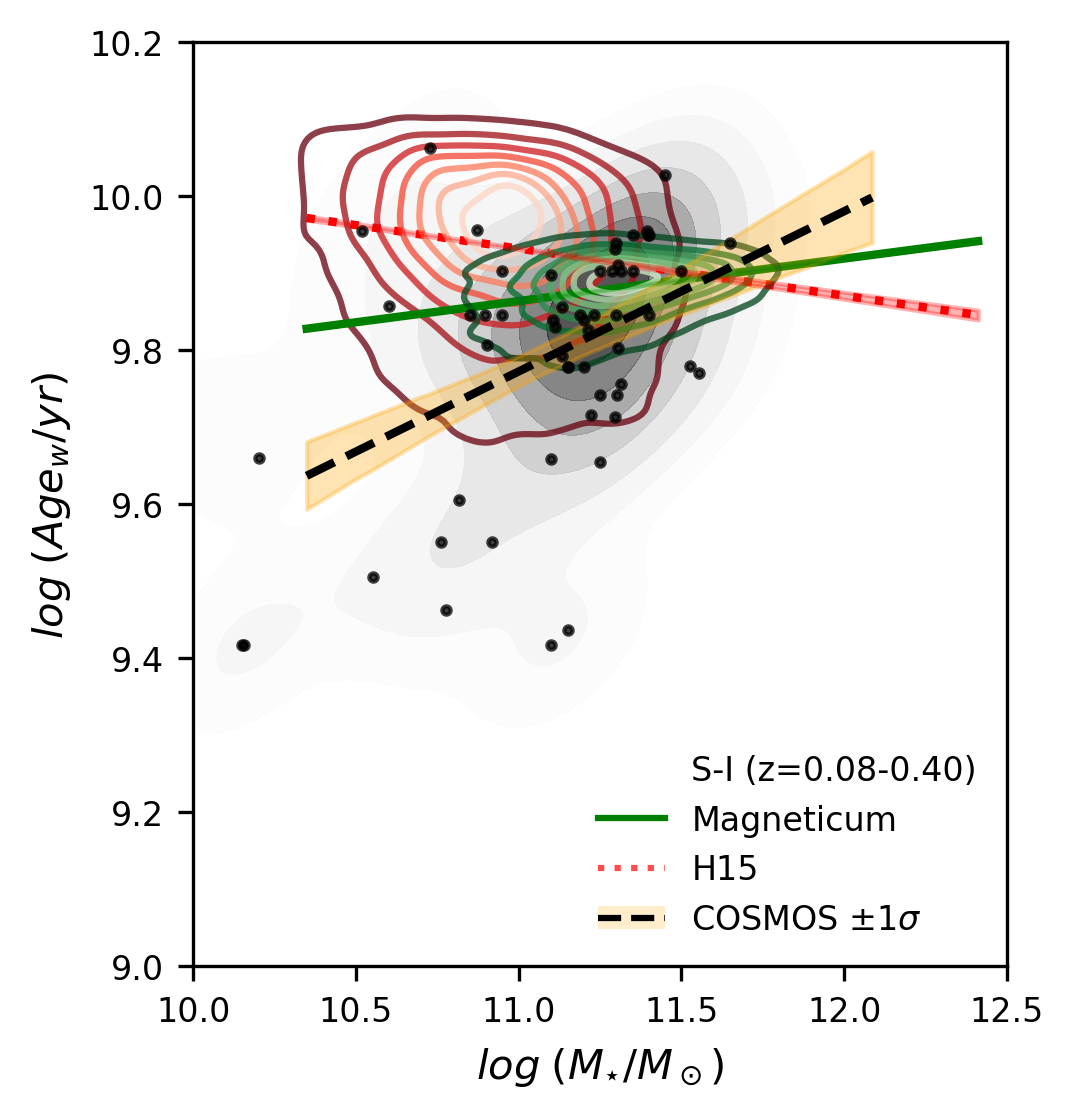}
 \includegraphics[width=0.3\textwidth]{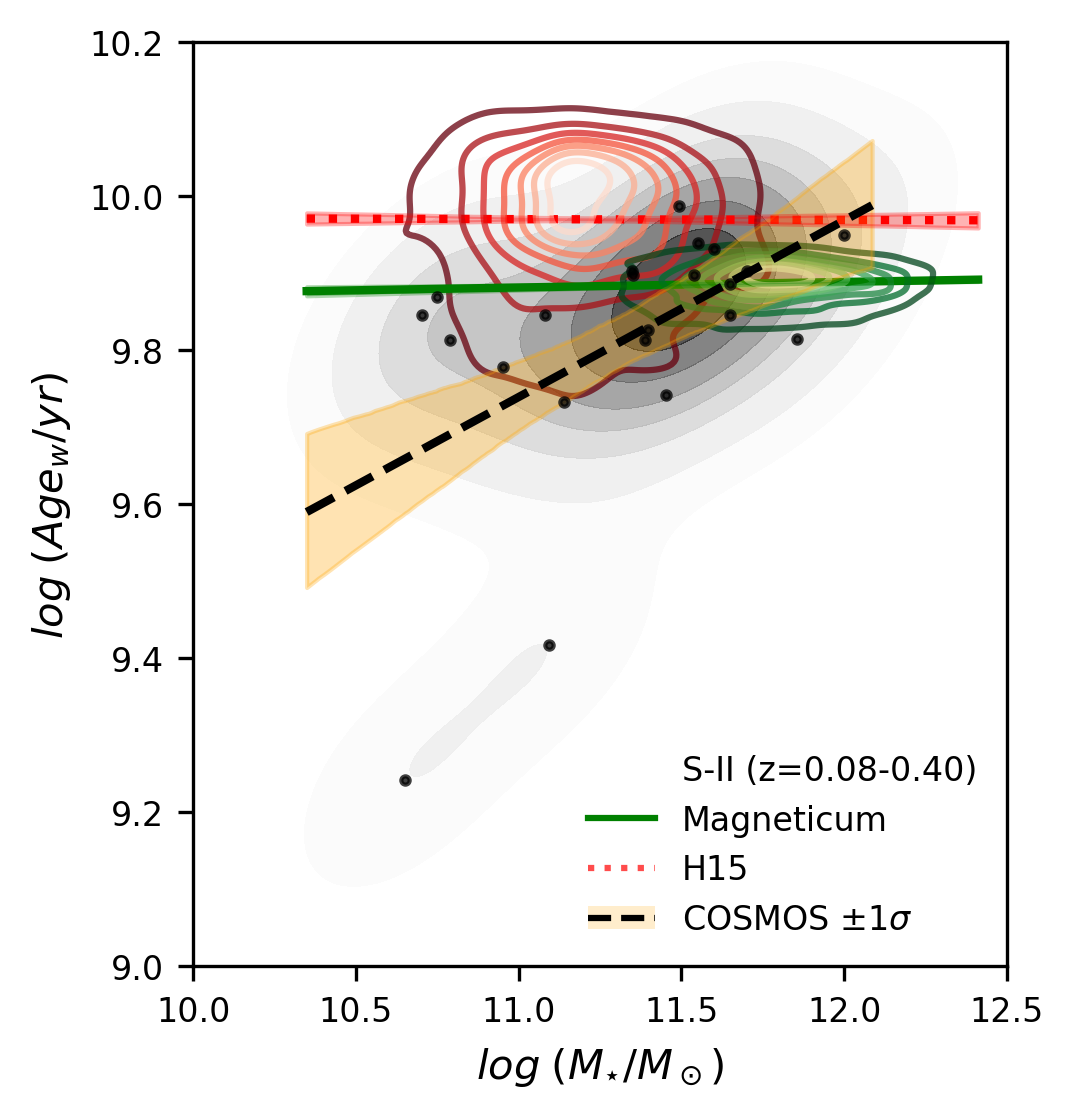}
 \includegraphics[width=0.3\textwidth]{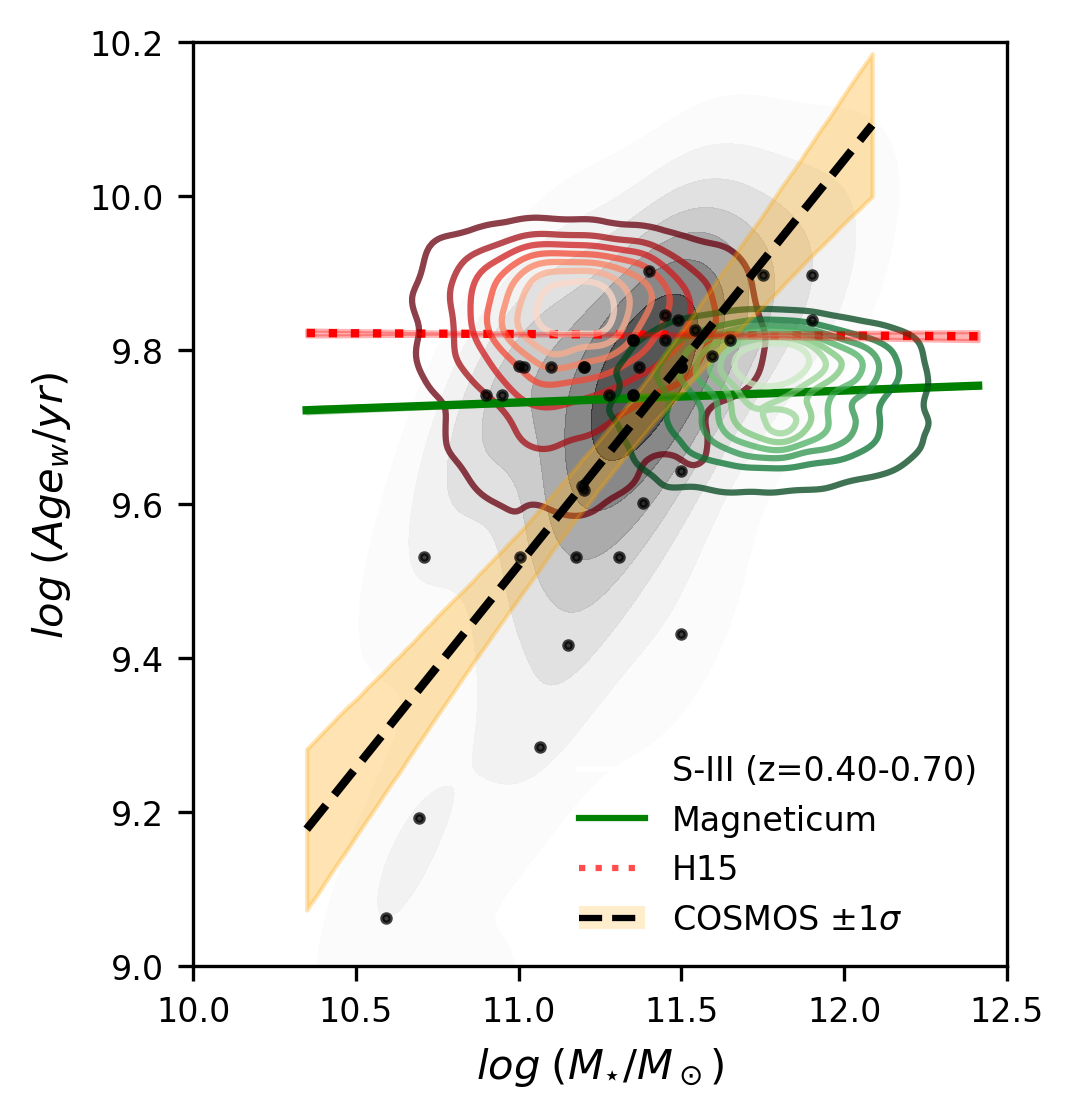}
 \includegraphics[width=0.3\textwidth]{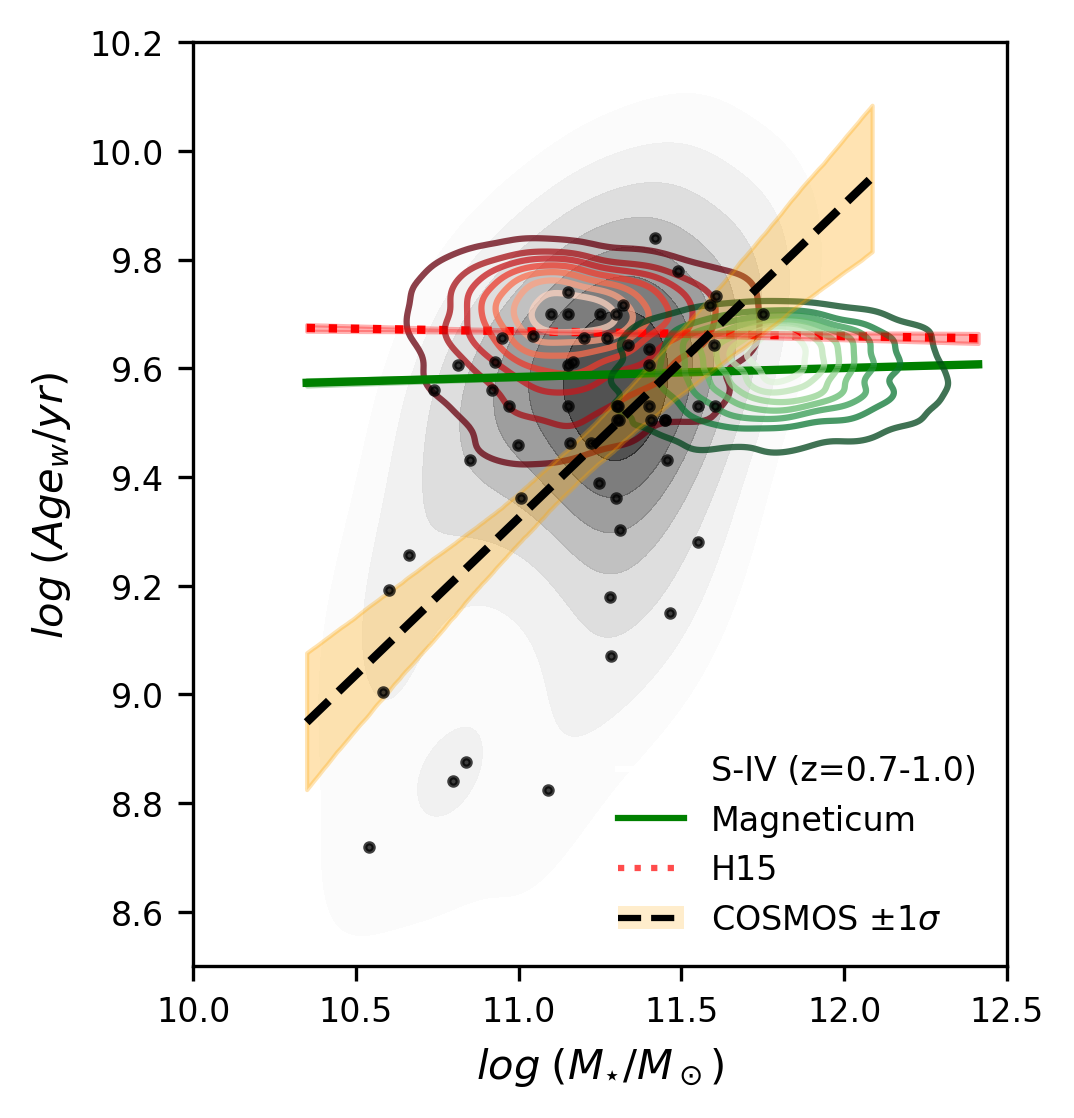}
 \includegraphics[width=0.3\textwidth]{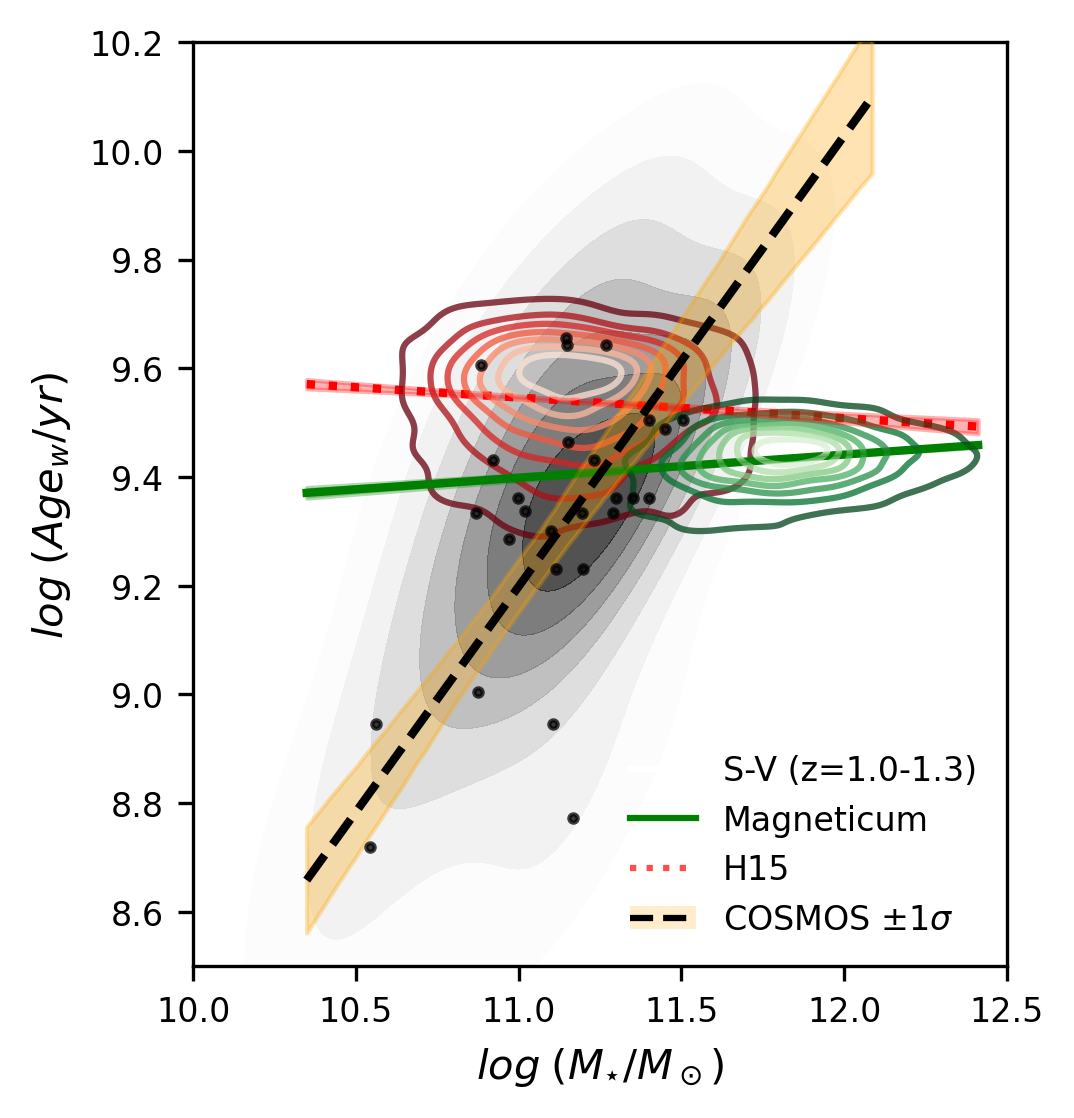}
    \caption[]{The correlation between $\log(\mathrm{Age_w/yr})$ and $\log(\mathrm{M}_{*}/\mathrm{M}_{\odot})$ for BGGs in the COSMOS field, H15, and Magneticum. Dashed, solid, and dotted lines indicate the optimal linear fit for all data points in each sub-sample.
}
      \label{Fig:sm_age}
       \end{figure*}
\begin{figure}[h]
 \centering
 \includegraphics[width=0.45\textwidth]{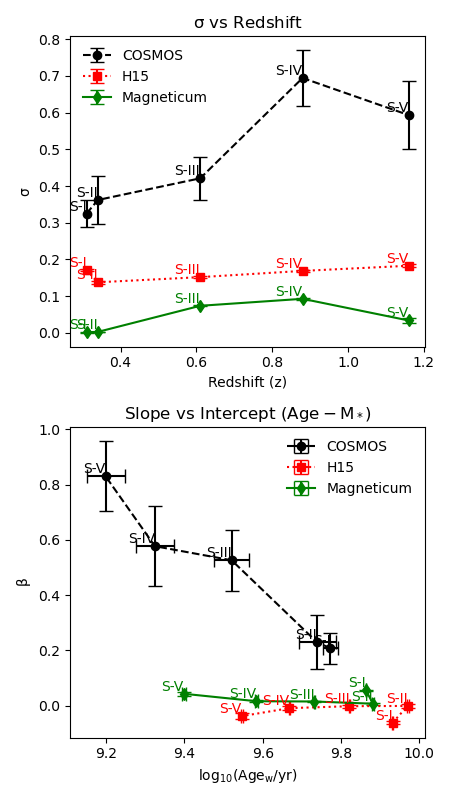}

    \caption[]{The best-fit results of the relationship between $\log(\mathrm{Age_w/yr})$ and $\log(\mathrm{M}_{*}/\mathrm{M}_{\odot})$ for BGGs in the COSMOS field, H15, and Magneticum.}
      \label{Fig:sm_age_fit_results}
       \end{figure}
Fig. \ref{Fig:sm_age} illustrates the best mean relationship between stellar mass and stellar age of the BGGs. Each section of Fig. \ref{Fig:sm_age} displays the observed stellar age of BGGs compared to stellar mass, marked with black circles (COSMOS). Future figures will follow the same color scheme.
     
To examine the patterns, we applied Eq. \ref{eq:linmix_eta}, using \texttt{linmix} linear regression to represent the correlation between $\log(\mathrm{Age_w/yr})$ and $\log(\mathrm{M_*})$ for BGGs across various samples (S-I to S-V). The \texttt{linmix} package employs Markov Chain Monte Carlo (MCMC) methods within a Bayesian framework to measure parameter estimate uncertainties. Prior to fitting, the stellar mass of the BGGs was normalized to $10^{11} M_\odot$. In Fig. \ref{Fig:sm_age}, the optimal linear fit to all data points (not merely the median values in each sub-sample) is shown with dashed, solid, and dotted lines in respective colors for both observations and models. The optimal fit parameters are provided in tab. \ref{tab3:age-mass}.

The following is a summary of our observations for the various sub-samples shown in each panel of Fig. \ref{Fig:sm_age}:

Within the S-I group, there is a gradual rise in the stellar age of BGGs as their stellar mass increases. However, this upward trend becomes less pronounced for BGGs in the S-II subset. Transitioning to S-III, a notable positive slope reveals a significantly increasing relationship between stellar age and stellar mass. For the S-IV and S-V groups, despite uncertainties, the trend's slope is consistent with that observed in S-III.

The Magneticum simulations show a better agreement with observations compared to the H15 SAM for subsamples S-I and S-II. Nonetheless, the slope in the Magneticum simulations is less steep, and BGGs are generally more massive than those observed, especially noticeable for subsamples S-III to S-V in the redshift range $z=0.4-1.3$.

Conversely, the H15 SAM simulation suggests that there is no observable correlation between stellar mass and age across all sub-samples. It even hints at a trend of decreasing ages as stellar mass increases.

As shown in Fig. \ref{Fig:sm_age_fit_results}, a detailed summary of the trends, including their slopes and intercepts, is provided to better understand the possible redshift evolution of the BGG stellar age and stellar mass relationship. These summaries are presented as functions of the median redshift for each subsample, with data points from S-II to S-V connected to illustrate their common halo mass range.

It is observed that the slope increases as the redshift decreases, and the intercept of the age-mass relationship for BGGs decreases (from S-I to S-V). This contradicts models that predict nearly stable slopes. These findings suggest a possible rejuvenation of the BGG around $z\sim1.0-0.6$, as evidenced by the stable slope for S-IV and S-III.

Furthermore, our findings show a notable rise in the intrinsic scatter of stellar age with increasing redshift. Similarly, the H15 SAM exhibits a related trend. On the other hand, the Magneticum simulation shows that there are no significant changes in the intrinsic scatter of the stellar age of BGGs with respect to redshift. We computed the average intrinsic scatter for each model and found that COSMOS possesses a higher mean value ($0.3072\pm 0.1$) compared to H15 ($0.1096$) and Magneticum ($0.007$). Calculating the fold change indicates that the values for COSMOS are, on average, approximately 2.8 times higher than those for H15.

To investigate the differences and similarities in intrinsic scatter between COSMOS and the models, we conducted a statistical analysis. The Kolmogorov-Smirnov (KS) test did not reveal a significant difference in the distribution between COSMOS and H15 ($p=0.357$), but a significant difference between COSMOS and Magneticum ($p=0.008$). This difference highlights the intricate nature of studying the redshift evolution of the characteristics of BGG, underlining that the evolution of BGG is dynamic and shaped by factors that change over time. These time-based variations might be related to changing physical processes such as mergers, star formation, or accretion, which influence the traits of BGGs.

Models struggle with accurately forecasting the aging of BGGs within the stellar mass range of $10^{10-11} M_\odot$, necessitating modifications. This challenge may be linked to the observed inefficiency of quenching in cluster galaxies at high redshift, as noted by \cite{kukstas2023}. The cause of this discrepancy at lower stellar masses remains unclear and is evident across various simulations, which differ considerably in resolution, feedback mechanisms, and hydrodynamic solvers. Future progress in simulations, featuring significantly higher resolution and intricate modeling of the cold interstellar medium, is anticipated to shed light on this issue and possibly resolve the tension at lower masses.

\subsection{The stellar age and SFR relation}

\begin{figure*}[h!]
  \centering
 \includegraphics[width=0.38\textwidth]{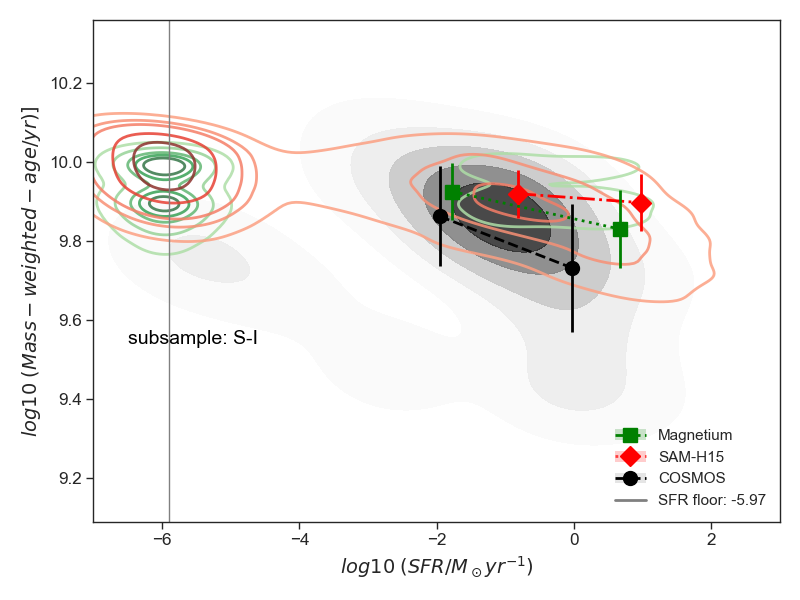}
 \includegraphics[width=0.38\textwidth]{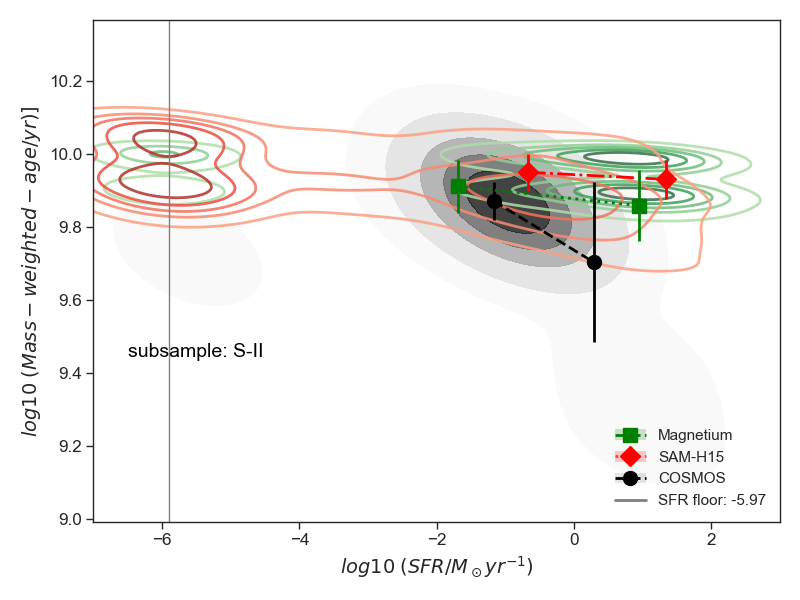}
  \includegraphics[width=0.38\textwidth]{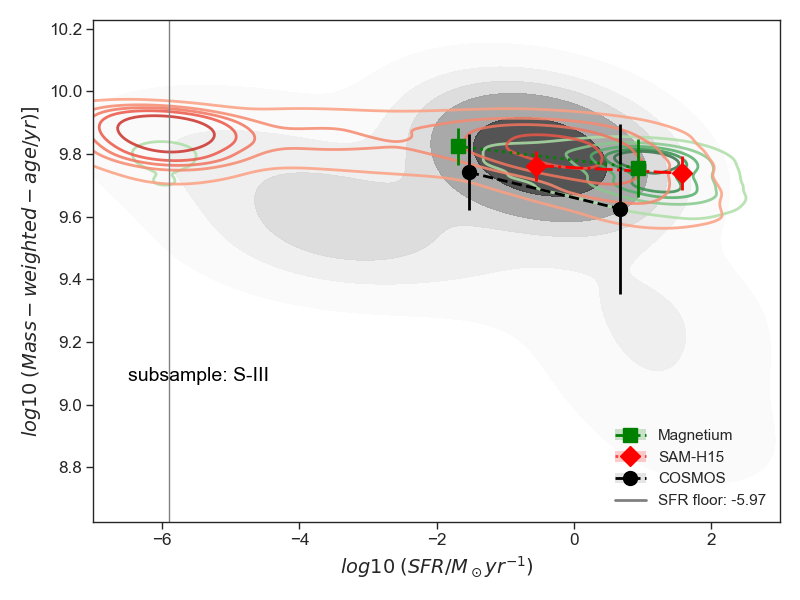}
  \includegraphics[width=0.38\textwidth]{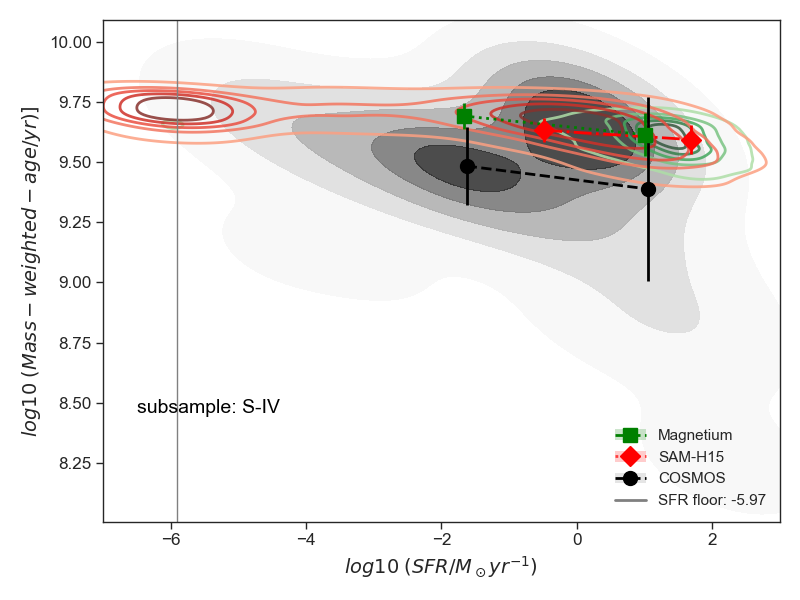}
  \includegraphics[width=0.38\textwidth]{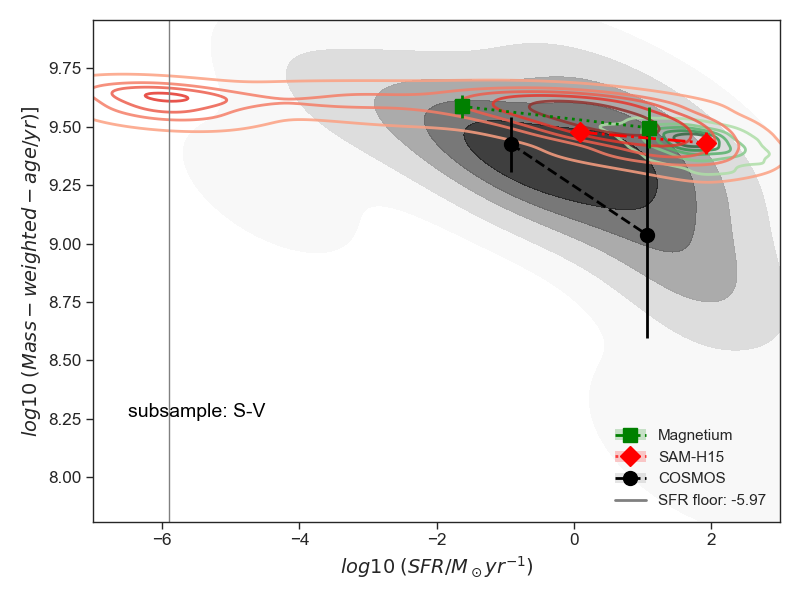}
    \includegraphics[width=0.38\textwidth]{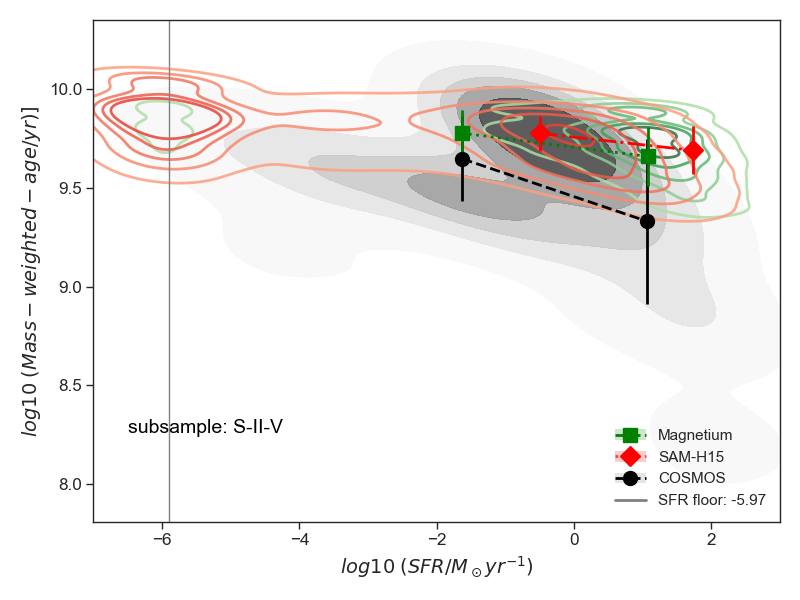}      
    \caption[]{The $\log(\mathrm{Age_w/yr})$ as a function of $\log(\mathrm{SFR/M}_{\odot}\mathrm{yr}^{-1})$ for BGGs in the observations (KDE density map, shaded grey area) for S-I to S-V subsamples. Red and green contours represent stellar age vs. stellar mass predictions by H15 and Magneticum. Median stellar age values in specific mass bins are indicated by black points (COSMOS), green diamonds (Magneticum), and red squares (H15). In the lower right panel, we present the relationship between stellar age and stellar mass for a combined sample of BGGs within S-II to S-V. The vertical grey line marks $\log(\mathrm{SFR/M}_{\odot}\mathrm{yr}^{-1}) = -5.97$, the observed SFR threshold. In models, galaxies with SFR=0 have been adjusted to the observed SFR threshold. For star-forming BGGs, $\log(\mathrm{Age_w/yr})$ shows a negative correlation with star formation rate with no significant evolution with increasing redshift, which is more pronounced in the observations compared to H15 and Magneticum simulation predictions.}
      \label{sfr_age}
       \end{figure*}

Figure \ref{sfr_age} shows the connection between the ages of BGGs, expressed as $\log_{10}(\mathrm{Age_w/yr})$, and their star formation rates, indicated as $\log_{10}(\mathrm{SFR/M}_{\odot}\mathrm{yr}^{-1})$. The plot uses KDE density contours to show the distribution relationship between these two physical quantities. Each panel presents results for a distinct subsample, labeled S-I through S-V.
The plot's gray-shaded region shows the observed relationship, whereas the red and green contours present the anticipated outcomes from H15 and Magneticum simulations. The data points signify the median age values for BGGs in various SFR bins: black dots for COSMOS data, green diamonds for Magneticum, and red squares for H15.
In the lower right panel, we combine the data from S-II through S-V to analyze the overall trend in the relationship between stellar age and SFR for BGGs within halos of the same mass, spanning a redshift range from z=0.08 to 1.3. A gray vertical line at $\log_{10}(\mathrm{SFR [M_\odot /yr]}) = -5.97$ represents the threshold for the observed star formation rates. Due to the use of a logarithmic scale, the SFR values for modeled BGGs with no ongoing star formation (SFR=0) were adjusted to this threshold. Hence, a peak is observed in both models at $\log_{10}(\mathrm{SFR/M}_{\odot}\mathrm{yr}^{-1}) = -5.97$, indicating these fully quenched BGGs.

For BGGs actively forming stars, there is an inverse relationship between star formation rate and median stellar age. This aligns with previous research, suggesting that galaxies with more vigorous star formation tend to host younger stellar populations \citep{Kennicutt1998,2004Brinchmann}, effectively adding younger members to the stellar populations. The predicted patterns align with observations of BGGs with little ($\mathrm{SFR} \sim 0.1 M_{\odot}\mathrm{yr}^{-1}$) to no star formation. However, the observed trend for BGGs with higher star-formation rates becomes more pronounced, implying that the connection between stellar age and star formation strengthens as star formation activity increases.
Additionally, KDE density maps generated from observational data and models illustrate two separate peaks, revealing distinct subpopulations within the BGG samples across various redshift bins. Our subsequent step requires a detailed examination of the BGG SFR distribution to obtain a further understanding of these findings.

Figure \ref{sfr_age} shows a negative correlation between SFR and the median stellar age, showing that BGGs with elevated SFR are associated with younger stellar populations. Importantly, the observed pattern becomes more pronounced for BGGs with elevated SFR, indicating an increasing link between stellar age and star formation activity. Furthermore, the KDE density maps show two distinct peaks in the BGG sample distribution, suggesting the presence of identifiable subpopulations in different redshift bins.

\subsection{The SFR distribution}
\begin{figure*}[h!]
     \centering
  \includegraphics[width=0.38\textwidth]{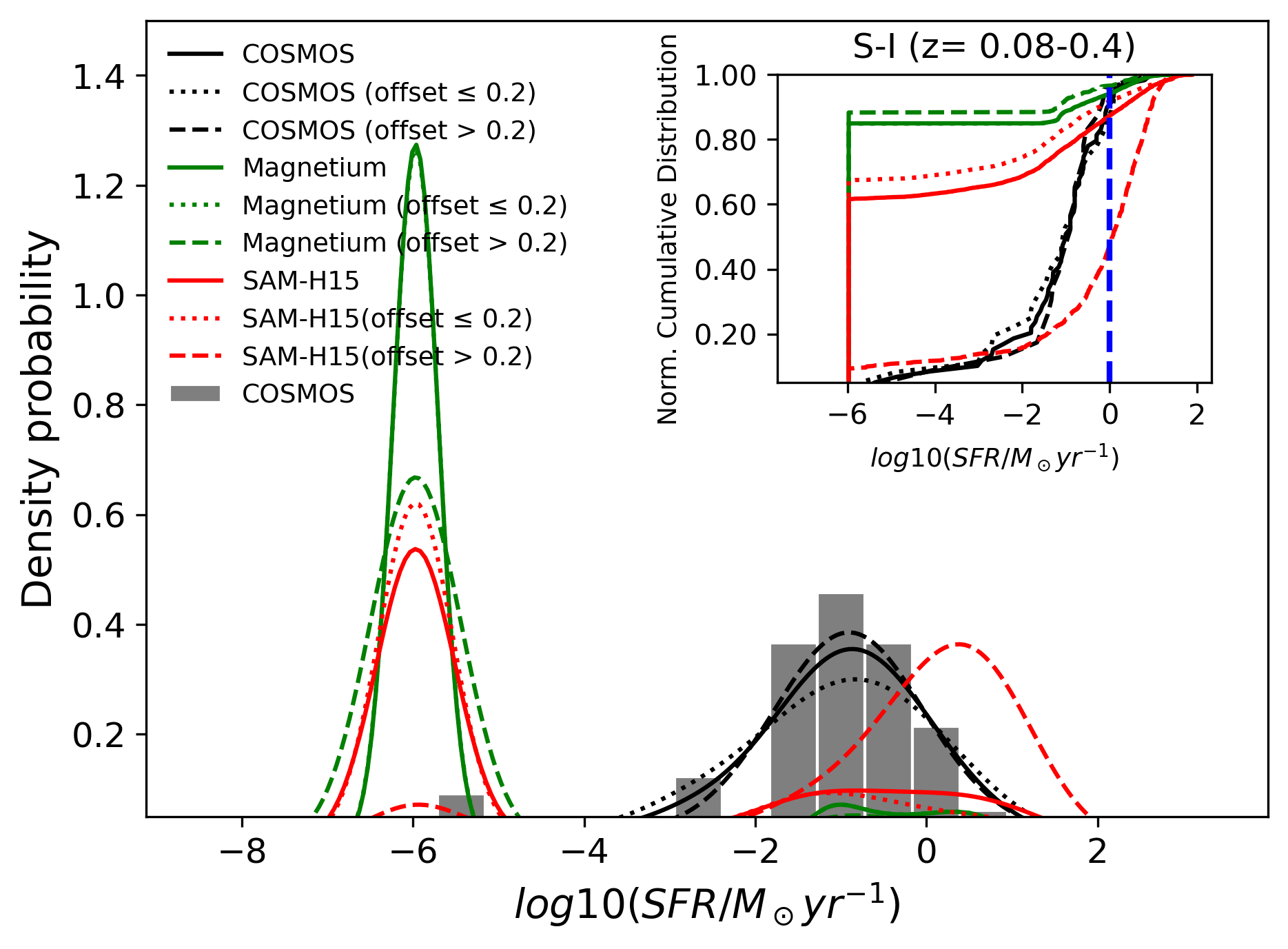} 
    \includegraphics[width=0.38\textwidth]{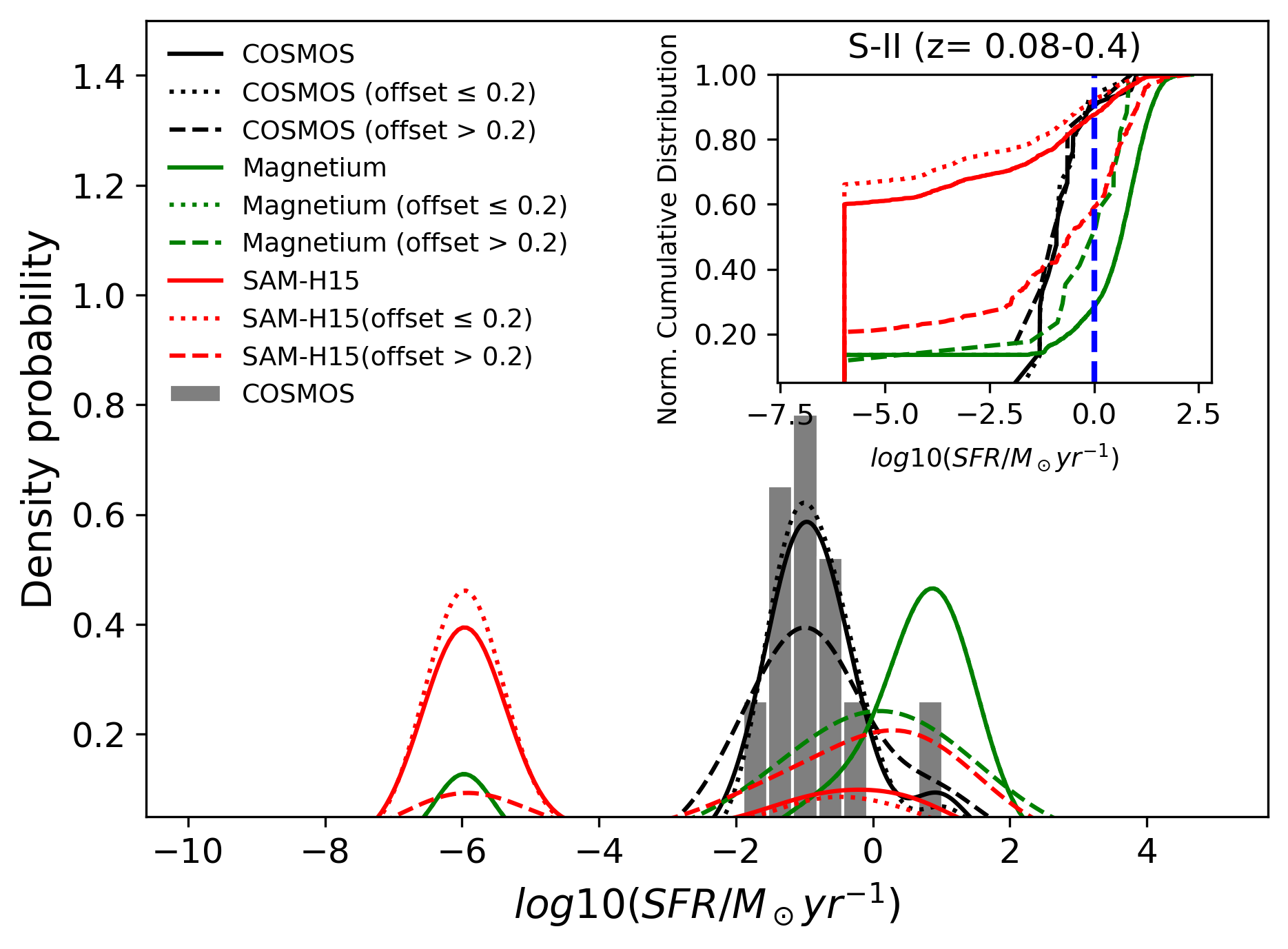} 
  \includegraphics[width=0.38\textwidth]{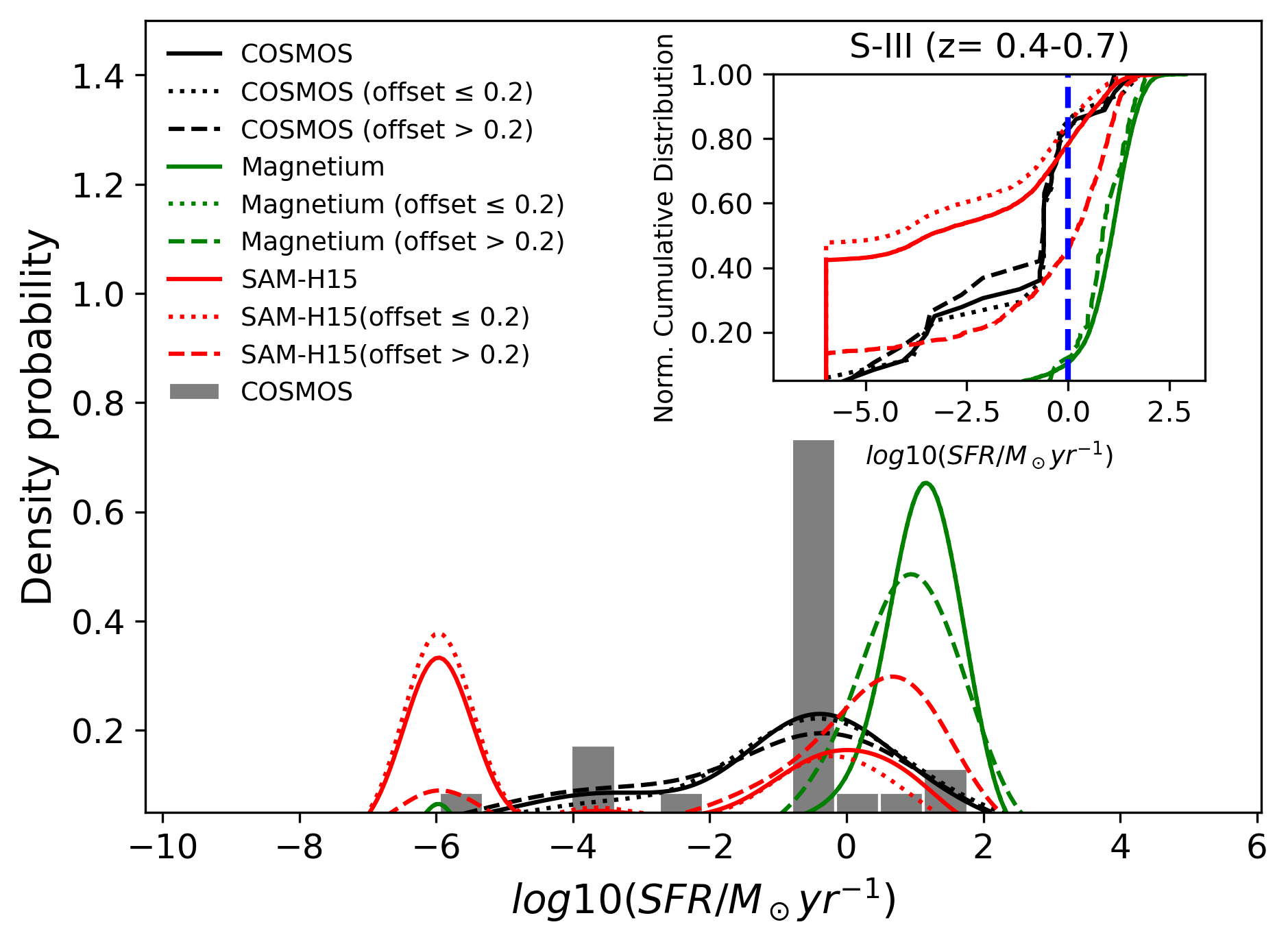}
\includegraphics[width=0.38\textwidth]{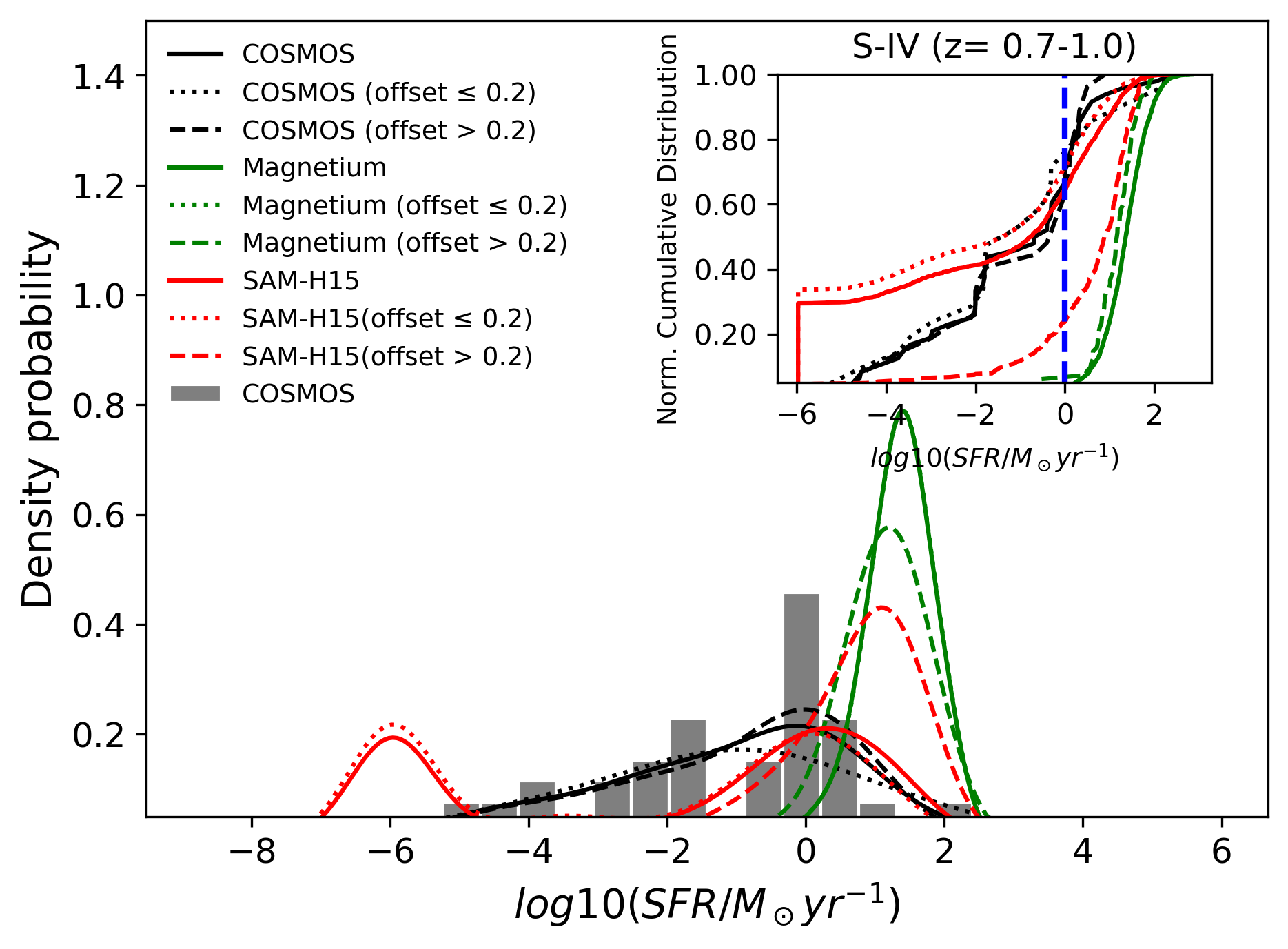}
  \includegraphics[width=0.38\textwidth]{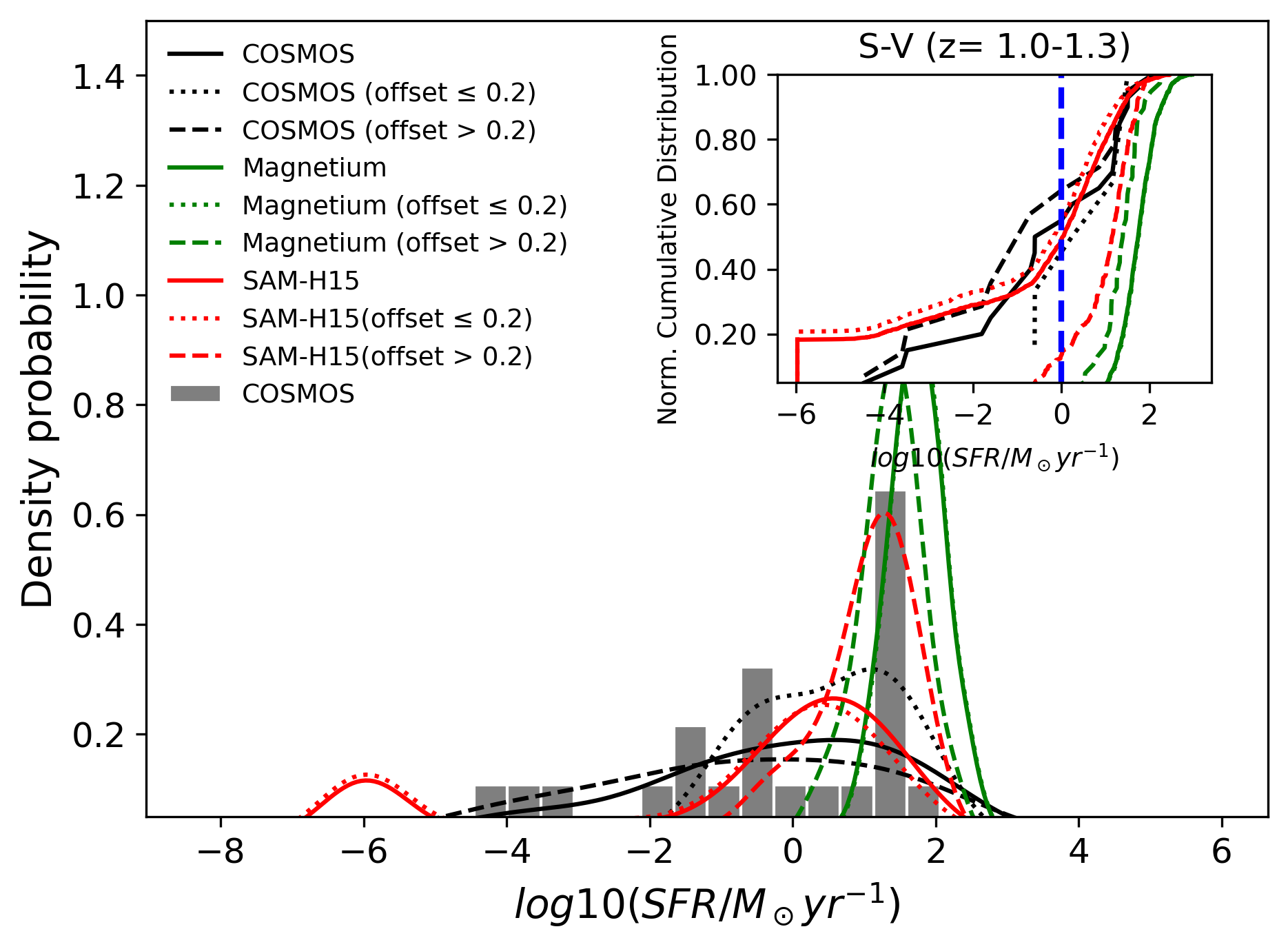}
   \includegraphics[width=0.38\textwidth]{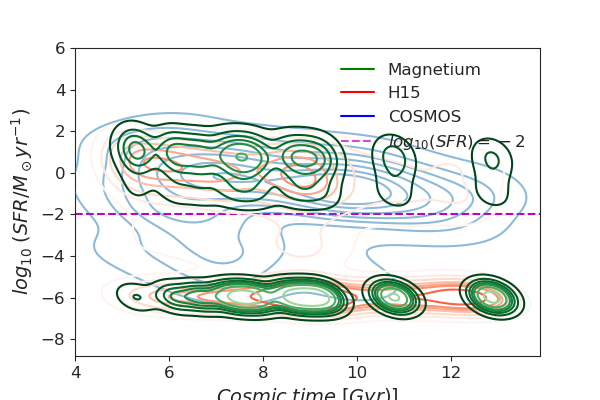}
    
	\caption[]{ The smoothed distribution of $\log(\mathrm{SFR/M}_{\odot}\mathrm{yr}^{-1})  $ for all BGGs within sub-samples of S-I to S-V  (solid black curve); Additionally, we illustrate the smoothed distributions for BGG sub-samples categorized by their proximity to the X-ray center: those with offset $\leq 0.2$ are represented by the dashed black curve, while those with offset $>0.2$ are shown with a dotted black curve. To have a comprehensive comparison with the prediction of SAM and hydrodynamical simulations, the smoothed SFR distribution of BGGs predicted by H15 and Magneticum are shown with red and green lines, respectively. The bottom right panel displays density maps of $\log(\mathrm{SFR/M}_{\odot}\mathrm{yr}^{-1})$ over cosmic time for all BGGs in both observations and models. The cumulative distribution of the SFR of galaxies in both observations and models is displayed within each panel of the figure as a subplot. }
	\label{sfr_dist_fig}
\end{figure*}

Examining Fig. \ref{sfr_dist_fig}, we observe the distribution of $\log(\mathrm{SFR/M}_{\odot}\mathrm{yr}^{-1})$ for all BGGs across subgroups S-I to S-V, represented by a solid black curve. Moreover, we provide an analysis of the smoothed distributions for BGG subgroups based on their distance from the group's X-ray center. The subgroups with offsets $\leq 0.2 R_{200}$ are shown with a dashed black curve, while those with offsets $> 0.2 R_{200}$ are indicated by a dotted black curve.

To compare with the predictions of SAM and hydrodynamical simulations, we include the smoothed SFR distribution of BGGs projected by H15 and Magneticum, shown by red and green lines, respectively. Additionally, within each panel of the figure, a subplot displays the normalized cumulative distribution of SFR for galaxies in both observational data and models, with the dashed blue line corresponding to $\log(\mathrm{SFR/M}_{\odot}\mathrm{yr}^{-1})=0$.

Across all subsamples, the SFR of the BGG in the observations shows no significant dependence on the projected offset from the center of the group. The distributions for all BGGs appear similar, regardless of whether they are centrally dominant or offset, with peaks at $\log(\mathrm{SFR/M}_{\odot}\mathrm{yr}^{-1})=-1$ for S-I, S-II and S-III, and at $0.0-0.5$ for S-IV and S-V. In the H15 SAM, offset BGGs generally exhibit higher star formation rates compared to central BGGs. For the Magneticum simulation, the central BGGs form the majority, and similar patterns are seen for all and central BGGs, except for slight deviations in S-II. Nevertheless, the magneticum simulation significantly overpredicts the SFR of BGGs in massive groups (S-II to S-V). In general, H15 more accurately forecasts the observed SFR distribution.

A further observation from the normalized CDF for S-I reveals that about 60\% of the BGGs in both the H15 SAM and Magneticum datasets show a zero SFR, signifying total quenching. Additionally, around $60\%$ of the observed BGGs have a cumulative probability of a randomly chosen BGG having an SFR of $0.1 M_{\odot}\mathrm{yr}^{-1}$.

The bottom right panel of Fig. \ref{sfr_dist_fig} displays density maps of $\mathrm{\log_{10}}(\mathrm{SFR/M}_{\odot}\mathrm{yr}^{-1})$ over cosmic time for the merged subsamples of BGGs from both observational data and models. From the patterns noticed in different datasets, it is evident that there are differences in the fractions of BGG with low SFR. According to our results, approximately $22.5\%$ of the observed BGGs exhibit $\mathrm{\log_{10}}(\mathrm{SFR/M}_{\odot}\mathrm{yr}^{-1}) <-2$ (indicated by the magenta horizontal line). In contrast, the Magneticum and H15 datasets suggest that $53.49\%$ and $25.15\%$ of BGGs, respectively, are galaxies with no star formation activity. In the subsequent subsection, we explore the positioning of BGGs within the main sequence of star-forming galaxies in cosmic history.

\subsection{BGGs on the main sequence of star-forming galaxies across cosmic times}

\begin{figure*}[h!]
     \centering
  \includegraphics[width=0.8\textwidth]{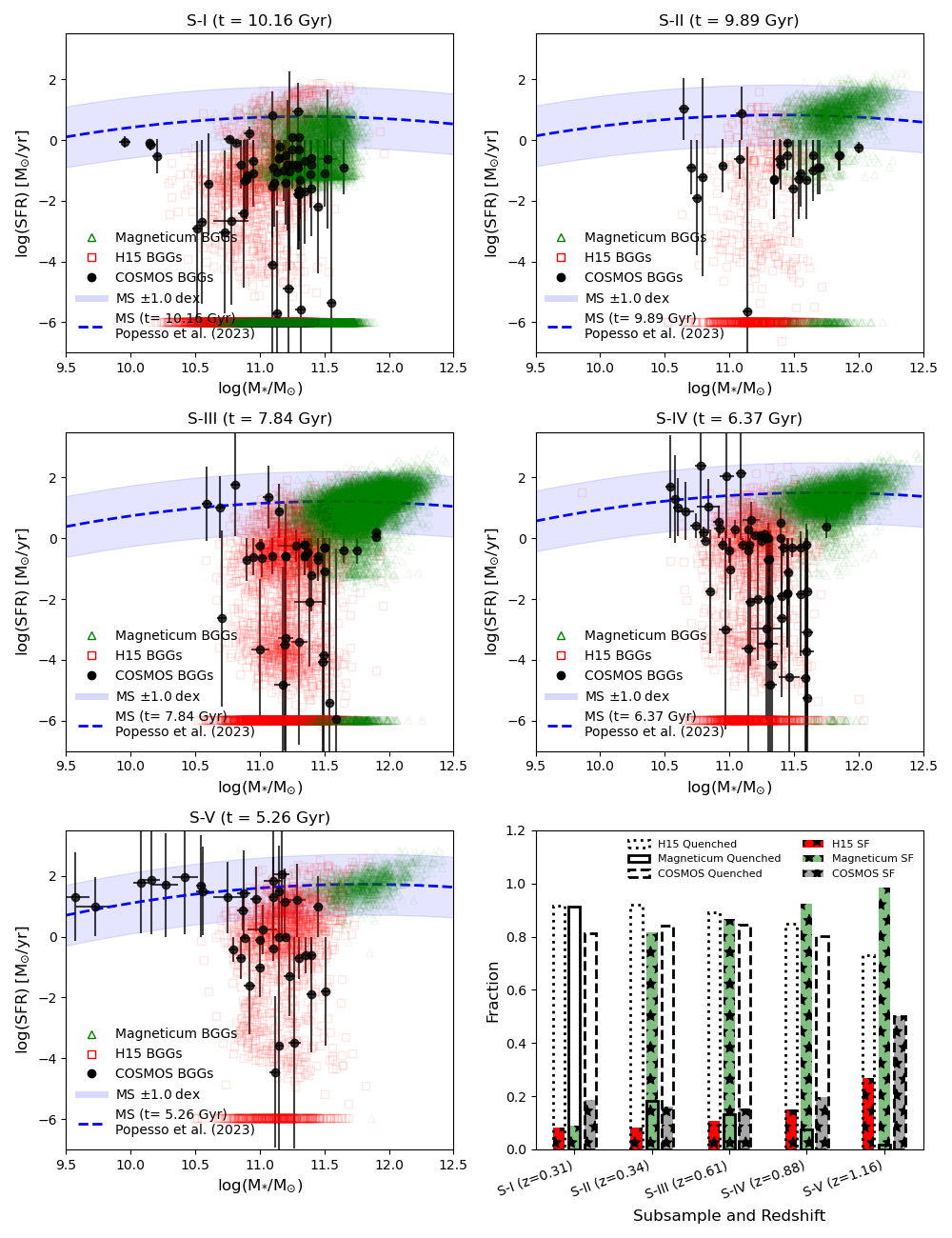}
\caption[]{ The main sequence (MS) of star-forming galaxies at five distinct cosmic epochs that match the median redshift of our group subsamples: S-I to S-V. This relationship (blue dashed line) is extracted from a recent study by \citealp{popesso2023main}, which also used COSMOS galaxies. The BGGs from COSMOS, H15, and Magneticum are superimposed on the MS in different panels. The shaded region indicates $\pm 1.0$ dex. The bottom-right panel illustrates the distribution of both star-forming (hatched with stars) and quenched BGGs. }
	\label{ms}
\end{figure*}

In Figure \ref{sfr_age} and \ref{sfr_dist_fig}, our analysis shows that the star SFR of BGGs is largely independent of their position relative to the halo center, indicating minimal environmental effect. For a more detailed understanding of the relationship between SFR and the stellar mass of BGGs, and its evolution with redshift, we illustrate the SFR versus stellar mass relationship, known as the Main Sequence (MS) of star-forming galaxies as presented in Fig. \ref{ms}.

\citet{popesso2023main} conducted an extensive review to investigate the evolution of the MS of star-forming galaxies over the widest range of redshift ($0 < z < 6$) and stellar mass ($10^{8.5} - 10^{11.5} M_\odot$) to date, finding consistent changes in MS shape and normalization over cosmic time, driven by the availability of cold gas in halos and black hole feedback, corroborated by IllustrisTNG simulations despite some high-mass inconsistencies. We used the MS relation by \citet{popesso2023main} at five different cosmic epochs that corresponded to the median redshift of our BGGs within the S-I to S-V subsamples in observations (blue line): 
\begin{equation}
\label{ms:popesso}
\log SFR(t, \log M_{\star}) = (a_1 t + b_1) \log M_{\star} + b_2 \log^2 M_{\star} + (b_0 + a_0 t).
\end{equation}

Where $a_0 = 0.20$, $a_1 = -0.034$, $b_0 = -26.134$, $b_1 = 4.722$, and $b_2 = -0.1925$. Here, $t$ is the cosmic time.

In the bottom-right panel of Fig. \ref{ms}, we present a summary of the fractions of star-forming (star-hatched) and quenched (unfilled bars) BGGs in each subset for COSMOS (dashed line bars), H15 SAM (dotted line bars) and the magneticum simulation (solid line bars), respectively. It is noted that approximately $\sim20\%$ of BGGs are regular star-forming galaxies within the redshift range of $z=0.08-0.7$. This proportion starts to increase gradually until $z=1.0$ and then increases abruptly. In the redshift range of $z=1.0-1.3$, the distribution of star-forming and quenched BGGs reaches an equal $50-50$ split. For S-I, both models predict that about $90\%$ of BGGs are quenched galaxies. For the remaining subsamples, the H15 SAM forecasted a fraction that is closer to the observed fraction of star-forming BGGs, but it is still 30\% lower than the observed value. Magneticum significantly overpredicts the fraction of star-forming BGGs in the S-II to S-IV subsamples by at least four times. 

\subsubsection{ The specific star formation rate  as a function of redshift.}
In Paper-I \citep{gozaliasl2016brightest}, we analyzed the distribution of the specific star formation rate (sSFR) of BGGs. In this work, our aim is to examine how the sSFR of these systems evolves with redshift. 
The sSFR parameter is used to classify BGGs as star-forming (SF) or quiescent. Although conventional methods often rely on color indices for such classifications, we acknowledge their limitations and potential misclassification \citep{Faber2007ApJ...665..265F, Lee2007ApJ...663L..69L, Strateva2001AJ....122.1861S}. 
Moreover, \cite{Wolf2005A&A...443..435W} identified a group of galaxies called dusty red galaxies. These galaxies appear red because of dust and have an intermediate age rather than the old age typical of regular early-type galaxies. \cite{Lee2015ApJ...810...90L} also noted that many red galaxies within the redshift range examined in our work continue to form stars, leading to potential misclassifications based solely on color. Notably, these red SF galaxies tend to have a higher dust content and gradually transition to a red quiescent population. Importantly, according to these studies, no red SF galaxy exhibits a sSFR greater than $10^{-8} yr^{-1}$.

The sSFR quantifies the rate at which a galaxy accumulates stellar mass through ongoing star-formation processes, with lower sSFR values indicating longer star-formation timescales. In other words, lower sSFR values mean that the galaxy forms stars more slowly over an extended period. Notably, the sSFR of galaxies exhibits a significant upturn from approximately $z \sim 0$ to $z \simeq 2-3$ and can be effectively described by a power-law relationship, as exemplified, for instance, by the equation presented by \citet{koyama2013}:

\begin{equation} \label{eq:koyama}
sSFR(z) [yr^{-1}] = 10^{-10} \times (1 + z)^{3}, 
\end{equation}

As a result, galaxies with sSFR higher than $sSFR(z)$ are categorized as SF galaxies, whereas those with sSFR lower than $sSFR(z)$ are considered quiescent or transitional. It is important to note that this classification method relies on measurements of both stellar mass and SFR.

\begin{table*}
    \centering
    \caption{Best Linear Fit Results for ln(sSFR) - ln(redshift) Relation}
    \begin{tabular}{cccccc}
        \hline \hline \\
        Sample Name & $\alpha$ & $\beta$ & $\sigma$ \\
        \hline
        COSMOS & $-27.1856 \pm 0.3841$ & $1.1980 \pm 0.4580$ & $4.1860 \pm 0.1972$ \\
        H15 SAM & $-31.8020 \pm 0.0761$ & $4.3307 \pm 0.0717$ & $9.4166 \pm 0.0363$ \\
        Magneticum & $-33.5946 \pm 0.1135$ & $6.7985 \pm 0.1672$ & $10.2581 \pm 0.0505$ \\

        \hline
    \end{tabular}
    \label{tab: ssfr-z}
\end{table*}

\begin{figure}[h!]
     \centering
  \includegraphics[width=0.48\textwidth]{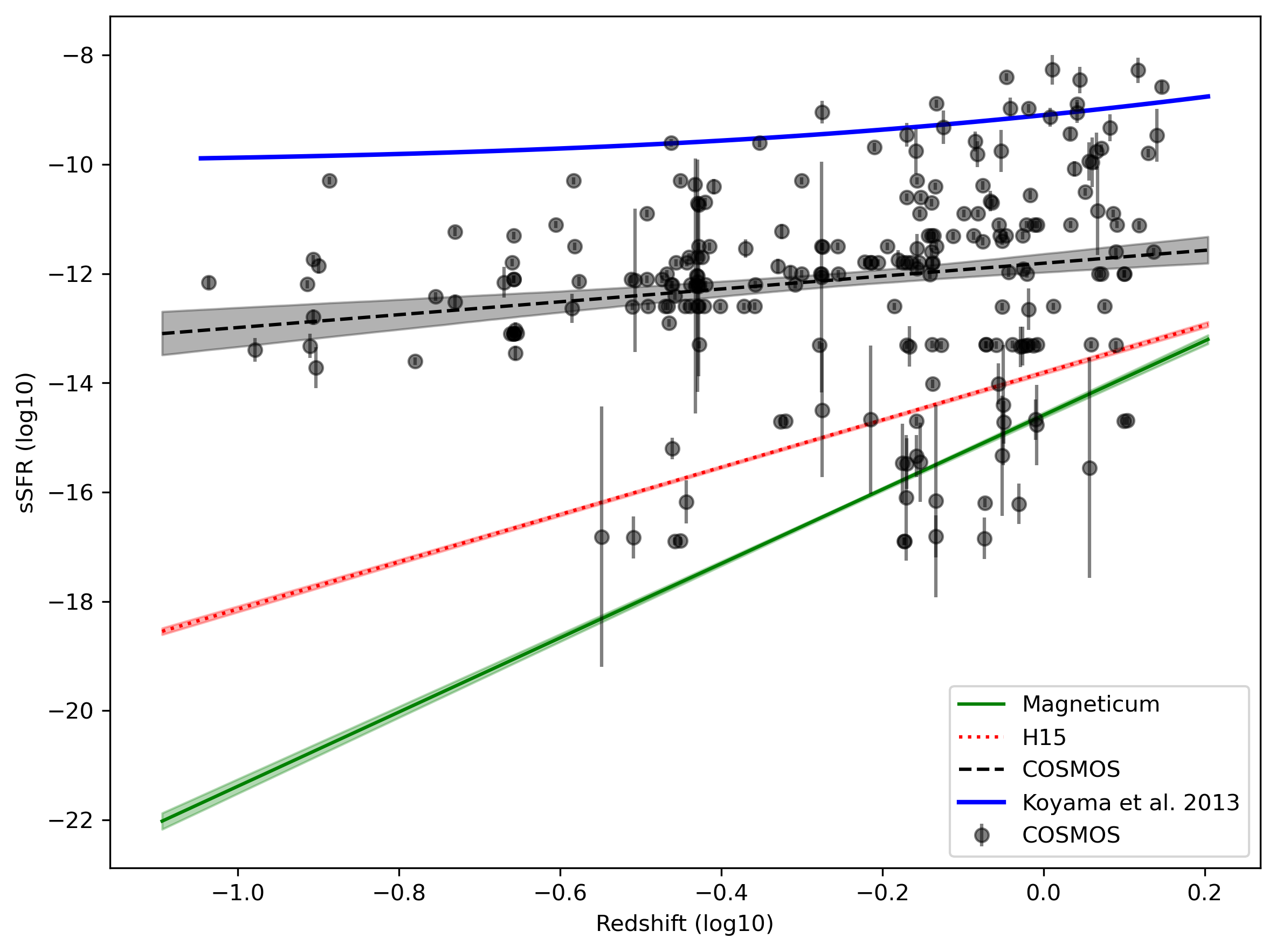}
\caption[]{The specific star formation rate (sSFR) versus redshift for BGGs from COSMOS is shown using black filled circles and a black line, while the H15 SAM is represented with a red line, and the Magneticum simulation is plotted with a green line. The sSFR-redshift relation from \citealp{koyama2013} is illustrated using a blue line.  }
	\label{ssfr_dist}
\end{figure}

In Fig. \ref{ssfr_dist}, the relationship between the sSFR of BGGs and their redshift is illustrated. Equation \ref{eq:koyama} from \cite{koyama2013} is depicted with a blue line. We employ Eq. \ref{eq:linmix_eta} to fit the data from the observations (represented by a black dashed line), the magneticum simulation (shown by a green solid line) and the SAM H15 (indicated by a red dotted line). Table \ref{tab: ssfr-z} provides the best-fit parameters in different datasets.
\cite{koyama2013} investigated how the correlation between galaxy sSFRs has evolved over the past 10 billion years, particularly by examining the influence of environmental factors. This research expanded on the environmental effects on the SFR and stellar mass relationship for star-forming galaxies up to redshift $z \sim 2$, identifying a rapid evolution of specific SFR with increasing redshift. Notably, despite this evolution, the relation between SFR and stellar mass remains consistently independent of the environment, showing a consistently minor difference (around 0.2 dex) between cluster and field star-forming galaxies throughout the cosmic history since $z \sim 2$. 

We applied equation \ref{eq:linmix_eta} to the BGG dataset and found our trend to be consistent with \cite{koyama2013}, despite observing a systematic deviation at given redshifts. We argue that this deviation is too substantial, hence we exclude this relationship when categorizing BGGs as either SF or quenched systems. According to the \cite{koyama2013} relation (equation \ref{eq:koyama}), the sSFR of BGGs is largely quiescent across both low and high redshift ranges, with over $90\%$ of BGGs classified as quiescent, irrespective of redshift. This contrasts with our findings in Paper I, and our results as shown in Fig. \ref{ms} as well as recent studies \citep[e.g.,]{Jung2022MNRAS.515...22J}. Utilizing the COSMOS Web survey data \citep{casey2023}, we are examining the sSFR as a function of redshift, and our preliminary results confirm the systematic deviation between our BGG $sSFR-z$ and that of \cite{koyama2013}.

Furthermore, it is important to note that both the H15 and Magneticum simulations predict steeper trends, typically underestimating the observed sSFR at certain redshifts.

To conclude, as illustrated in Fig. \ref{ssfr_dist}, our results imply that the proportion of quiescent and star-forming BGGs evolves gradually over time. This suggests that the mass of the galaxy might have a greater influence on quenching BGGs compared to environmental factors. 
 \subsection{The stellar age-halo mass relation}
 
 \begin{figure*}[h!]
  \centering
  \includegraphics[width=0.38\textwidth]{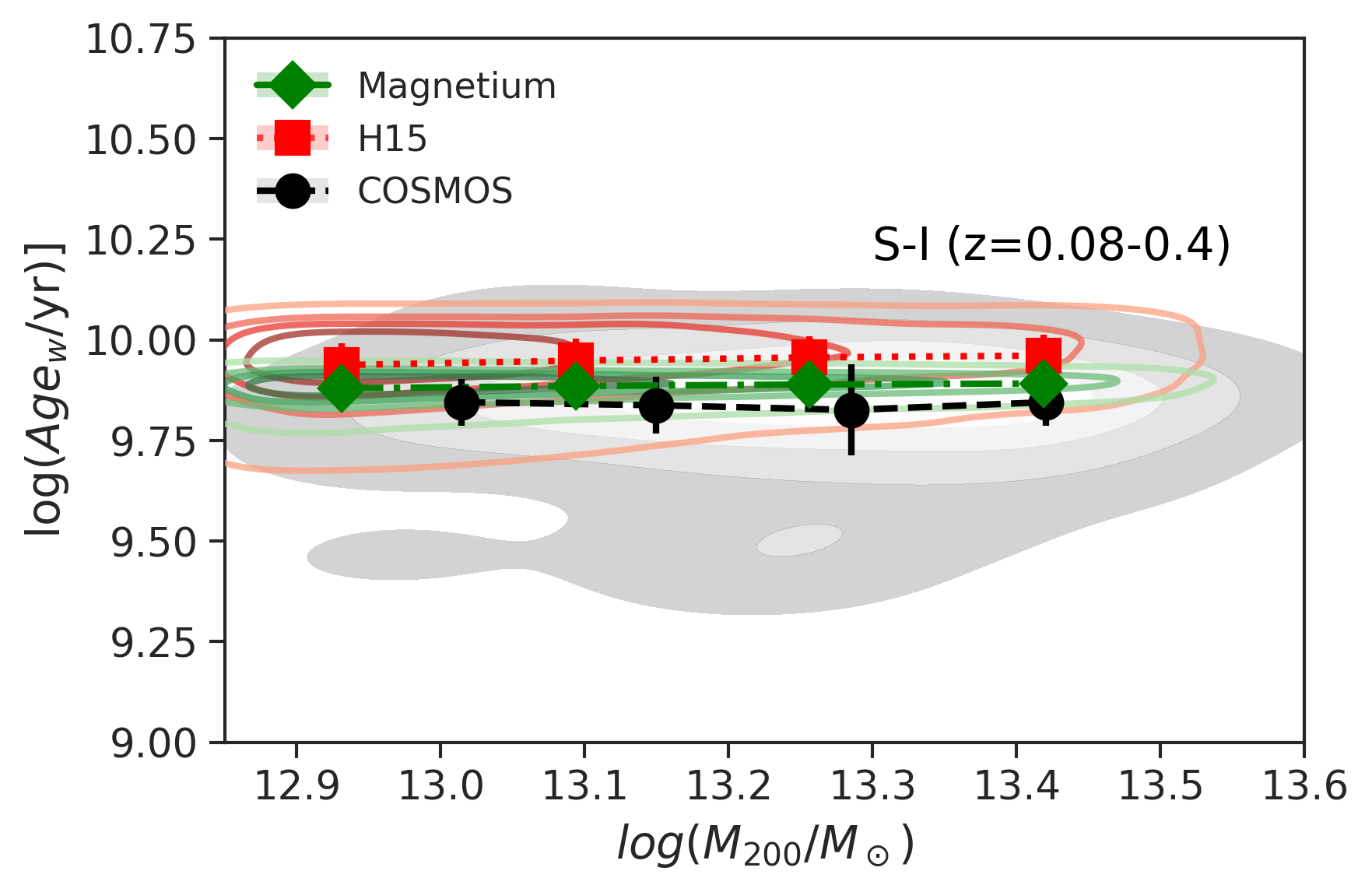}
  \includegraphics[width=0.38\textwidth]{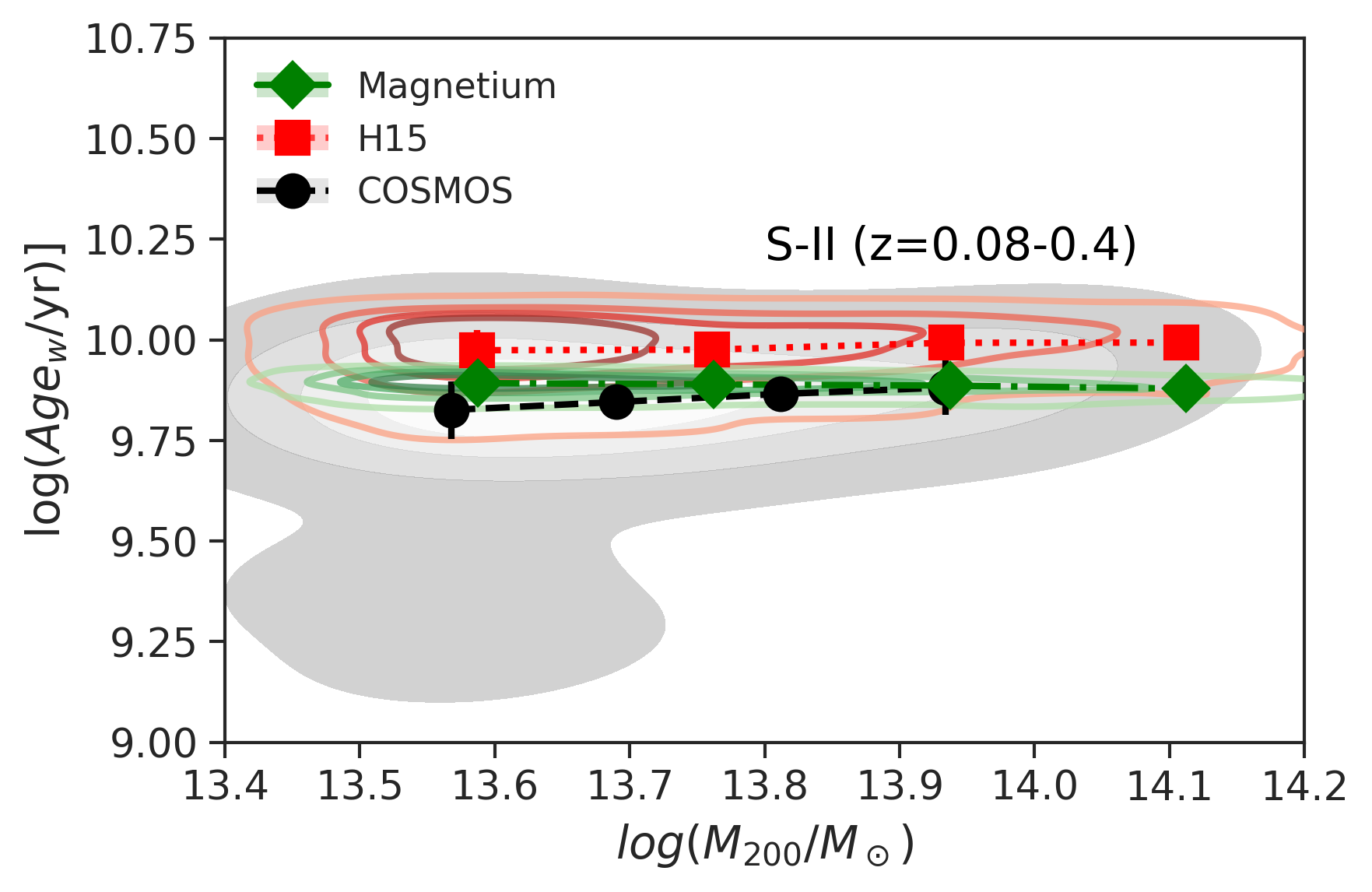}
  \includegraphics[width=0.38\textwidth]{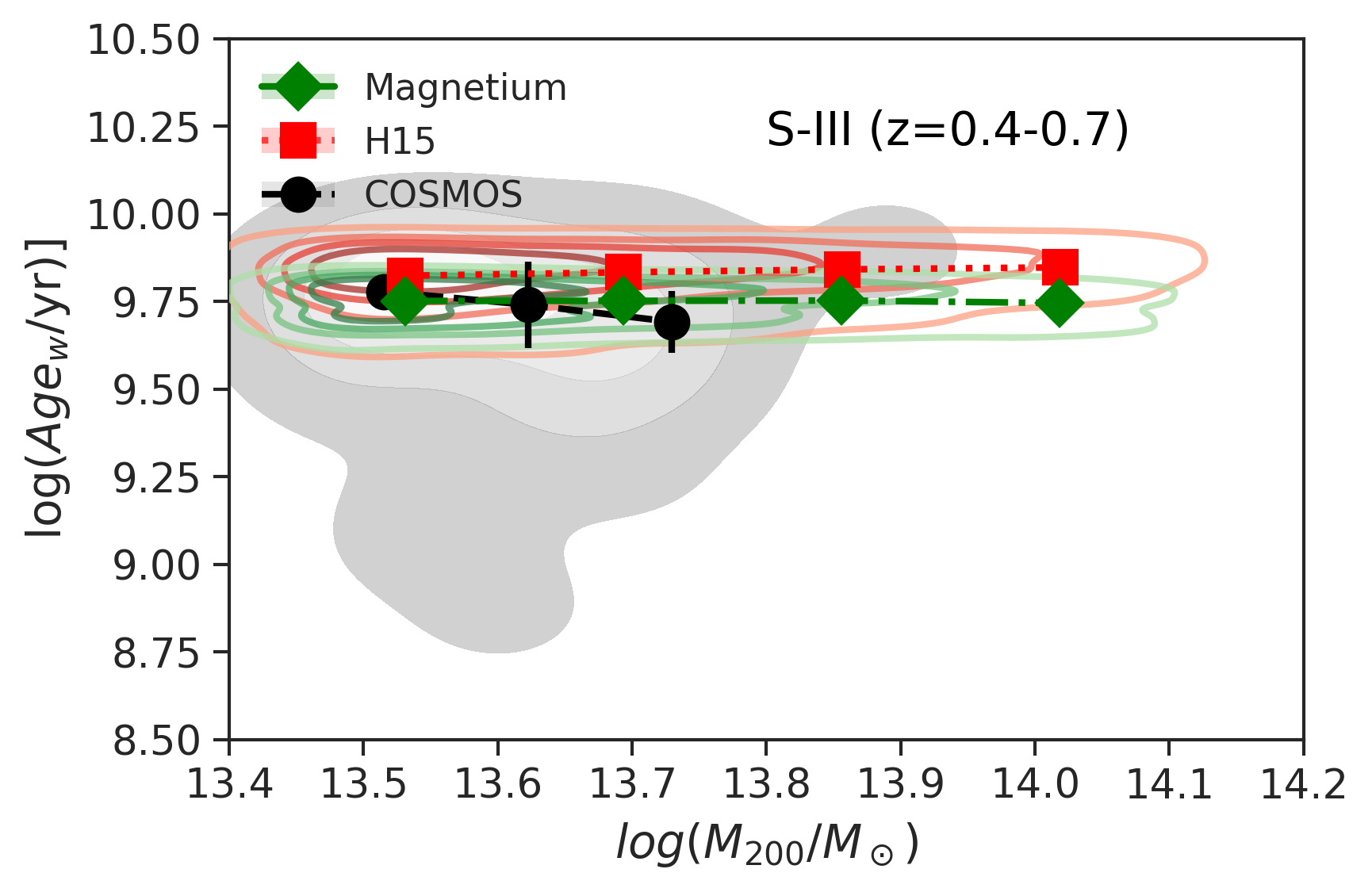}
  \includegraphics[width=0.38\textwidth]{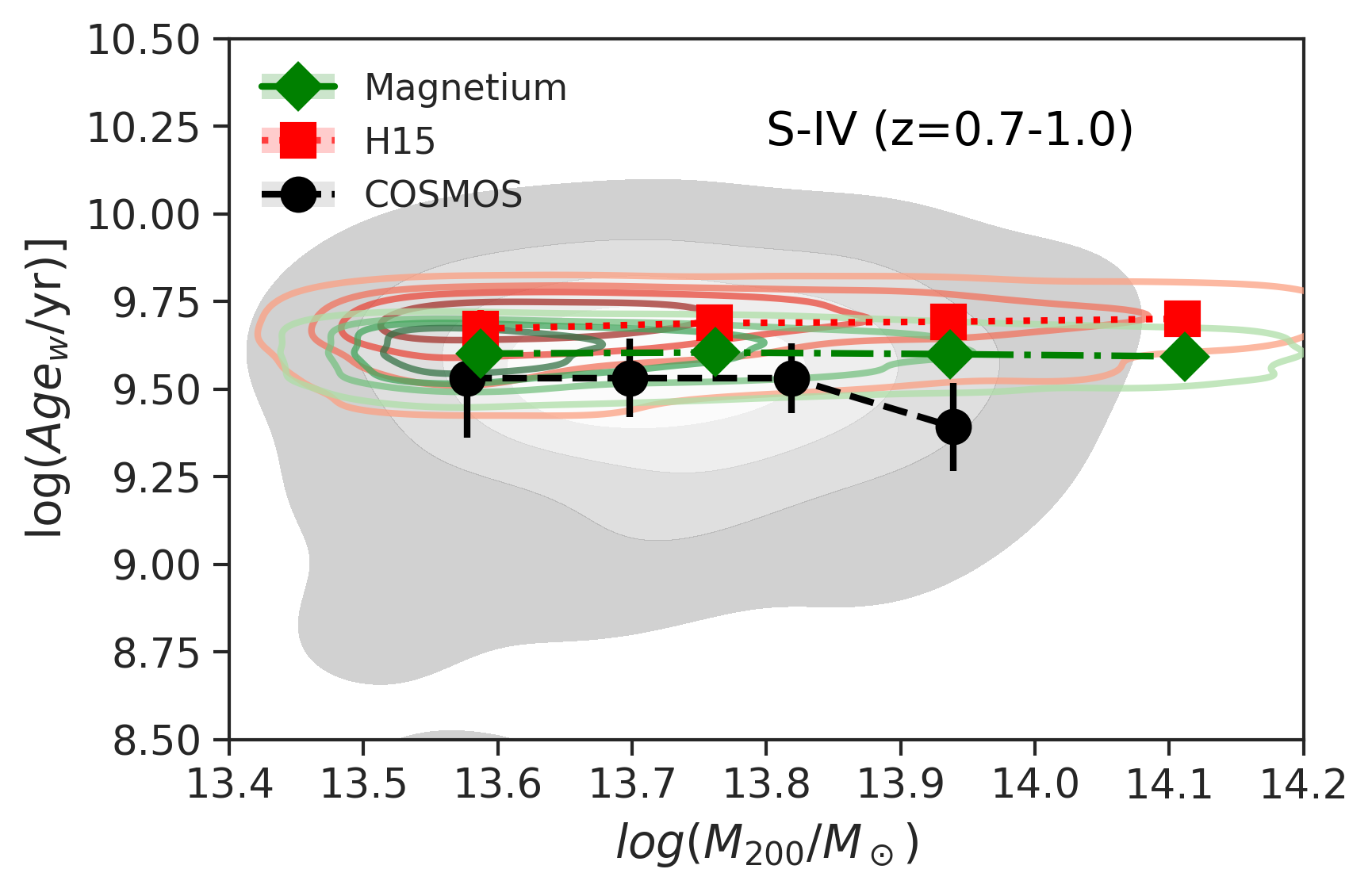}
   \includegraphics[width=0.38\textwidth]{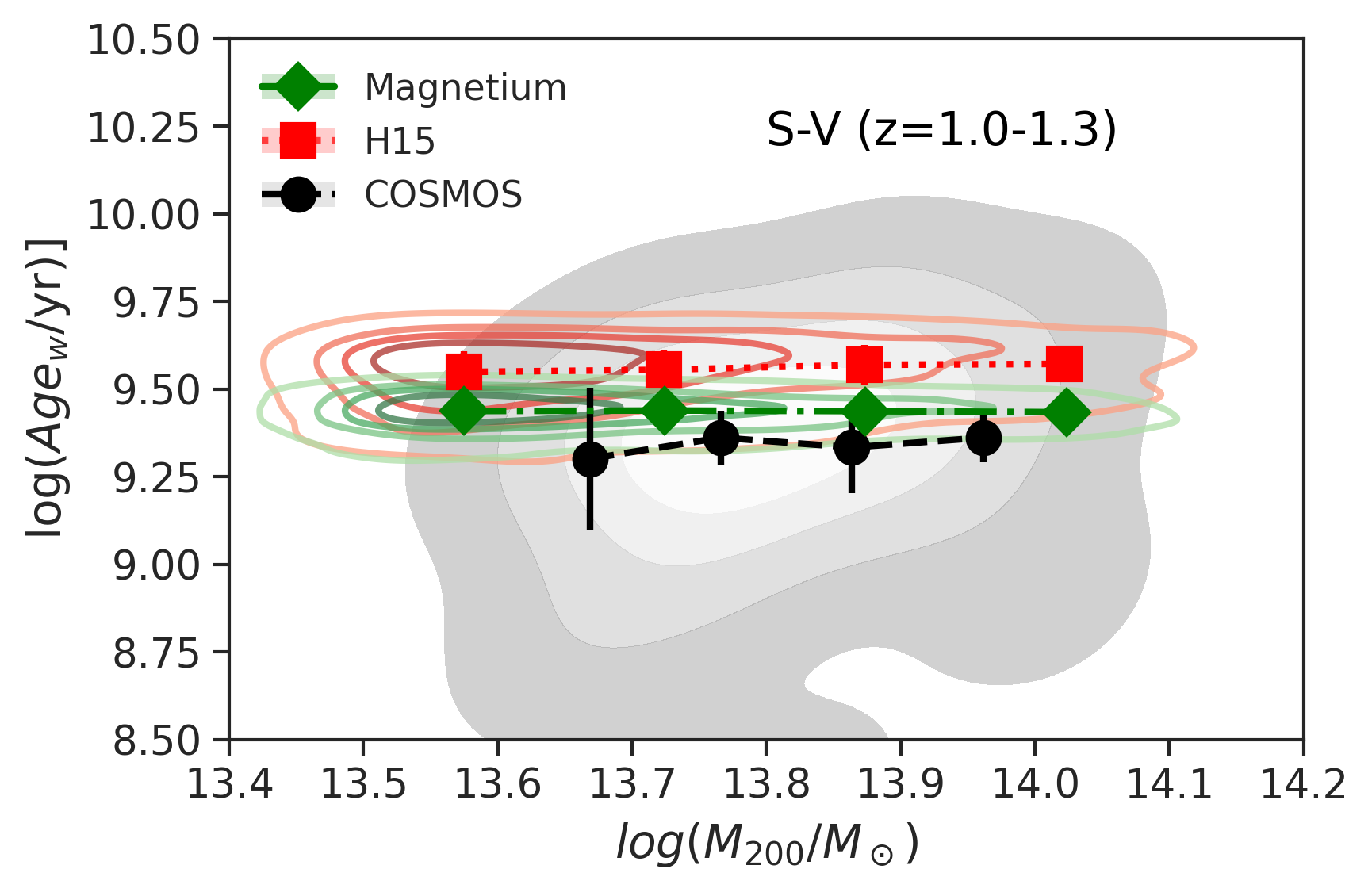}
	\caption[]{The $\log(\mathrm{Age_w/yr})$ of the BGGs as a function of the halo mass of groups  ($\log(\mathrm{M}_{200}/\mathrm{M}_{\odot})$) in the observations (shaded KDE density maps), Magneticum (green contours), and H15 SAM (red contours) for S-I to S-V subsamples. Black points, green diamonds, and red squares present the median stellar age at given halo masses in the observations, Magneticum and H15, respectively. The bottom right panel illustrates the stellar age of BGGe versus group mass for the combined sample of BGGs in S-II to S-V. All trends indicate no noticeable relationship between the stellar age of BGGs and the halo mass of their host groups.  }
	\label{m200_age}
\end{figure*}

Figure \ref{m200_age} shows an analysis of how the halo mass influences the stellar age in BGGs. Observed BGGs are shown with black contours and circles, the H15 study is represented with red contours and squares, and Magneticum simulations are depicted with green contours and diamonds. The data points indicate the median stellar age, with error bars illustrating the absolute median deviation. For BGGs classified within the S-I to S-II groups, there is a slow but gradual positive correlation between stellar age and halo mass. Specifically, the model predictions show a very slight positive correlation, which is consistent regardless of redshift. Essentially, we find no strong dependence of the stellar age of BGGs on their halo mass, albeit with some uncertainty. This finding is supported by \cite{pasquali2010ages}, who similarly identified no substantial relationship between the stellar age of massive central galaxies and the halo mass of their associated groups.
\begin{figure}[h!]
\centering
  \includegraphics[width=0.4\textwidth]{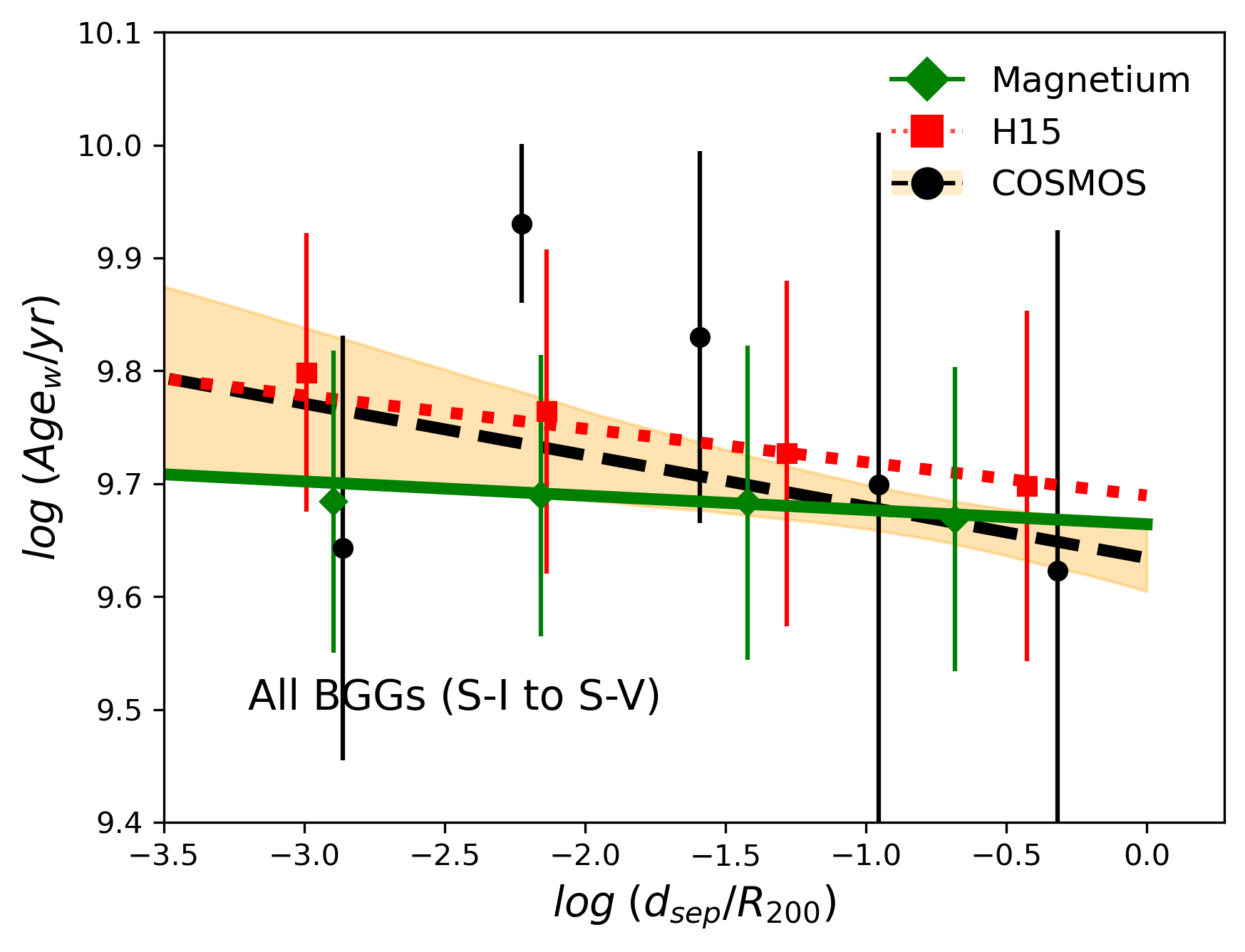}
  \includegraphics[width=0.4\textwidth]{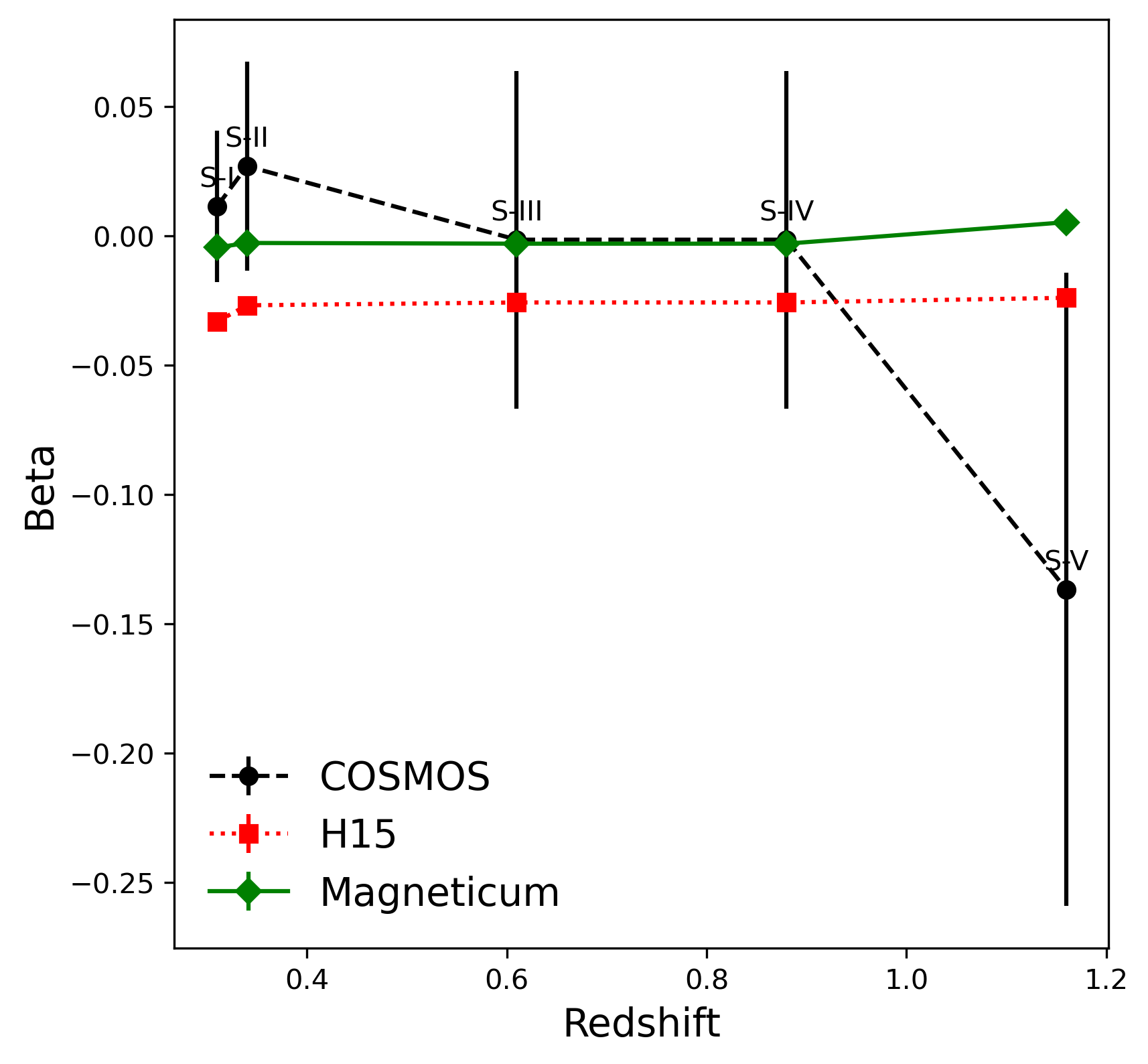}
	\caption[]{The relation between the $log(Age_w/yr)$ as a function of the BGG offset from the group center. The data points represent the median values, and the dashed black, dotted red, and solid green lines illustrate the best linear relation to the data from COSMOS, H15 SAM, and Magneticum simulation, respectively. }
	\label{Fig:age_offset_fullsample}
\end{figure}

\begin{table*}[h!]
    \centering
    \caption{Best linear fit parameters for $\mathrm{ln}(\mathrm{age_w/yr})$ vs. $\mathrm{ln}(\mathrm{d}_{\mathrm{sep}}/\mathrm{R}_{200})$ relationships for BGGs in different subsamples (S-I to S-V).}
    \begin{tabular}{ccccccc}
        \hline
        Sample & Subsample & $\alpha$ & $\beta$ & $\sigma$ \\
        \hline\\
        S-I & COSMOS & $22.727 \pm 0.073$ & $0.011 \pm 0.029$ & $0.570 \pm 0.190$ \\
            & H15 & $22.615 \pm 0.006$ & $-0.033 \pm 0.001$ & $0.381 \pm 0.045$ \\
            & Magneticum & $22.737 \pm 0.003$ & $-0.005 \pm 0.001$ & $0.078 \pm 0.028$ \\
        \hline\\
        S-II & COSMOS & $22.798 \pm 0.131$ & $0.027 \pm 0.040$ & $0.565 \pm 0.247$ \\
            & H15 & $22.753 \pm 0.009$ & $-0.027 \pm 0.001$ & $0.327 \pm 0.053$ \\
            & Magneticum & $22.757 \pm 0.006$ & $-0.003 \pm 0.002$ & $0.055 \pm 0.028$ \\
        \hline\\
        S-III & COSMOS & $22.413 \pm 0.125$ & $-0.001 \pm 0.065$ & $0.686 \pm 0.245$ \\
             & H15 & $22.421 \pm 0.005$ & $-0.026 \pm 0.001$ & $0.354 \pm 0.042$ \\
             & Magneticum & $22.427 \pm 0.005$ & $-0.003 \pm 0.001$ & $0.272 \pm 0.032$ \\
        \hline\\
        S-IV & COSMOS & $22.116 \pm 0.141$ & $0.046 \pm 0.058$ & $0.780 \pm 0.261$ \\
            & H15 & $22.051 \pm 0.008$ & $-0.028 \pm 0.001$ & $0.379 \pm 0.061$ \\
            & Magneticum & $22.105 \pm 0.007$ & $0.002 \pm 0.002$ & $0.306 \pm 0.045$ \\
        \hline\\
        S-V & COSMOS & $21.392 \pm 0.220$ & $-0.137 \pm 0.122$ & $0.891 \pm 0.339$ \\
           & H15 & $21.784 \pm 0.013$ & $-0.024 \pm 0.002$ & $0.409 \pm 0.064$ \\
           & Magneticum & $21.737 \pm 0.012$ & $0.005 \pm 0.004$ & $0.192 \pm 0.071$ \\
        \hline\\
combined (S-I to S-V) & COSMOS & $22.183 \pm 0.069$ & $-0.046 \pm 0.029$ & $0.681 \pm 0.209$ \\
            & H15 & $22.311 \pm 0.005$ & $-0.030 \pm 0.001$ & $0.356 \pm 0.033$ \\
            & Magneticum & $22.252 \pm 0.009$ & $-0.013 \pm 0.003$ & $0.302 \pm 0.032$ \\
        \hline   \\
    \end{tabular}\label{tab:age_offset}
\end{table*}
 \begin{figure}[h!]
\centering
  \includegraphics[width=0.4\textwidth]{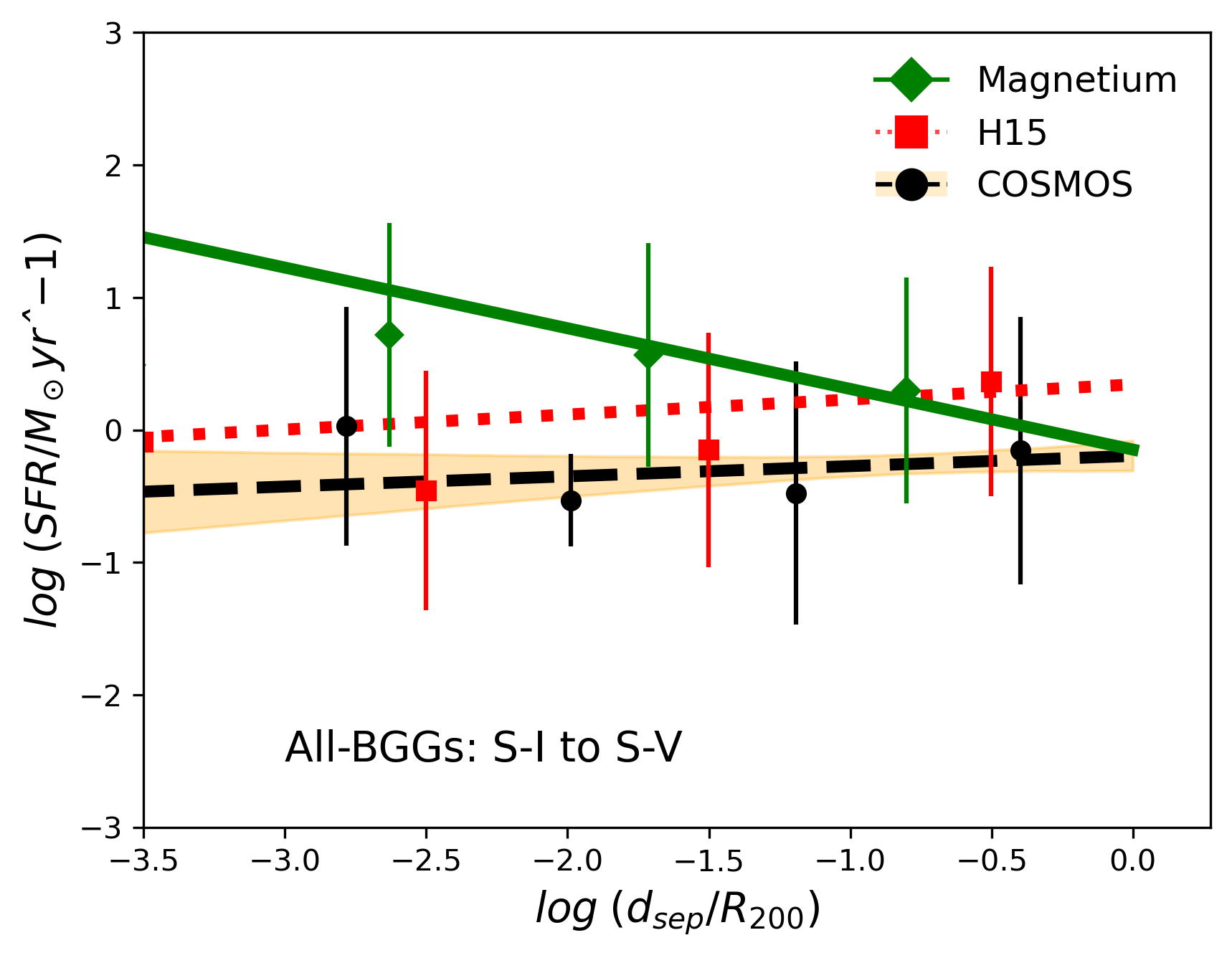}
  \includegraphics[width=0.4\textwidth]{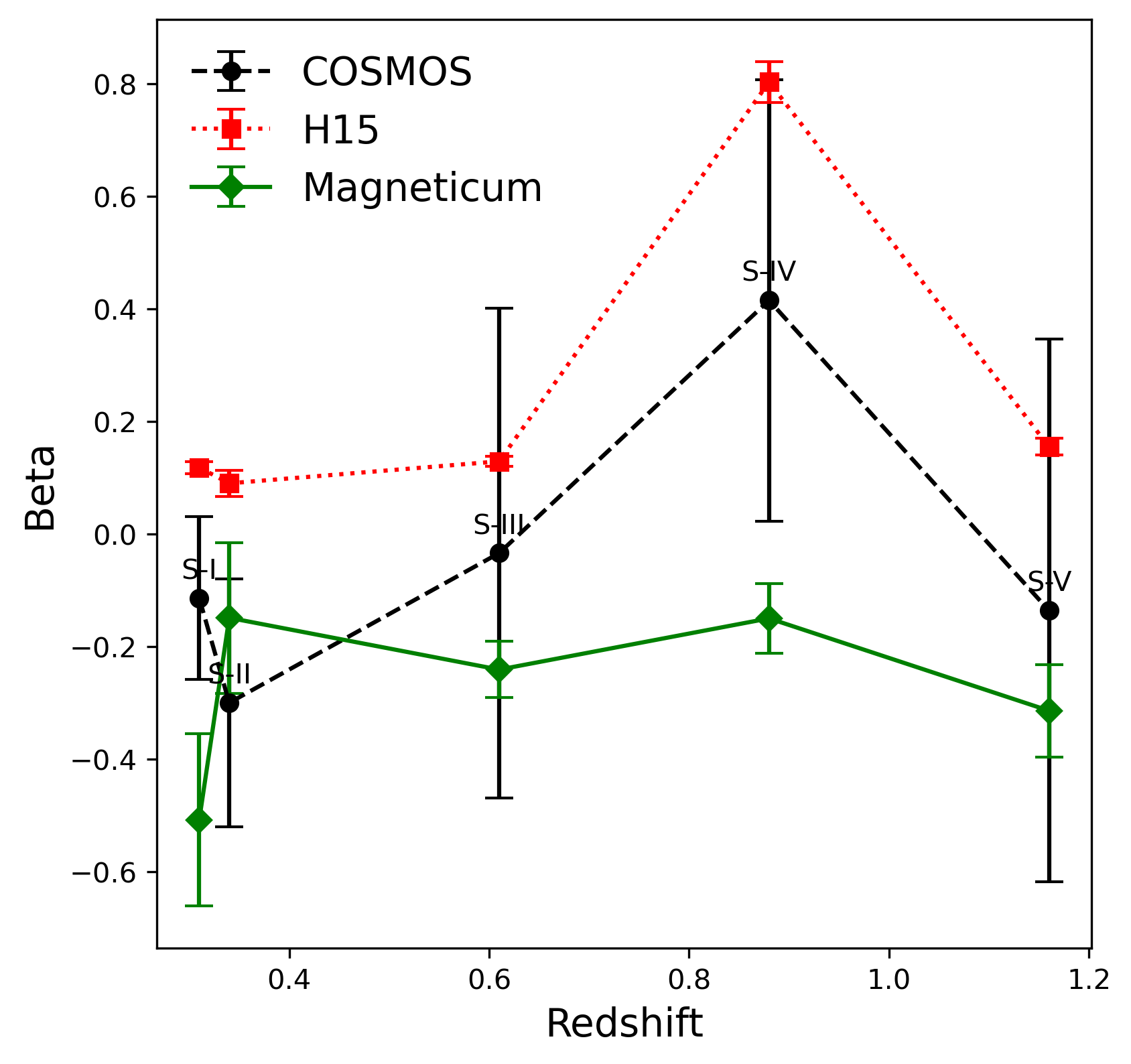}
	\caption[]{The relation between the SFR as a function of the BGG offset from the group center. The data points represent the median values, and the dashed black, dotted red, and solid green lines illustrate the best linear fit to the data from COSMOS, H15 SAM, and Magneticum simulation, respectively. }
	\label{Fig:sfr_offset_fullsample}
\end{figure}

\begin{table*}[h!]
    \centering
    \caption{Best linear fit parameters for $\mathrm{ln}(\mathrm{SFR/M_\odot \mathrm{yr}^{-1}})$ vs. $\mathrm{ln}(\mathrm{d}_{\mathrm{sep}}/\mathrm{R}_{200})$ relationships for BGGs in different subsamples (S-I to S-V).}
    \begin{tabular}{ccccccc}
        \hline
        Sample & Subsample & $\alpha$ & $\beta$ & $\sigma$ \\
        \hline\\
        S-I & COSMOS & $-1.972 \pm 0.356$ & $-0.114 \pm 0.145$ & $1.538 \pm 0.171$ \\
            & H15 & $0.012 \pm 0.085$ & $0.118 \pm 0.011$ & $2.162 \pm 0.032$ \\
            & Magneticum & $-2.088 \pm 0.469$ & $-0.508 \pm 0.153$ & $1.682 \pm 0.067$ \\
        \hline\\
        S-II & COSMOS & $-2.613 \pm 0.726$ & $-0.300 \pm 0.220$ & $1.775 \pm 0.329$ \\
            & H15 & $-0.065 \pm 0.156$ & $0.090 \pm 0.023$ & $2.028 \pm 0.066$ \\
            & Magneticum & $1.258 \pm 0.445$ & $-0.149 \pm 0.134$ & $1.579 \pm 0.067$ \\
        \hline\\
        S-III & COSMOS & $-0.475 \pm 0.706$ & $-0.034 \pm 0.435$ & $1.910 \pm 0.318$ \\
             & H15 & $0.684 \pm 0.067$ & $0.129 \pm 0.009$ & $1.911 \pm 0.025$ \\
             & Magneticum & $1.482 \pm 0.168$ & $-0.241 \pm 0.050$ & $1.458 \pm 0.029$ \\
        \hline\\
        S-IV & COSMOS & $0.299 \pm 0.785$ & $0.415 \pm 0.392$ & $2.524 \pm 0.338$ \\
            & H15 & $1.292 \pm 0.290$ & $0.803 \pm 0.036$ & $6.103 \pm 0.078$ \\
            & Magneticum & $2.564 \pm 0.207$ & $-0.150 \pm 0.062$ & $1.189 \pm 0.035$ \\
        \hline\\
        S-V & COSMOS & $0.632 \pm 0.915$ & $-0.136 \pm 0.482$ & $3.214 \pm 0.503$ \\
           & H15 & $2.142 \pm 0.122$ & $0.155 \pm 0.015$ & $1.875 \pm 0.038$ \\
           & Magneticum & $2.988 \pm 0.281$ & $-0.314 \pm 0.082$ & $0.854 \pm 0.057$ \\
        \hline\\
combined (S-I to S-V) & COSMOS & $-0.451 \pm 0.261$ & $0.077 \pm 0.112$ & $2.321 \pm 0.122$ \\
            & H15 & $0.790 \pm 0.034$ & $0.113 \pm 0.004$ & $2.108 \pm 0.012$ \\
            & Magneticum & $-0.342 \pm 0.082$ & $-0.458 \pm 0.027$ & $1.942 \pm 0.014$ \\
        \hline   \\
        \end{tabular}
\label{tab:sfr-offset}
\end{table*}

\subsection{Variations in stellar ages relative to BGG displacement from X-ray center}
A potential explanation for BGG formation is the "inside-out" model, which posits that star formation occurs later in the galaxy's outskirts compared to its core. Unlike BCGs in densely packed clusters, BGGs tend to show more continuous star formation rather than predominantly showing AGN activity. Additionally, BGGs are frequently found to be displaced from the center of their host dark-matter halo \citep[e.g.,]{Oliva_Altamirano_2014}.

The inside-out formation theory proposes that BGGs expand by steadily accumulating material from their surrounding environment, resulting in a noticeable gradient in both the SFR and stellar age throughout the galaxy.

To verify the validity of this inside-out scenario, we investigated the correlation between the age and SFR of the BGGs and their displacement from the center of the group. According to the predictions of the inside-out model, we expect a specific trend: as the BGG's distance from the group center grows, its SFR should increase, while the galaxy's age should concurrently decrease.

\subsubsection{The Age - offset relation}
To assess the connection between the stellar age of BGGs and their displacement from group centers, we combine all sub-samples spanning different redshifts and halo masses. Figure \ref{Fig:age_offset_fullsample} shows $\mathrm{log(Age_w)}$ versus the displacement of BGG from the center of the group, merging the subsamples (SI to SV) from COSMOS, H15, and Magneticum. Each point indicates the median value, with dashed black, dotted red, and solid green lines representing the best-fit linear relationships for COSMOS, H15 SAM, and the Magneticum simulation, respectively. In COSMOS, the group center is identified by the system's X-ray peak position. The trends indicate an inverse relationship between the BGG's age and its distance from the group center. Taking into account the observed uncertainties, the trend observed is in agreement with the modeled predictions. 

We have analyzed the relationship between the BGG offset from the group center and $\mathrm{log(Age_w)}$ for all sub-samples (S-I to S-V) and present the best-fit parameters in Table \ref{tab:age_offset}. The lower panel of Fig. \ref{Fig:age_offset_fullsample} illustrates the slopes as a function of the redshift for S-I to S-V. The only notable difference is observed for S-V, which is our highest redshift subsample; however, the uncertainty is significantly large.

\subsubsection{The SFR-offset relation}
We have illustrated the relationship between the offset BGG from the center of the group and ($\mathrm{SFR}$) for the combined subsamples (S-I to S-V) in Fig. \ref{Fig:sfr_offset_fullsample}. The relationship for individual sub-samples is determined. The lower panel of Fig. \ref{Fig:sfr_offset_fullsample} summarizes the slopes against the redshift from S-I to S-V. The \texttt{linmix} method provides the optimal fit parameters for both the observed and simulated relationships, considering individual and combined subsamples, as detailed in the tab. \ref{tab:sfr-offset}. Regarding the SFR-offset relationship, COSMOS and H15 BGGs show a slightly positive slope, whereas BGGs from the Magneticum simulations show a negative slope.

The age of BGGs shows a slightly negative relationship with their distance from the halo's center. This implies that BGGs nearer to the halo's core are generally older, while those further out are younger. This pattern could indicate continued or delayed star formation at greater distances, leading to younger stellar populations. Additionally, the negative slope suggests that the group / cluster environment influences the ages of BGG. Denser regions near the X-ray peak might have experienced early, rapid star formation, whereas outer regions assembled more slowly.

The conclusions of this research are largely in line with the findings of \cite{Edwards_2019}. This study explored the formation and development of nearby BCGs through integral spectroscopy performed on 23 BCGs. Interestingly, the X-ray luminosity of most of the hosting halos in their data set matches the traits noted in our examined groups. By investigating from the core to the intracluster light (ICL), the research indicates that BCG cores developed rapidly and early, whereas the outer regions and ICL formed later through minor mergers. The ICL shows younger ages and lower metallicities in contrast to the BCG cores. In most cases, the velocity dispersion profiles indicate either an increasing or a flat pattern. The outcomes align with the notion that recent star formation in BCGs is a secondary aspect associated with the cool core condition of the host cluster.
\section{Discussion: Review of BGG Evolution} \label{discussion}
\subsection{Stellar Properties of BGGs: Insights from Paper I and II}
Estimating stellar age and mass is a task fraught with inherent uncertainties. In this study, we employ the SED fitting technique to determine galaxies' stellar mass and age of galaxies. Although alternative methods are acknowledged to have been developed to obtain these properties with increased precision \citep{gallazzi2005ages}, the SED fitting offers a practical and cost-effective approach to the analysis of large galaxy samples. For example, it is a valuable tool for investigating the evolution of mean or mean stellar properties \citep{walcher2011fitting}. To fully understand the estimation process, we direct the reader to the details in \cite[Paper I]{gozaliasl2016brightest} and \cite{ilbert2010galaxy,ilbert2013mass}.

\begin{figure*}
\centering
  \includegraphics[width=0.75\textwidth]{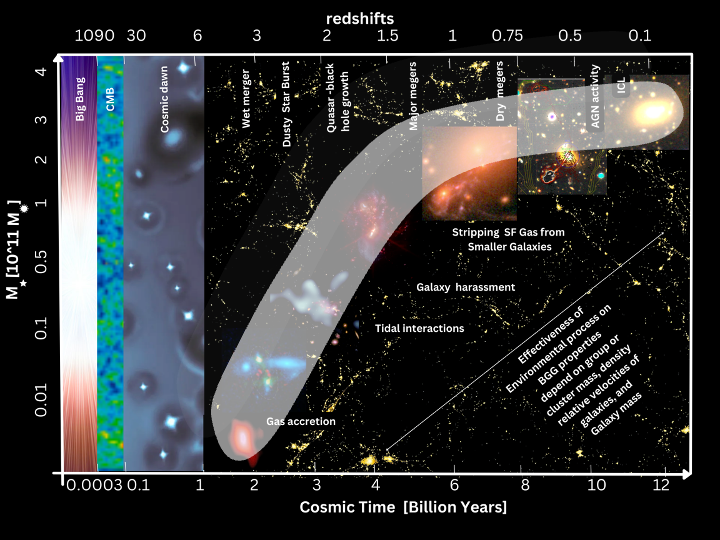}
  \includegraphics[width=0.75\textwidth]{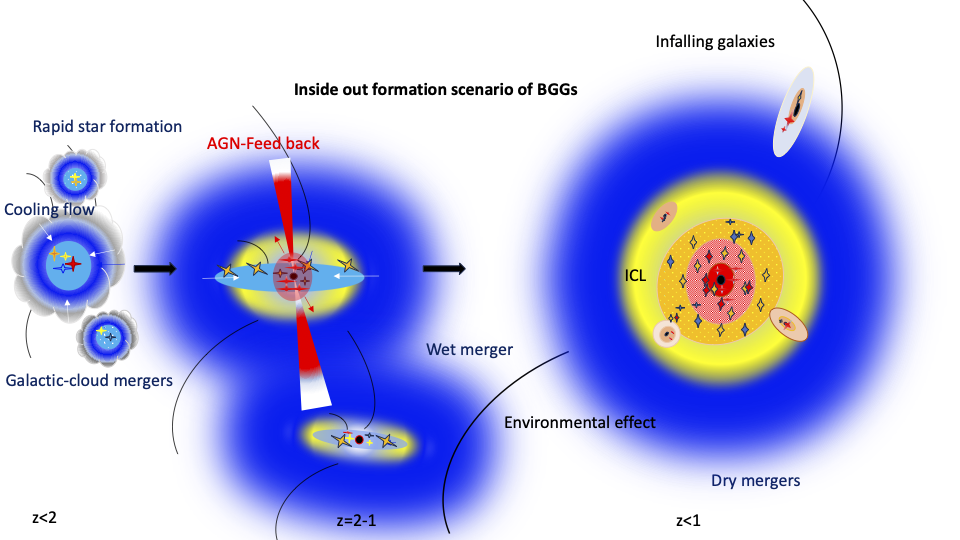}
	\caption{An artistic portrayal capturing the intricate processes contributing to mass assembly and shaping the growth and evolution of BGGs. Understanding the evolution of massive galaxies critically hinges on discerning the nuanced interplay of external and internal physical processes over cosmic time. The lower panel synthesizes the evolutionary trajectory of BGGs, drawing insights from both observational data and simulations. Image credits: ESA/Christophe Carreau, NASA/JPL-Caltech/UC Irvine/STScI/Keck/NRAO/SAO. (Sources: \citealp{Ding2023Natur.621...51D, feldmann2010, Di2023Natur.615..809D, Kelly2015Sci...347.1123K, gozaliasl2016brightest,gozaliasl2018brightest})}
	\label{BGG}
\end{figure*}

This study is part of a comprehensive series of investigations of stellar properties of BGGs. It is the third in the series, following our previous two studies: \cite[Paper I]{gozaliasl2016brightest} and \cite[Paper II]{gozaliasl2018brightest}. 
 In Paper I, we investigated the distribution and evolution of the stellar mass, sSFR, and the MS of BGGs ($SFR-M_*$). In agreement with earlier studies \citep[e.g.,]{lin2013stellar}, we observed a significant growth in the stellar mass of BGGs, indicating an increase by a factor of approximately 2  between redshifts 1.3 and 0.1. Figure \ref{BGG} provides an artistic representation of BGG growth and the principal processes driving their formation. We utilized results from various comprehensive observational and simulation studies \citep{Ding2023Natur.621...51D, feldmann2010, Di2023Natur.615..809D, Kelly2015Sci...347.1123K, gozaliasl2016brightest}, with a specific emphasis on the stabilization observed in stellar mass growth rates below $z=0.4$. Moreover, our study detected a varied set of BGGs, with more than $20\%$ exhibiting indications of star formation, as supported by the recent research conducted by \cite{Jung2022MNRAS.515...22J}.

Building on these results, Paper II expanded the investigation to include an in-depth analysis of the star formation rate, stellar mass, and the relative contributions of BGG to the baryon budget in $f^{BGG}_{b,200}$. We discover that $f^{BGG}_{b,200}$ increases 2.5 times over $z=0.08-1.3$. This research also includes the impact of halo mass growth on the stellar mass assembly of BGGs, demonstrating that this factor is more impactful than in situ star formation in BGG growth. The identification of two separate groups of quiescent and star-forming BGGs revealed the dual nature of their evolutionary trajectories. Additionally, the relationship between BGG mass and halo mass highlighted the hierarchical development of central galaxies along with their host halos. 

 The findings detailed in this study emphasize the dynamic evolution of BGGs and the intricate interplay of various physical processes over cosmic epochs. Notably, our detection of a trend towards younger and intermediate stellar ages within BGG age distributions indicates the presence of star-forming or late-type central galaxies \citep[see Fig. A.7][]{Gozaliasl2014gap}, thereby challenging the traditional view of BGGs as inactive systems.
\subsection{Influence of Environment on BGG Evolution}
A significant element of the evolution of BGG that stands out in our research is the importance of environmental factors. Our prior studies \citep{gozaliasl2016brightest,gozaliasl2018brightest} suggested that the group environment impacts star formation activity in BGGs, with certain low-mass groups acting as intermediaries between low-density areas and rich galaxy clusters. According to \cite{gozaliasl2020}, our research introduces a temporal aspect, showing that BGGs in low-mass groups demonstrate unrestricted dynamic behavior even at the core of the group. 

% HYPERREF
Upon exploring the relationship between stellar age and SFR, we found that BGGs with higher SFR tend to have younger stellar populations, which confirms previous research (e.g. \citealt{trager2000stellar,kuntschner2002early,thomas2010environment}). Interestingly, examining the relationship between age and halo mass reveals nuanced weak correlations between stellar age and halo mass. 
Likewise, \cite{Loubser_2009} investigated possible correlations between derived parameters and the internal characteristics of galaxies (such as velocity dispersion, rotation, luminosity) and those of host clusters (including density, mass, distance from BCG to X-ray peak, presence of cooling flows) to discern whether BCG properties are more influenced by their internal features or by those of host clusters. The findings point to a minimal dependence of single-SSP parameters on the mass or luminosity of galaxies, as well as on the mass or density of the host systems.
Consistent with the findings of \cite{Edwards_2019}, our results also partially support the "inside-out" formation model for BGGs. This model suggests that these galaxies typically form their cores early, while the outer regions and the ICL assemble more recently, potentially as a result of minor mergers and subsequent star formation.

\citet{Einasto2023arXiv231101868E} examined SDSS DR10 data to investigate group galaxies and BGGs within the cosmic web. Their results showed a division between high-luminosity, rich groups/clusters and low-luminosity, poor groups, affecting the fraction of red and quenched BGGs. The study underscored the unique influence of superclusters on group evolution. In \citep{Darragh2019MNRAS.489.5695D}, we also showed that greater cosmic connectivity correlates with increased group mass, altering BGG characteristics and implying a relationship between connectivity and the BGG quenching.
 
Galaxy groups and clusters evolve through galaxy infall and mergers \citep{2009ApJ...690.1292B, 2009MNRAS.400..937M, 2018MNRAS.477.4931H, 2020MNRAS.498.3852B}. Galaxy evolution involves stellar mass growth, stochastic SFR fluctuations, and eventual quenching, influenced by internal and external processes \citep{2006PASP..118..517B, 2020A&A...633A..70P, Tacchella_2020, Patel_2023}. The stellar mass grows through mergers and gas infall, with the mass quenching determined by the dark-matter halo mass. Internal processes include gas ejection through stellar winds, supernovae, and AGN feedback \citep{2006MNRAS.372..265M, 2006MNRAS.365...11C, 2019MNRAS.485.3446H, Vulcani_2021}, more effective in massive galaxies like BGGs \citep{2020ApJ...889..156C} and higher redshifts. External processes, driven by environmental quenching within groups or clusters, involve gas removal due to ram pressure \citep{1972ApJ...176....1G, 2019MNRAS.483.1042Y, Kolcu_2022, Herzog_2022, 2023arXiv230907037Z}, viscous stripping \citep{1982MNRAS.198.1007N}, starvation, cosmic web removal \citep{2019OJAp....2E...7A, 2019A&A...621A.131M, Winkel_2021, Herzog_2022} and harassment \citep{1996Natur.379..613M}. The effectiveness of these processes on the assembly of the galaxy depends on the density of the environment, the mass of the galaxy, and the orbital properties, with the greatest impact on less massive galaxies.

Single massive galaxies with no massive pairs are more influenced by their dark-matter halos and the intergalactic medium, with processes such as gas inflow disruption leading to star formation quenching \citep{Haywood:2016un, Bell:2017vy, Di-Matteo:2019wt, 2022A&A...666A.170Q, 2016A&A...588A..79L, 2022ApJ...927..124M, 2019OJAp....2E...7A}. BGGs in low-mass groups experience altered interactions due to close neighboring galaxies, affecting their star formation through mergers or interactions \citep{Jung2022MNRAS.515...22J, 2021A&A...651A..56G, 2020MNRAS.493.4950S}. Star-forming BGGs can also obtain gas from the surrounding medium or stripped satellites, and the cessation of gas supply leads to cosmic web detachment \citep{2019OJAp....2E...7A,Jung2022MNRAS.515...22J}. In very poor groups, increased gas supply and interactions with nearby galaxies may trigger star formation and subsequent quenching \citep{1982ApJ...255..382H, 2018MNRAS.481.2458D,Einasto2023arXiv231101868E}. The local environment may play an important role in the formation of BGG in the faintest groups \citep{Einasto2023arXiv231101868E}. Differences in the local and global environments of red and blue spirals suggest that the local environment may play a more significant role than the entire group in the shaping of their BGGs \citep{2022JCAP...03..024S}. The accelerated evolution of groups and single galaxies in superclusters may be linked to a larger amount of gas in filaments in globally high-density regions near clusters \citep{2021A&A...646A.156T}. The findings emphasize the crucial role of high-density environments in the evolution of BGGs and their hosting groups.
\subsection{Predictions from Simulations}

\citet{Jung2022MNRAS.515...22J} and \citet{2023MNRAS.525.5677S} analyzed the properties of central galaxies of group-sized haloes from Romulus simulations and the circumgalactic medium around them. \citet{2023MNRAS.525.5677S} suggested that the presence of cold gas and gaseous disks, rejuvenations, and ongoing star formation in BGGs of groups indicates that BGGs receive an inflow of gas from their surroundings. Gas flows onto the central BGG via filamentary cooling flows and infalling cold gas. They describe two pathways by which the gas surrounding the BGGs cools: filamentary cooling inflows and condensations forming from rapidly cooling density perturbations, mainly seeded by orbiting substructures. In nearby groups, AGN feedback, stripping, and both gas-rich and gas-poor mergers in the history of BGGs are also crucial \citep{2017MNRAS.472.1482O, 2018A&A...618A.126O, 2021IAUS..359..180K, Loubser2022}. The significance of accretion in the evolution of BGGs in groups is noted in \citet{gozaliasl2020}. 

The Horizon Run 5 simulation suggests that in the densest regions of the cosmic web, where present-day clusters form, galaxies may have initiated their formation earlier than in other regions \citep{2022ApJ...937...15P, 2023arXiv230411911P}. This idea finds support in observations of extremely high redshift galaxies within forming protoclusters \citep{Hashimoto_2023, 2023Natur.616..266L}. Additionally, \citet{2023MNRAS.523.3201D} proposes that the massive, very early ($z \sim 10$) progenitor galaxies in these environments could form through feedback-free starbursts, with respect to stellar feedback.  \cite{Rennehan2024manhattanarxiv} proposes a similar picture, except that the feedback-free fashion is with respect to AGN feedback, and at the galaxy group and cluster scale at $z \sim 3$.  At that scale, and in a strong matter overdensity, the free fall times are so short that AGN feedback cannot overcome the cooling flow leading to extremely high star formation rates.

The star formation quenching process is a primary driver of the aging of galaxies, leading to an increase in its stellar age.  The results of the TNG and SIMBA simulations on the cluster scale indicates $91\%$ of the BCGs are quenched \citep{Oppenheimer2021Univ....7..209O}. At group scales, the quenched fraction drops more in simulations (e.g., in Illustris, C-Eagle, Eagle-Ref, Romulus-C, TNG300) than in observations. The differences between simulations can be attributed to the varying models for the interaction between AGN feedback and cooling procedures. However, we find that in the H15 SAM and Magneticum simulation, the fraction of quenched BGGs is more than observed, and all BGGs are approximately quenched systems when applying the sSFR threshold by \cite{koyama2013}. Using ROMULUS simulations, \citet{Jung2022MNRAS.515...22J} stated that efficient gas cooling from the CGM is an essential prerequisite for star formation in galaxy groups, while ram pressure gas stripping from gas-rich satellites significantly supports suitable gas cooling flows. Galaxies in groups are often preprocessed (quenched) before joining clusters, indicating a large-scale environmental dependence on galaxy properties \citep{Oppenheimer2021Univ....7..209O, 2022A&A...668A..69E}. The distinctions between the processes in the rich and poor groups were also emphasized by \citet{Jung2022MNRAS.515...22J}. In rich clusters, BCGs lie in the center of the cluster. Their evolution is affected by mergers, galactic cannibalism, dynamic friction, and other processes \citep[see, e.g.,][for a detailed discussion on the formation of BGGs of rich clusters]{Marini2021MNRAS.507.5780M}. In poor groups, the galaxy velocities are low and the mergers of galaxy and strong tidal interactions are effective \citep{Jung2022MNRAS.515...22J}.

The phase before in-fall also holds significant importance. A higher environmental density is associated with a more pronounced decline in star formation. Specifically, group centrals experience a more substantial decline compared to group satellites, which, in turn, undergo more pronounced quenching than isolated galaxies before in-fall. Moreover, there is an observable trend in which high stellar mass galaxies tend to undergo stronger quenching before infall than their low stellar mass counterparts \citep{lotz2019MNRAS.488.5370L}. In fact, \cite{lotz2019MNRAS.488.5370L} also presents three fundamental conclusions in this context. First, they highlight that the primary quenching mechanism in galaxy clusters is ram pressure stripping. Secondly, ram-pressure stripping proves to be sufficiently effective, quenching the majority of star-forming satellite galaxies within approximately 1 Gyr during their initial passage. Third, the quenching effect of ram-pressure stripping is observed to preferentially target radial star-forming satellite galaxies.

The two recent studies \cite{Rhea-Silvia2023arXiv231016089R} and \cite{Spzila2024arxiv} analyze a distinctive population of massive quiescent galaxies beyond the redshift z = 3. Characterized by early quenching at z = 6, these galaxies present a significant puzzle in our understanding of early cosmic galaxy formation (see e.g. \citealt{deGraaff2024arxiv, Setton2024arxiv}).  They both replicate the properties of these massive quenched galaxies at high redshifts, mirroring observed number densities. Intriguingly, their simulations revealed that by z = 2, a notable fraction undergoes accretion or rejuvenation processes, challenging the conventional ideas of galaxy evolution. Contrary to expectations, these massive quenched galaxies predominantly inhabit side nodes of approximately the Milky Way halo mass, not the most massive cosmic web nodes. Furthermore, the study in \cite{Rennehan2024manhattanarxiv} shows that a similar process occurs at the higher mass end of our observed sample; rare fluctuations (at a given mass scale) lead to accelerated galaxy evolution, forming massive (sometimes quenched) galaxies quickly in the early Universe.  That echoes the theory put forth in \cite{Rennehan2020MNRAS.493.4607R}, who show that downsizing in the cores of galaxy clusters leads to rapid BCG formation.  They further extrapolate their theory that, at every mass scale (including groups), there is a continuous distribution of core formation epochs, throughout cosmic time leading to a distribution of star formation rates across all massive galaxies.   These discoveries highlight the intricate nature of their formation and evolution, challenging the prevailing paradigms of massive galaxy assembly in the early Universe.
\subsection{Inside-out formation process and assembly scenario for BGG}
In the lower panel of Figure \ref{BGG}, we synthesize the evolutionary trajectory of the BGGs, drawing insights from both observational data and simulations \citet{Ding2023Natur.621...51D, feldmann2010, Di2023Natur.615..809D, Kelly2015Sci...347.1123K, gozaliasl2016brightest,gozaliasl2018brightest}. BGGs initiate their formation at the density peak of hosting halos, with approximately 50\% of their stars generated by rapid cooling, star formation, and merging with galactic clouds by redshift z=2 as predicted by SAM. Although the majority of BGGs transform into typical elliptical galaxies, low-mass groups may harbor red spiral BGGs that do not undergo a major merger. The growth of the central black hole marks the onset of star formation quenching. Any residual star formation beyond this point is sustained by major mergers and the extraction of gas and stars from neighboring galaxies, particularly evident by z=1.0. Concurrently, intragroup light undergoes significant evolution, with outer envelopes exhibiting younger and intermediate stars, a result of stripping and formation from disrupted neighboring satellites. 

In line with the findings of \citep[e.g.,]{Edwards_2019}, our results (Figures \ref{Fig:age_offset_fullsample} and \ref{Fig:sfr_offset_fullsample}) suggest an inside-out formation and assembly scenario for BGGFig.. Between redshifts z=1.0 and z=0.5, environmental quenching becomes notably more effective, leading to a substantial decline in star formation during this period. Dry mergers and connectivity to local filaments facilitate the late-stage growth of BGGs.

\section{Conclusions and summary}

Following our previous studies exploring the evolution of the stellar properties of BGGs and their connection to hosting halos \citep[][Paper I and II]{gozaliasl2016brightest,gozaliasl2018brightest}, this paper represents the third study focusing on the evolution of stellar ages in BGGs over 9 billion years ($z=0.08-1.30$). Our investigation is based on a carefully selected sample of 246 BGGs obtained from our X-ray galaxy groups catalog \citep{gozaliasl2019}, with halo masses ranging from $10^{12.8}$ to $10^{14}M_{\odot}$ within a 2-square-degree area in the COSMOS field.

We have comprehensively investigated the connections between stellar age, stellar mass, star SFR, and halo mass. We analyze the relationships $SFR- M_*$ and $sSFR-z$. Our observational results are compared with the predictions of two different models: H15 SAM based on the Millennium Simulation and Magneticum hydrodynamical simulations. Our primary observations and conclusions are as follows.
Our results demonstrate a robust stellar age determination, particularly in the context of massive galaxies ($M_*> 10^{10}M_\odot$) using the Le phare SED fitting code. \sloppy The reliability and cost-effectiveness of age determination using this software, using photometry and SED fitting techniques, are analogous to the confidence placed in photometric redshifts. As large-scale surveys continue to advance, the potential for substantial statistical improvements in accuracy is highly promising. 

Stellar age distributions in the observations consistently skew toward younger and intermediate ages. Although no significant distinctions are observed in age distributions between BGGs with varying offsets from the group centers in observations and the magneticum simulation, a contrast is observed in H15 SAM, where offset BGGs tend to be younger than their central counterparts. The normalized cumulative distribution of stellar ages across different panels illustrates that observed BGGs are generally younger, Magneticum's BGGs exhibit intermediate ages, and H15 features the oldest BGGs.

Disparities between model predictions and observed data are apparent in two specific cases: the subsample of BGGs within low-mass and low-redshift groups (S-I) and the highest-redshift massive groups (S-V). However, for the remaining subsamples (S-II to S-IV), the models consistently correspond to the observed stellar ages of the BGGs. Notably, BGGs in low-mass groups serve as a potential link between low-density environments and rich galaxy clusters, indicating a reduced influence of environmental factors on BGGs in low-mass groups compared to those in highly massive halos. This hypothesis finds support in our previous research, such as our examination of BGG dynamics in low-mass groups, where we illustrated the unrestrained behavior of the dynamics of these systems even when located in the center of the group \citep{gozaliasl2020}.

Our KDE density map of BGG stellar ages as a function of cosmic time and redshift, shown in the lower right panel of Fig. \ref{Fig:age-dist-z}, displays a significant overlap between modeled and observed BGG ages over approximately the last nine billion years of the Universe's age. It is important to note that there is an evident scatter in the observed BGG ages at a specific cosmic time, which we attribute, in part, to measurement uncertainties. With the advent of forthcoming surveys such as Euclid and 4MOST, we anticipate opportunities to optimize our measurements and refine our methods, potentially reducing the observed age scatter and enhancing the accuracy of our age determinations.

Analysis of age-mass relationships among BGGs reveals intriguing insights into their evolution. The observed trends, particularly the moderate to mild positive relationships in the lower redshift sub-sample \(z<0.4\) and the trend becomes stronger with increasing redshift. The models encounter difficulties in precisely predicting the behavior of BGGs within the stellar mass range of $10^{10-11} M_\odot$, which requires adjustments. This concern has also been highlighted by \cite{kukstas2023}.   

The analysis of the stellar age and SFR relationship across different subsamples (S-I to S-V) shows a negative correlation, which aligns with prior research, indicating that galaxies with more star formation tend to host younger stellar populations. The comparison with H15 and Magneticum simulations showcases the models' ability to capture these trends, with Magneticum closely aligning with COSMOS observations. This trend becomes more pronounced at higher SFRs in observations, emphasizing the strengthening connection between stellar age and star-formation activity.  \sloppy
Turning to the smoothed distribution of $\log(\mathrm{SFR/M}_{\odot}\mathrm{yr}^{-1})$ in Figure \ref{sfr_dist_fig}, we explore the SFR distributions of the BGGs.  At $z<0.4$, the normalized CDF for the SFR highlights that approximately 60\% of the BGG exhibit a zero SFR, indicating full quenching, and H15 and Magneticum model this behavior. 

 We employ the relation $SFR-M_*$ at different cosmic time by \cite{popesso2023main} and explore the location of BGGs with respect to MS of star-forming galaxies to identify quiescent and SF types and find that about $20\%$ of the BGGs are classified as normal star-forming systems, and this fraction increases to $50\%$ at $z=1.0-1.3$. In particular, the H15 models anticipate a significantly larger fraction of quiescent BGGs in all subsamples, but they more closely align with observations compared to the magneticum simulation. Both models do not accurately capture the S-I star-forming BGG fraction. The Magneticum simulation predicts a number of star-forming BGGs in massive groups that is at least four times as high as observed at redshifts $z = 0.08-1.0$. Furthermore, we notice that the sSFR of BGGs changes at a pace comparable to that \cite{koyama2013} reported for all group galaxies. However, we contend that the relation \cite{koyama2013} generally overestimates sSFR at a given redshift.  Although the models in our research forecast a sharper decline in the sSFR of the BGG, this discrepancy with observational data highlights the need for further investigation. To better understand these patterns, we are conducting this analysis with a more deep dataset from the COSMOS Web survey (Gozaliasl in prep.).

 We investigated the correlation between stellar age and halo mass in BGGs, contrasting observational data with H15 and Magneticum simulations. In agreement with models, the findings indicate that stellar age increases very gradually with halo mass in BGGs. Considering observational uncertainties, we do not detect a significant dependence. 

In conclusion, we argue that the evolutionary trajectory of BGGs differs significantly from that of both individual massive galaxies and BCGs. It appears that environmental effects are not as pronounced as observed in rich clusters. Considering their unique characteristics, it is advisable to regard BGGs as distinct populations that serve as a bridge in the evolutionary spectrum between individual massive galaxies, such as our Milky Way, and highly massive BCGs found in rich clusters.

\begin{acknowledgements}
The authors thank the referee for insightful comments. GG acknowledges support from Prof. Maarit Korpi-Lagg and the Ministry of Education and Culture Global Program pilot project USA (975812001-T31302). GG also acknowledges the contribution of Dr. Francesco Montanari to the project. EV acknowledges support from the Carl Zeiss Stiftung with the project code KODAR. GEM acknowledges the Villum Fonden research grant 13160, Gas to stars, stars to dust: trace star formation across cosmic time grant 37440, The Hidden Cosmos and the Cosmic Dawn of the Danish National Research Foundation through grant DNRF140. We used the data from the Millennium simulation, and the web application providing on-line access to them was constructed as part of the activities of the German Astrophysics Virtual Observatory. KD acknowledges the support for the COMPLEX project of the European Research Council (ERC) under the European Union’s Horizon 2020 research and innovation program grant agreement ERC-2019-AdG 882679. WC is supported by the Atracci\'{o}n de Talento Contract no. 2020-T1 / TIC-19882 was granted by the Comunidad de Madrid in Spain, and the science research grants were from the China Manned Space Project. He also thanks the Ministerio de Ciencia e Innovación (Spain) for financial support under Project grant PID2021-122603NB-C21 and HORIZON EUROPE Marie Sklodowska-Curie Actions for supporting the LACEGAL-III project with grant number 101086388. DR is supported by the Simons Foundation.
\end{acknowledgements}
 
\bibliographystyle{aa}
\bibliography{bibli}

\end{document}